\documentclass[11pt]{article}
\usepackage[top=1in, bottom=0.7in,left=1in, right=1in]{geometry}
\usepackage{amssymb,amsmath}
\usepackage{times}
\usepackage{amsthm}
\usepackage[dvips]{graphicx}
\usepackage{subfigure}
\usepackage{float}
\usepackage{setspace}
\usepackage[centerlast,small,bf]{caption}
\newtheorem{thm}{Theorem}[section]
\newtheorem{lem}[thm]{Lemma}

\begin{document}
\date{}

\title{A simple certifying algorithm for 3-edge-connectivity}

\author{
Yung H. Tsin\footnote{School of Computer Science,  University of Windsor, Windsor, Ontario, Canada, N9B 3P4; peter@uwindsor.ca. Research supported by U of Windsor, GRF grant \#815160.}\\
}

\maketitle \thispagestyle{empty}

\begin{abstract}
A linear-time certifying algorithm for 3-edge-connectivity is presented.
Given a connected undirected graph $G$,
 if $G$ is 3-edge-connected,
 the algorithm generates a construction sequence as a positive certificate for $G$.
Otherwise, the algorithm decomposes $G$ into its 3-edge-connected components and
generates a construction sequence for each of them as well as the bridges and a cactus representation of the cut-pairs in $G$ as negative certificates.
  All of these are done by making only one pass over $G$ using an innovative graph contraction  technique.
Moreover, the graph need not be 2-edge-connected.
The currently best-known algorithm is                  
more complicated as it makes multiple passes           
over $G$ and uses involved reduction and perturbation techniques
rather than just basic graph-theoretic techniques.

\vspace{30pt}

\end{abstract}

\doublespacing

\newpage

\noindent

\section{Introduction}

A major problem in software engineering is correctness of software.
Even after an algorithm is proven correct by its designer,
its implementation as a program may still contain bugs (implementation errors).
This is particularly true for complex algorithms as
their implementations tend to be error-prone.
Consequently, it is difficult to tell if an output generated by a program is correct or has been compromised by a bug.

McConnell et al.~\cite{MMNS11} addressed this problem by introducing certifying algorithms.
   A certifying algorithm is an algorithm that generates
a certificate along with its output for each input.
   The certificate is an evidence that can be
used by an authentication
algorithm to verify the correctness of the output.
An authentication
algorithm is a separate algorithm that takes the input, the output, and the certificate to
verify (independently of the algorithm) whether the output is correct.
     Certifying algorithms
are of great value in practice as the user can verify the correctness of the output
they received for each input
regardless of whether the program is bug-free.
  Certifying algorithms have been extensively used in the software
library LEDA~\cite{MN98} after a user discovered an error in a program for testing graph planarity.
The inclusion of these algorithms greatly improved the reliability of the library.

  A graph recognition problem is a graph-theoretic problem that checks for membership of a class of graphs.
 A certifying algorithm for a graph recognition problem generates a \emph{positive certificate} if the answer is `\emph{yes}' and  a \emph{negative certificate} if the answer is `\emph{no}'.
  Efficient recognition algorithms exist for many graph recognition problems~\cite{CDH13,EMS12,HJ05,KN06,KMMS06,MNS17,NP07,Sch12}.
   However, only a few are certifying.

  In this paper, we address the 3-edge-connectivity problem
which has applications in a wide variety of areas~\cite{Ch14,Con11,Co06,De06,LT13,Pa10}.
    A number of linear-time algorithms had been proposed~\cite{GI91,NI92,TWO92,T07,T09}.
    However, none of them is certifying.
 Recently, Mehlhorn, Neumann and Schmidt~\cite{MNS17} presented a linear-time certifying algorithm for testing 3-edge-connectivity.
   Their algorithm returns a Mader construction sequence as a positive certificate if the graph is 3-edge-connected and  returns a cut-pair as a negative certificate otherwise.
    The key idea underlying their algorithm is as follows.

 Given a 2-edge-connected undirected graph $G=(V,E)$,
   the algorithm first makes one pass over $G$ to decompose $G$ into a collection of chains,
   $C_1, C_2, \ldots, C_{|E|-|V|+1}$,
   which are cycles or paths.
   Then, starting from the graph $G_c$ which is a $K_2^3$-subdivision consisting of $C_1$ and $C_2$,
    the remaining chains are added one at a time to $G_c$ to generate a Mader construction sequence for $G$.
   The chains are classified as \emph{interlacing} or \emph{nested}.
   An interlacing chain can be added immediately to the current $G_c$ if its terminating vertices are in $G_c$.
    Each nested chain is combined with a group of chains to form a segment so that once
the chain is added to the current $G_c$, the remaining chains in the segment become interlacing and
 can also be added.
      The algorithm proceeds in phases. In Phase $i, 1 \le i \le |E|-|V|+1$,
the set of chains whose starting end-vertex is on $C_i$ are determined and all the interlacing chains that can be added into the current $G_c$
    are added.
   For the segments that are determined by nested chains,
    a correct order to add them to the current $G_c$ is to be determined.
%
%
This is achieved by reducing the problem to a problem on intervals based on the notion of overlap graph as follows:

   First  an overlap graph based on the segments is generated.
   Then by associating each segment with a set of intervals, a graph on the intervals is constructed.
   Then,
 using a method of Olariu and Zomaya~\cite{OZ96} and a perturbation technique  for  handling intervals having common endpoint, a spanning forest in the  graph
 is constructed.
    The ordering exists and hence the algorithm can carry on if the spanning forest is a spanning tree.
Otherwise, the algorithm terminates execution and outputs a cut-pair as a negative certificate that the given graph is not 3-edge-connected.
Mehlhom et al. also explained how to extend their algorithm to generate a cactus representation of the cut-pairs if $G$ is not 3-edge-connected.

Although the ideas underlying the algorithm are elegant,
the algorithm is rather complicated owing to the  multiple passes it makes over the input graph and the non-trivial reduction and perturbation techniques it uses.
For instance, the algorithm requires that the input graph is 2-edge-connected.
Therefore, to use their algorithm, an application programmer must first implement a bridge-connected-component algorithm to decompose the input graph into a set of 2-edge-connected components and then runs the certifying algorithm on each of them.
This would increase the workload of the programmer and induce substantial run-time overhead.
The reduction technique also requires data structures to be created during run-time to represent the overlap graphs.
Straightly speaking, Mehlhorn et al. only presented a certifying algorithm for testing if
a 2-edge-connected graph is 3-edge-connected. It produces only a cut-pair if the answer is negative.
Although they did explain how to extend the algorithm to generate a construction sequence for each 3-edge-connected-component and a cactus representation of the cut-pairs, no pseudo-codes are provided. Hence,
it would not be easy for an application programmer to
implement the extension correctly.
For generating a construction sequence for each 3-edge-connected-component, it is likely that they
first implement an existing non-certifying 3-edge-connected-component algorithm to generate  the 3-edge-connected components of the input graph and then uses the certifying   algorithm of Mehlhorn et al. to generate a construction sequence for each 3-edge-connected-component, making yet one more pass over the input graph.





    In this paper, we present a simpler algorithm as an alternative.
   Specifically, we present a linear-time certifying algorithm
that makes only one pass over the input graph $G$ (not necessarily 2-edge-connected) to seamlessly generate
all 3-edge-connected components of the graph each with a Mader construction sequence as well as
 all the bridges and
a cactus representation of the cut-pairs.
 Clearly, if there is only one Mader construction sequence,  $G$ is 3-edge-connected with the construction sequence as a positive certificate;
if there are more than one Mader construction sequences,
 $G$ is not 3-edge-connected and the cactus representation or any cut-pair in it is a negative certificate. Moreover, the Mader construction sequences associated with each 3-edge-connected component serves as a positive certificate for that 3-edge-connected component.
  The algorithm is a simple extension of the 3-edge-connected-component algorithm of Tsin~\cite{T07}.

Since our algorithm is a one-pass algorithm and does not reduce any of the subtasks (generating 3-edge-connected components, generating
Mader construction sequences, generating bridges, generating cactus representation of cut-pairs)
to other computational problems,
the various parts of our algorithm for solving the subtasks thus constitute a cohesive algorithm
that is  self-contained, conceptually simple, and easy to implement.
Hence,
it is more likely used by experimental scientists and application programmers.

\section{Some basic definitions and facts}

The definitions of the graph-theoretic concepts used in this article are standard and
  can be found in many textbooks or references such as~\cite{Ev, Tr}.
  However, to make the article self-contained, we give some of the important definitions below.

An undirected graph is represented by $G=(V,E)$, where $V$ is the vertex set and $E$ is the edge set.
An edge $e$ with $u$ and $v$ as end-vertices is represented by $e=(u,v)$.
The graph may contain parallel edges (two or more edges sharing the same pair of end vertices).
The \emph{degree} of a vertex $u$ in $G$, denoted by $deg_G(u)$, is the number of edges with $u$ as an end-vertex.
A \emph{path} $P$ in $G$ is a sequence of alternating vertices
   and edges,
   $u_0 e_1 u_1 e_2 u_2 \ldots e_k u_k$, such that $u_i \in V, 0 \le i
   \le k,$ $e_i  = (u_{i-1}, u_i), 1 \le i \le k$, where $u_i, 0 \le i
   \le k,$ are distinct with the exception that $u_0$ and $u_k$ may be identical.
    The edges $e_i, 1 \le i \le k,$ could be omitted if no
    confusion could occur as a result.
     The path is a \emph{null path} if $k=0$ and is a \emph{cycle} if  $u_0 = u_k$.
     The path is called an $u_0-u_k$ \emph{path} with vertices $u_0$ and $u_k$ as \emph{terminating vertices}
and $u_i, 1 \le i <k$, as \emph{internal} vertices.
    If the path $P$ is given an orientation from $u_0$ to $u_k$, then $u_0$ is the \emph{source}, denoted by $s(P)$, and $u_k$ is the \emph{sink}, denoted by $t(P)$, of $P$ and the path $P$ is also represented by $u_0 \rightsquigarrow_G u_k$.
         The graph $G$ is \emph{connected} if $\forall u,v \in V$, there is a $u-v$ path
     in it. It is \emph{disconnected} otherwise.
     Let $G = (V,E)$ be a connected graph.
  An edge is a \emph{bridge} in $G$ if removing it from $G$ results in a
  disconnected graph.
   The graph $G$ is \emph{2-edge-connected}  if it has no
   bridge.
   A \emph{cut-pair} of $G$ is a pair of edges whose removal
   results in a disconnected graph and neither is a bridge. A \emph{cut-edge} is an edge in a cut-pair.
     $G$ is \emph{3-edge-connected} if it is bridgeless and has no cut-pair.
     A \emph{3-edge-connected component} (abbreviated $\emph{\textbf{3ecc}}$) of $G$ is a maximal subset $U \subseteq V$ such that $\forall u, v \in U, u \neq v$, there exists three edge-disjoint $u - v$ paths in $G$.
 A graph $G' =(V', E')$ is a \emph{subgraph} of $G$ if $V' \subseteq V$ and $E' \subseteq E$.
  Let $U \subseteq V$, the \emph{subgraph of $G$ induced by $U$}, denoted by $G_{\langle U \rangle}$, is the maximal subgraph of $G$ whose vertex set is $U$.
     Let $D \subseteq E$, $G \setminus D$ denotes the graph resulting from $G$ after
     the edges in $D$ are removed.


  \emph{Depth-first search} (abbreviated \emph{dfs}) augmented with vertex labeling is a powerful graph
  traversal technique first introduced by Tarjan~\cite{Tr}.
     When a \emph{dfs} is performed over a graph, each
     vertex $w$
     is assigned a \emph{depth-first number}, $dfs(w)$, such that
     $dfs(w)=k$ if vertex $w$ is the $k$th vertex visited by the search for the first time.
      The search also partitions the edge set into two types of edges,
      \emph{tree-edge} and \emph{back-edge}
and gives each edge an orientation.
      With the orientation taken into consideration, a tree-edge $e = (u,v)$ is denoted by $u \rightarrow v$ where  $dfs(u) < dfs(v)$
      and a back-edge $e = (u,v)$ is denoted by $v \curvearrowleft u$, where $dfs(v) < dfs(u)$.
      In the former case,  $u$ is the \emph{parent} of
       $v$, denoted by $u=parent(v)$, while $v$ is a \emph{child} of $u$.
      In the latter case, $(v \curvearrowleft u)$ is an \emph{incoming back-edge} of $v$ and
       an \emph{outgoing back-edge} of $u$.
In either case, $u$ is the \emph{tail}, denoted by $s(e)$, while $v$ is the \emph{head}, denoted by $t(e)$, of the edge.
         The tree edges form a directed spanning tree $T = (V, E_T)$ of
$G$ rooted at the vertex $r$ from which the search begins.
          A path from vertex $u$ to vertex $v$ in $T$ is denoted
          by $u \rightsquigarrow_T v$.
   Vertex $u$ is an \emph{ancestor} of vertex $v$, denoted by $u \preceq v$, if and only if $u$ is a vertex on $r \rightsquigarrow_T v$.
   Vertex $u$ is a \emph{proper ancestor} of $v$, denoted by $u \prec v$, if $u \preceq v$ and $u \neq v$.
 Vertex $v$ is a (\emph{proper}) \emph{descendant} of vertex $u$ if and only if vertex $u$ is an  (proper) ancestor of vertex $v$.
   When a depth-first search reaches a vertex $u$, vertex $u$ is
   called the \emph{current vertex} of the search.
    The \emph{subtree} of $T$ \emph{rooted at vertex} $w$, denoted by
    $T_w$, is the subtree containing all the descendants of $w$.


\noindent $\forall w \in V, lowpt(w)  = \min (  \{dfs(w)\} \cup \{ dfs(u) \ | \ (u
\curvearrowleft w) \in E \setminus E_T \}  \cup \{ lowpt(u) \ | \ (w
\rightarrow u) \in  E_T \}  )$.~\cite{Tr}


A \emph{subdivision} is an operation that replaces an edge $(u,v)$ with a
path $uxv$, where $x$ is a new vertex. A \emph{subdivision of a graph} $G$ is a
graph resulting from applying zero or more subdivision operations on $G$.
The graph $K^3_2$ is the graph consisting of two vertices and three parallel edges.

\section{A certifying algorithm for 3-edge-connectivity}

Mehlhorn et al.~\cite{MNS17} showed that a construction sequence, called \emph{Mader construction sequence},
can be used as a positive certificate for 3-edge-connected graphs. It is based on the following generalization of Mader's Theorem~\cite{Ma78}.

\begin{thm}\label{Mander}
 Every subdivision of a non-trivial 3-edge-connected graph (and no other graph) can be constructed from a subdivision of a $K^3_2$ using the following three operations:
    \vspace{-8pt}
 \begin{description}
   \item[$(i)$] adding a path connecting two branch vertices (vertices of degree at least three);
   \vspace{-6pt}
   \item[$(ii)$] adding a path connecting a branch vertex and a non-branch vertex;
   \vspace{-6pt}
   \item[$(iii)$] adding a path connecting two non-branch vertices lying on distinct links
   (maximal paths whose internal vertices are of degree two).

 \end{description}
 \vspace{-5pt}
   In each of the above three cases, the path is called a \emph{Mader path} and its internal vertices are new vertices.
\end{thm}

In this section, we show that the 3-edge-connected-component algorithm of~\cite{T07} can be easily extended to generate, in addition to the 3-edge-connected components, a Mader construction sequence for each of them, and all the bridges as well as a cactus representation of the cut-pairs if the given graph is not 3-edge-connected.
All these are generated seamlessly with one depth-first search over the given graph $G$.

  As with Mehlhorn et al., our Mader paths are ears of an ear decomposition of the 2-edge-connected components of
the input graph. We use a ear decomposition of 2-edge-connected graphs used in~\cite{T11}.
Mehlhorn et al.~\cite{MNS17} uses a decomposition method of Schmidt~\cite{Sch13b}.

    Let $T=(V,E_{T})$  be a depth-first search tree of a 2-edge-connected graph $G$.
    Using the depth-first search numbers of the vertices, we can
     rank the back-edges as follows.

\noindent   \textbf{Definition:}
 Let $(q \curvearrowleft p)$ and $(y \curvearrowleft x)$ be two back-edges.
     Then   $(q \curvearrowleft p)$  is  \emph{lexicographically smaller than} $(y \curvearrowleft x)$, denoted by $(q \curvearrowleft p) \lessdot (y \curvearrowleft x)$,
  if and only if

  (i) $dfs(q)  < dfs(y)$, or

  (ii) $q = y$
        and
          $dfs(p)  < dfs(x)$
        such that  $p$  is not an ancestor of  $x$, or

  (iii) $q = y$  and  $p$  is a descendant of $x$, or

  (iv) $q = y$ and $p = x$ and $(q \curvearrowleft p)$ is encountered before $(y \curvearrowleft x)$
during the $\mathit{dfs}$. (parallel edges)


   Since every tree-edge is the parent edge of a unique vertex $u$,
   every tree-edge can be represented by $(parent(u) \rightarrow u)$.
    Using the back-edges, we can partition the edges of $G$
  into edge-disjoint paths such that
every path contains exactly one back-edge as follows:
   for each tree-edge   $(parent(u) \rightarrow u)$,
 we associate with it the back-edge $(y \curvearrowleft x)$
  with the  lowest rank in lexicographical order such that
        $x$ is a descendent of $u$
  while $y$ is a proper ancestor of $u$.
  The back-edge exists because $G$ is 2-edge-connected.
  It is easily verified that the back-edge  $(y \curvearrowleft x)$ and all the
tree-edges associated with it form a path
 $yxv_{1} \cdots v_{p}v$   such that
$v \rightsquigarrow_T x$. 
  The path is called an \emph{ear} and is denoted by $P_{y \curvearrowleft x}: yxv_{1} \cdots v_{p}v$.
  Note that  $P_{y \curvearrowleft x}$ is given an orientation from $y$ to $v$;
i.e. $P_{y \curvearrowleft x}$ is an $y \rightsquigarrow_G v$ such that $s(P_{y \curvearrowleft x}) = y$ and $t(P_{y \curvearrowleft x}) = v$.
 Furthermore, if $(y \curvearrowleft x)$ has the rank $i$ lexicographically,
we shall also denote the ear by $P_{i}:yxv_{1} \cdots v_{p}v$.
     As a result, the ears can be ranked lexicographically as
     $P_1, P_2, \ldots, P_{m-n+1}$ which is called an \emph{ear-decomposition} of $G$.
   An ear is \emph{non-trivial} if it contains at least one tree-edge and is  \emph{trivial} otherwise.
    Note that for a back-edge $(y \curvearrowleft v)$,
$s(y \curvearrowleft v) = v$ and $t(y \curvearrowleft v) = y$
but when it is considered as a trivial ear $P_{y \curvearrowleft v}$, then $s(P_{y \curvearrowleft v}) = y$ and $t(P_{y \curvearrowleft v}) = v$.

\begin{lem}\label{lem1}
Let $G=(V,E)$ be a 2-edge-connected graph. $G$ has an ear-decomposition,
 $P_1, P_2,\ldots , P_{|E|-|V|+1}$ in which $P_1$ is a cycle and
  $P_i, 2 \le i \le |E|-|V|+1$ is a path or a cycle of which each terminating vertex lies
on some ear $P_j, 1 \le j < i,$ and
no internal vertex lies on any ear $P_j, 1 \le j < i$.
\end{lem}

\noindent \textbf{Proof:}
 Immediate from the definition of $P_{i}$ and the 2-edge-connectivity of $G$.
 $\blacksquare$

Contrary to Mehlhorn et. al, our ear-decomposition is not generated explicitly.
Instead. it is generated by labeling every edge $e \in E$ with the back edge that determines the ear containing e. This back edge, denoted
by $ear(e)$, is determined during the depth-first search based on the following recursive definition:


\vspace{12pt}
\noindent $ear(e) = \left\{
  \begin{array}{ll}
    e & \hbox{if $e \in E \backslash E_T$;} \\
    \min_{\lessdot} (\{ f \mid f =(u \curvearrowleft w) \in E \backslash E_T \} \cup \\ \hspace{0.4in} \{ ear(f) \mid f = (w \rightarrow u) \in E_T\} ), & \hbox{if $e = (parent(w) \rightarrow w) \in E_T$ }
  \end{array}
\right.$

\vspace{9pt}
\noindent  The ears are generated only when the construction sequence is constructed at the end.

 In the sequel, let $e = (v,w) \in E$,
if $e \in E_T$ ($e \in E \setminus E_T$, respectively), we shall use $ear((v,w))$ and $ear(v \rightarrow w)$ ($ear(w \curvearrowleft v)$, respectively) interchangeably.

\subsection{A high-level description}

\hspace{15pt}
Since our algorithm is based on
 \textbf{Algorithm} 3-edge-connectivity of~\cite{T07},
we first briefly review the algorithm.

Starting with the input graph $G = (V, E)$, the graph is
gradually transformed so that vertices that have been confirmed to be belonging to the same $3ecc$ are merged into one vertex, called a \emph{supervertex}.
Each supervertex is represented by a vertex $w \in V$ and a set $\sigma(w) (\subseteq V)$
consisting of vertices that have been confirmed
to be belonging to the same $3ecc$
as $w$. Initially, each vertex $w$ is regarded as a supervertex with $\sigma(w) = \{w\}$.
When two supervertices $w$ and $u$ are known to be belonging to the same
$3ecc$, they are merged into a single supervertex with one of them, say $w$, absorbing the other resulting in $\sigma(w) := \sigma(w) \cup \sigma(u)$.
   When that happens, the edges incident on $u$ become edges incident on $w$ (the latter are called \emph{embodiment}s of the former).
When a supervertex containing all  vertices of a $3ecc$ is formed,
it must be \emph{of degree one
 or two} in the transformed graph (corresponding to a bridge or a cut-pair is found)\footnote{In~\cite{T07}, it is pointed out that \textbf{Algorithm} 3-edge-connectivity can be easily modified to handle non-2-edge-connected graphs.}. When this condition is detected,
the supervertex is separated (\emph{ejected}) from the graph to become an isolated supervertex.
At the end, the graph is transformed into a collection
of isolated supervertices each of which contains the vertices of a distinct $3ecc$ of $G$.

 \begin{figure}
 \centering
 \includegraphics[width=4in]{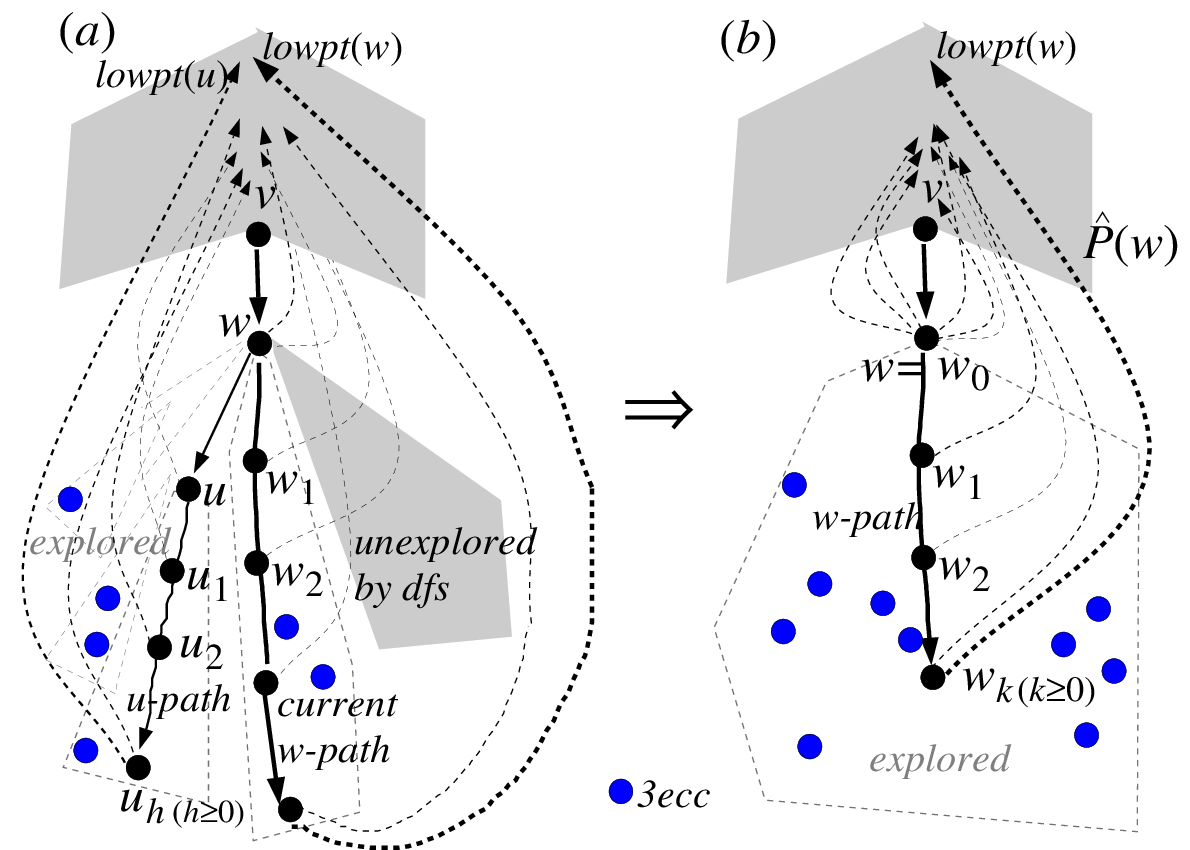}
 \caption[ ]{$(a)$ The current $w$-path and an $u$-path. \ $(b)$ When $\mathit{dfs}$ backtracks from vertex $w$ to vertex $v$.}
 \end{figure}\label{Fig1.eps}

The transformation is carried out by performing a depth-first search over $G$,
starting from an arbitrary vertex $r$.
     At each vertex $w \in V$, when the search backtracks from a child $u$, let $\hat{G}_u$ be the graph to which $G$ has been transformed at that point of time.
The subgraph of $G$ induced by the vertex set of $T_u$, $G_{\langle V_{T_u} \rangle}$, has been transformed into a set of isolated supervertices (each of which corresponds to a distinct $3ecc$ of $G$)
and a path of supervertices, $P_u: (u=)u_0 u_1 u_2 \ldots u_k$, called the $u$-\emph{path}.
 The $u$-path has the properties summarised in the following lemma.

\vspace{-3pt}
\begin{lem} \label{Tsin09}

\emph{[Lemma 6 of~\cite{T09}]} ~
    Let $P_u: (u=)u_0 u_1 u_2 \ldots u_k$  (Figure~1).

 \begin{description}
    \vspace{-6pt}
   \item[$(i)$] $deg_{\hat{G}_u}(u_0) \ge 1$  and $deg_{\hat{G}_u}(u_i) \ge 3, 1 \le i \le k$;

   \vspace{-6pt}
   \item[$(ii)$]  for each back-edge $f = (x \curvearrowleft u_i), 0 \le i \le k,$
      $x \preceq w$ (i.e. $x$ lies on the $r \rightsquigarrow_T w$ tree-path);

    \vspace{-6pt}
   \item[$(iii)$] $\exists (z \curvearrowleft u_k)$ such that $dfs(z) = lowpt(u)$. 
 \end{description}
\end{lem}

\vspace{-6pt}
 If $deg_{\hat{G}_u}(u_0) = 1$, then $(w,u)$ is a bridge, $k = 0$, and $\sigma(u)$ is a $3ecc$ of $G$.
   The supervertex $u$ is ejected from the $u$-path to become an isolated supervertex and the
   $u$-path becomes a null path.

 If $deg_{\hat{G}_u}(u_0) = 2$, then $\{(w \rightarrow u), (u \rightarrow u_1)\}$ or
  $\{(w \rightarrow u), (z \curvearrowleft u)\}$, where $z = t(ear(w \rightarrow u))$, is a cut-pair
 implying $\sigma(u)$ is a $3ecc$ of $G$.
   The supervertex $u$ is ejected from the $u$-path to become an isolated supervertex and the
   $u$-path is shorten to $u_1 u_2 \ldots u_k$ in the former case  or a null path in the latter case.
   Next, if $lowpt(w) \le lowpt(u)$,  then no edge on the $u$-path can be a cut-edge which implies that the vertices in the supervertices on the $u$-path must all belong to the same $3ecc$ as $w$.
 The supervertices are thus absorbed by $w$.
 Likewise, if $lowpt(w) > lowpt(u)$, then the vertices in the supervertices on the current $w$-path must all belong to the same $3ecc$ as $w$ and
 the supervertices are absorbed by $w$; moreover, $lowpt(w)$ is updated to $lowpt(u)$
and the $u$-path becomes the current $w$-path.

   When an outgoing back-edge of $w$, $(z \curvearrowleft w)$, with $dfs(z) < lowpt(w)$
is encountered, vertex $w$ absorbs the current $w$-path because all the supervertices on it belong to the same $3ecc$ as $w$; $lowpt(w)$ and the $w$-path are then updated to $dfs(z)$ and the null path $w$, respectively.

When an incoming back-edge of $w$, $(w \curvearrowleft x)$, is encountered,
let $x \in \sigma(w_{h})$, where $w_{h}$ is a supervertex on the current $w$-path
 $w w_{1} w_{2} \ldots w_k (k \ge h)$,
 then no edge on the path, $w w_1 \ldots w_{h}$, can be a cut-edge. As a result, the vertices in $\sigma(w_i), 1 \le i \le h$,
  must all belong to the same $3ecc$ as $w$.
   The supervertices $w_i, 1 \le i \le {h},$ are thus absorbed by $w$ and
the current $w$-path is shortened to $w w_{h+1} w_{h+2} \ldots w_k$.

When the adjacency list of $w$ is completely processed, if $w \neq r$,
the depth-first search backtracks to the parent vertex of $w$.
Otherwise, the input graph $G$ has been transformed into a collection of isolated supervertices each of which contains the vertices of a
distinct $3ecc$ of $G$.

  Nagamochi et al.~\cite{NI08} pointed out that the graph resulting from $G$
 by contracting every $3ecc$ into a supervertex
 is a cactus representation of the cut-pairs in $G$.
 Hence, the algorithm can be easily extended to generate a cactus representation of $G$ as a byproduct.

It remains to explain how to modify the algorithm so that along with generating the $3ecc$s, it also  generates
a Mader construction sequence for each $3ecc$, the bridges and a cactus representation for the cut-pairs
by making only one $\mathit{dfs}$ over the given graph.
For clarity, we shall consider these two tasks separately.

\subsection{Generating construction sequences (positive certificates)}


Since each $3ecc$ is not a subgraph but a subset of vertices,
before discussing generating a Mader construction sequence for a $3ecc$,
we must address the following question first: ``The Mader construction sequence for a $3ecc$ is generated based on what set of edges?''
   Although the edge set of the subgraph of $G$ induced by $\sigma(w)$, i.e.
$G_{\langle \sigma(w) \rangle}$, appears to be the correct answer,
  unfortunately, it is not as  $G_{\langle \sigma(w) \rangle}$ may not be 3-edge-connected.
 This is because
if two vertices in $\sigma(w)$ are connected by exactly three edge-disjoint paths in $G$ and one of the paths uses edges outside $G_{\langle \sigma(w) \rangle}$ through the cut-pair incident on $G_{\langle \sigma(w) \rangle}$,
there are only two edge-disjoint paths connecting them in $G_{\langle \sigma(w) \rangle}$.
Therefore, the paths that use edges outside $G_{\langle \sigma(w) \rangle}$ must be accounted for
when $\sigma(w)$, hence $G_{\langle \sigma(w) \rangle}$, is separated from $G$.
   It is easily verified that any path connecting two vertices in $G_{\langle \sigma(w) \rangle}$ that uses edges outside $G_{\langle \sigma(w) \rangle}$ must use the cut-pair incident on $G_{\langle \sigma(w) \rangle}$.
   We can thus replace all these paths by a \emph{virtual edge} connecting the two end-vertices
 of the cut-pair in $G_{\langle \sigma(w) \rangle}$~(Figure 2).


 \begin{figure}
\centering
\includegraphics[width=4.5in]{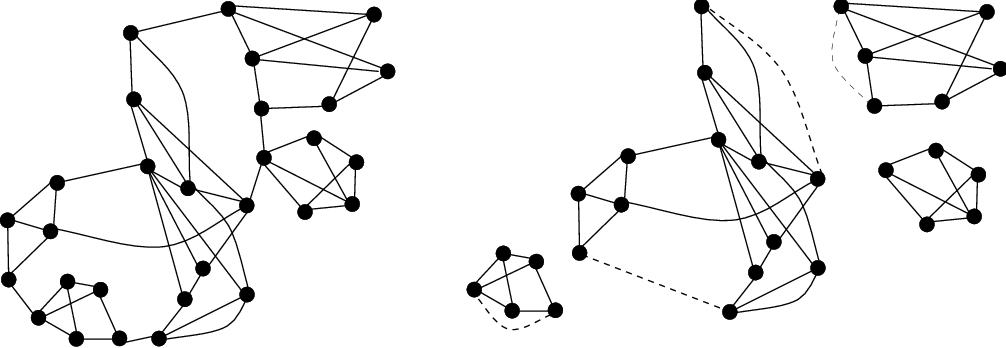}

\caption[ ]{
 A graph and the augmented subgraphs, $\acute{G}_{\langle \sigma(w) \rangle}, w \in V,$ induced by its 3-edge-connected components.
}
\end{figure}\label{Fig2-0.eps}



\begin{lem}\label{CS-for-3ecc}
 Let $\sigma(w)$ be a $3ecc$ of $G=(V,E)$ and $\{(v,w),(\ddot{w},d)\}$ be the cut-pair with $w,\ddot{w} \in \sigma(w)$.

Let
  $\acute{G}_{\langle \sigma(w) \rangle} = \left\{
                                             \begin{array}{ll}
                                               G_{\langle \sigma(w) \rangle}
\text{ and a new edge } (w, \ddot{w}) \notin E, & \hbox{if $w \neq \ddot{w}$;} \\
      G_{\langle \sigma(w) \rangle}, & \hbox{if $w = \ddot{w}$.}
                                             \end{array}
                                           \right.$
   Then, $\acute{G}_{\langle \sigma(w) \rangle}$ is 3-edge-connected.



\end{lem}
\vspace{-6pt}
\noindent \textbf{Proof:}
  Let $u, v \in \sigma(w)$. Then, there exist three edge-disjoint $u-v$ paths in $G$.
If none of the paths uses edges outside $G_{\langle \sigma(w) \rangle}$,
then the three paths are edge-disjoint $u-v$ paths in $G_{\langle \sigma(w) \rangle}$, and hence in $\acute{G}_{\langle \sigma(w) \rangle}$.

Suppose one of the $u-v$ paths uses edges outside $G_{\langle \sigma(w) \rangle}$.
   This path must contain the cut-edges $(v,w)$ and $(\ddot{w},d)$.
   Clearly, the edges on the path that are not in  $G_{\langle \sigma(w) \rangle}$ all lie on the section connecting $w$ and $\ddot{w}$.

If $w = \ddot{w}$, discarding the section results in an $u-v$ path in $G_{\langle \sigma(w) \rangle}$. Hence, there are three edge-disjoint $u-v$ paths in $G_{\langle \sigma(w) \rangle}$, and hence in $\acute{G}_{\langle \sigma(w) \rangle}$.
If $w \neq \ddot{w}$,
replacing the section with the new edge $(w,\ddot{w})$ results in
 a $u-v$ path in $\acute{G}_{\langle \sigma(w) \rangle}$.
 Again, there are three edge-disjoint $u-v$ paths in $\acute{G}_{\langle \sigma(w) \rangle}$

Hence, $\acute{G}_{\langle \sigma(w) \rangle}$ is 3-edge-connected.
\ \ \ \ \  $\blacksquare$

Hence, a Mader construction sequence for $\sigma(w)$ is constructed based on the edge set of $\acute{G}_{\langle \sigma(w) \rangle}$.

Since a Mader construction sequence starts with a $K_2^3$-subdivision,
we consider how to determine the $K_2^3$-subdivision next.

  If $|\sigma(w)| = 1$, no construction sequence is to be constructed.
Let $|\sigma(w)| > 1$.


\vspace{9pt}
\noindent
\textbf{Definition:}  $\forall w \in V$,
let $\hat{\mathcal{P}}_w = \{ P_{ear(e)} \mid e = (w \rightarrow u) \in E_T \vee e = (u \curvearrowleft w) \in E \setminus E_T \}$ and $\hat{P}(w) = \min_{\lessdot} \hat{\mathcal{P}}_w$.
When the $\mathit{dfs}$ backtracks from vertex $w$ to its parent vertex or execution of the $\mathit{dfs}$ terminates at $w (= r)$,
 let $\hat{\delta}(w) = \{ P \mid P \text{ is an ear } \wedge t(P) \in \sigma(w) \}$,
$\delta(w) = \hat{\delta}(w) \setminus \{ \hat{P}(w) \}$ and
$\overleftarrow{P}(w) = \min_{\lessdot}\delta(w)$.

Intuitively, $\hat{\mathcal{P}}_w$ is the set of ears that contain a child edge or an outgoing back-edge of $w$,
$\hat{\delta}(w)$ is the set of ears whose sink is in $\sigma(w)$
(Note that $\sigma(w)$ contains only those vertices that are determined to be 3-edge-connected to $w$ up to the time when the $\mathit{dfs}$ backtracks from $w$).
$\hat{\mathcal{P}}(w)$ is the lexicographically smallest ear in $\hat{\mathcal{P}}_w$.
$\overleftarrow{P}(w)$ is the lexicographically smallest ear in $\hat{\delta}(w)$ excluding the ear $\hat{\mathcal{P}}(w)$.
If $w \neq r$ and the parent edge of $w$ is not a bridge,
$\delta(w) = \hat{\delta}(w)$,
$\hat{P}(w)$ is the ear containing the parent edge of $w$, and
$\overleftarrow{P}(w)$ is the lexicographically smallest ear with the sink in $\sigma(w)$.
Otherwise,
$\delta(w) = \hat{\delta}(w) \setminus \{\hat{P}(w)\}$,
$\hat{P}(w)$ is the lexicographically smallest ear with the sink and source in $\sigma(w)$ (it is thus a cysle), and
$\overleftarrow{P}(w)$ is the lexicographically second smallest ear with the sink in $\sigma(w)$.


Gabow~\cite{Gab00} classified depth-first-search-based graph connectivity algorithms into two types: $lowpoint$-based and path-based. While the algorithm~\cite{T07} based on which  our algorithm is developed is $lowpoint$-based, our certifying algorithm is path-based.
This is because it uses ears (paths) to generate a Mader construction sequence.
 Clearly, $lowpt(w)$ can be redefined in terms of ears.

\begin{lem}\label{ears-and-lowpt}
Let $w \in V$ such that $|\sigma(w)|>1$,
\vspace{-6pt}
\begin{description}
  \item[$(a)$] $dfs( s(\hat{P}(w)) ) = lowpt(w)$; (Figure 1$(b)$)

\vspace{-6pt}
  \item[$(b)$]  
   $t(\overleftarrow{P}(w))$ lies on $\hat{P}(w)$.

\vspace{-6pt}
  \item[$(c)$]   Let $w \neq r$ and $(v \rightarrow w)$ be its parent edge which is not a bridge.
  Then $\hat{P}(w) = P_{ear(v \rightarrow w)}$.
\end{description}
\end{lem}

\noindent \textbf{Proof:}

\vspace{-6pt}
\begin{description}
  \item[$(a)$] By induction on the height of $w$ in the $\mathit{dfs}$ tree.

\vspace{-3pt}
  \item[$(b)$] Suppose  $t(\overleftarrow{P}(w))$ does not lie on $\hat{P}(w)$.
    Let $u = t(\overleftarrow{P}(w))$ and $v \rightarrow u$ be the parent edge of $u$.
Then $P_{ear(v \rightarrow u)} \lessdot \overleftarrow{P}(w)$.
Clearly,  $t(P_{ear(v \rightarrow u)})$ lies on $w \rightsquigarrow_T u$.
As  $u \in \sigma(w) \Rightarrow t(P_{ear(v \rightarrow u)}) \in \sigma(w)$,
 $P_{ear(v \rightarrow u)} \in \delta(w)$ which contradicts that $\overleftarrow{P}(w)$ is the lexicographically smallest  ear in $\delta(w)$.

\vspace{-3pt}
  \item[$(c)$] Immediate from the definitions of $\hat{P}(w)$ and $ear(v \rightarrow w)$.
                                                      \ \ \ \ \ $\blacksquare$
\end{description}

Note that by Lemma~\ref{ears-and-lowpt}$(a)$, the $w$-path is created from $\hat{P}(w)$ through the eject and absorb operation.

 \begin{figure}
 \centering
 \includegraphics[width=6in]{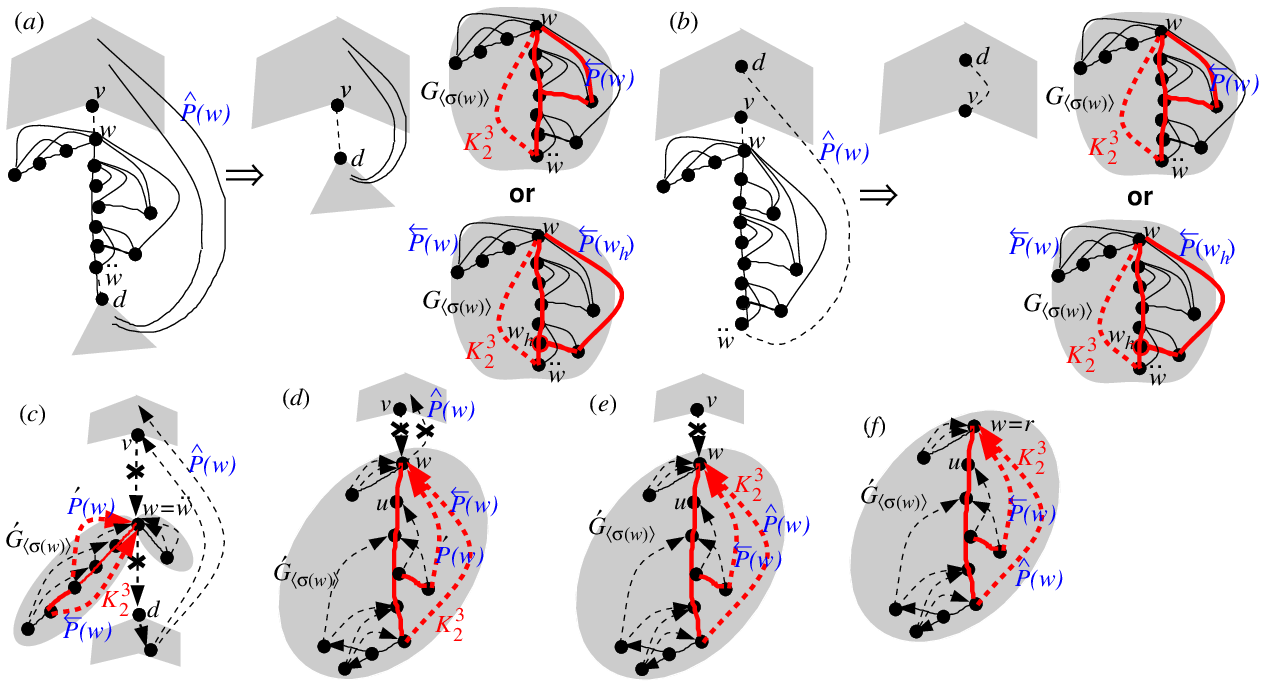}
 \caption[ ]{$(a)$ A cut-pair $\{(v \rightarrow w),(\ddot{w} \rightarrow d)\}, w \neq \ddot{w}$;
    $(b)$ a cut-pair $\{(v \rightarrow w),(d \curvearrowleft \ddot{w})\}, w \neq \ddot{w}$, and\\
\hspace{7pt} $\acute{G}_{\langle \sigma(w) \rangle} =
G_{\langle \sigma(w) \rangle} \cup \{(w,\ddot{w})\}$, where $(w,\ddot{w})$
is a virtual edge represented by a dotted line.
$(c)$ A cut-pair $\{(v \rightarrow w),(\ddot{w} \rightarrow d)\}, w = \ddot{w}$;
    $(d)$ a cut-pair $\{(v \rightarrow w),(d \curvearrowleft \ddot{w})\}, w = \ddot{w}$,
 $(e)$ parent edge of $w$ is a bridge; $(f)$ $w = r$.\\
\hspace{-120pt} In the last four cases,
$\acute{G}_{\langle \sigma(w) \rangle} = G_{\langle \sigma(w) \rangle}$.
The thick lines  form the $K_2^3$-subdivision.  }
 \end{figure}\label{Fig3-0.eps}


The $K_2^3$-subdivision is constructed as follows:

\vspace{-9pt}
 \begin{description}
   \item[$(i)$] Let $\{e,e'\}$ be the cut-pair separating $\sigma(w)$ from the rest of the graph:
 in the course of executing \textbf{Algorithm} 3-edge-connectivity,
 let $e$ become $(v \rightarrow w)$ while  $e'$ become either $(\ddot{w} \rightarrow d)$ or
$(d \curvearrowleft \ddot{w})$  such that $w, \ddot{w} \in \sigma(w)$ and $w \preceq \ddot{w}$.

\vspace{-9pt}
\begin{description}
  \item[$(a)$] $w \neq \ddot{w}$:~(Figure 3$(a),(b)$)
 if $t(\overleftarrow{P}(w)) \neq w$, the $K_2^3$-subdivision consists of the virtual edge $(w,\ddot{w})$,
the tree-path $w \rightsquigarrow_T \ddot{w}$ and $\overleftarrow{P}(w)$;
    if $t(\overleftarrow{P}(w)) = w$, the $K_2^3$-subdivision consists of the virtual edge $(w,\ddot{w})$,
the tree-path $w \rightsquigarrow_T \ddot{w}$ and $\overleftarrow{P}(w_h)$, where
 $w_h$ is the lowest vertex on the $w$-path absorbed by $w$.

  \item[$(b)$] $w = \ddot{w}$:~(Figure 3$(c),(d)$)
   the $K_2^3$-subdivision consists of $\overleftarrow{P}(w)$
(a cycle) and $\acute{P}(w) = \min_{\lessdot} \delta(w) \setminus \{\overleftarrow{P}(w)\}$ which is the lexicographically second smallest ear with sink in $\sigma(w)$.

\end{description}

\vspace{-15pt}
   \item[$(ii)$] Let $(v,w)$ be a bridge:~(Figure 3$(e)$)
     The $K_2^3$-subdivision consists of $\hat{P}(w)$ (a cycle) and $\overleftarrow{P}(w)$.

\vspace{-9pt}
\item[$(iii)$] $w = r$:~(Figure 3$(f)$)
The $K_2^3$-subdivision consists of $\hat{P}(w)$ (a cycle) and $\overleftarrow{P}(w)$.

 \end{description}


 It remains to determine an ordering for the remaining ears in $\acute{G}_{\langle \sigma(w) \rangle}$ so that they can be added to the construction sequence as mader paths after the $K_2^3$-subdivision. The key idea is to maintain the following invariant at every vertex during the $\mathit{dfs}$.


\noindent \textbf{Definition:}
   Let $\delta$ be a set of ears. A \emph{construction sequence for} $\delta$, denoted by $\mathcal{CS}_{\delta}$, is an
ordering, $\pi$, of the ears in $\delta$
   such that once the first ear in $\pi$ is added to the construction sequence under construction,
   the remaining ears can be added to the construction sequence as Mader paths by following their order in $\pi$.

\begin{description}
  \item[Invariant] ``When the $\mathit{dfs}$ backtracks from a vertex $w (\neq r)$ to its parent vertex,
   a construction sequence, $\mathcal{CS}_{\delta(w)}$, has been determined with $\overleftarrow{P}(w)$ as the first element.''

   Note that in this context, $\sigma(w)$ consists of all the vertices that have been absorbed by $w$ so far; it becomes a $3ecc$ only after $w$ is ejected.
  We call $\overleftarrow{P}(w)$ the \emph{anchor} of $\delta(w)$.
\end{description}


To generate the construction sequences for the $3ecc$s based on the above discussion,
 we modify \textbf{Algorithm} 3-edge-connectivity of~\cite{T07} as follows.
 Note that we only outline the key idea. The correctness will be proven later on.

\vspace{-9pt}
\begin{description}

 \item[$(a)$] $w$ is a leaf in $T$:
       If $deg_G(w) \le 2$, $\sigma(w) = \{w\}$ is a $3ecc$ and no construction sequence is to be constructed.
   Let $deg_G(w) > 2$.
  Then $\delta(w) = \{ (u \curvearrowleft w) \mid P_{ear(u \curvearrowleft w)} = (u \curvearrowleft w) \}$ which is the set of trivial ears whose sink is $w$.
Let $(u_1 \curvearrowleft w),  (u_2 \curvearrowleft w), \ldots, (u_{|\delta(w)|-1} \curvearrowleft w)$ be the ears in $\delta(w) \setminus \{\overleftarrow{P}(w)\}$ in an arbitrary order.
Then
  $CS_{\delta(w)} = \overleftarrow{P}(w) \ (u_1 \curvearrowleft w) \ (u_2 \curvearrowleft w)  \ldots (u_{|\delta(w)|-1} \curvearrowleft w)$
and the invariant holds for $\delta(w)$.


  \item[$(b)$]
      $w$ is an internal vertex of $T$:
Let $\{ u_i \mid 1 \le i \le \tilde{m} \}$ be the set of vertices in $L[w]$ such that
either $P_{ear(w \rightarrow u_i)} \in \hat{P}_w \setminus \{\hat{P}(w)\}$ or
$P_{(u_i \curvearrowleft w)} \in \hat{P}_w \setminus \{\hat{P}(w)\}, 1 \le i \le \tilde{m}$.

For $u_i, 1 \le i \le \tilde{m},$
$(i)$ If $P_{(u_i \curvearrowleft w)} \in \hat{P}_w \setminus \{\hat{P}(w)\}$,
 then let $\delta_{u_i} = \{(u_i \curvearrowleft w)\}$ and
$\mathcal{CS}_{\delta_{u_i}} = (u_i \curvearrowleft w)$.

$(ii)$ If $P_{ear(w \rightarrow u_i)} \in \hat{P}_w \setminus \{\hat{P}(w)\}$,
when the $\mathit{dfs}$ backtracks from $u_i$ to $w$,
   if $deg_{\hat{G}_{u_i}(u_i)} \le 2$,
    $\sigma(u_i)$ is a $3ecc$.
 The supervertex $u_i$ is ejected from $\hat{G}_{u_i}$
   and $\mathcal{CS}_{\delta(u_i)}$ is used to generate a Mader construction sequence for
$\acute{G}_{\langle \sigma(u_i) \rangle}$.
 If $\mathcal{P}_u$ becomes $nil$, then $P_{ear(w \rightarrow u_i)}$ is replaced by
a virtual back-edge $(d \curvearrowleft w)$, where $d = t(ear(w \rightarrow u_i)),$
which becomes Case $(i)$.
Hence, let $\delta_{u_i} = \{ (d \curvearrowleft w) \}$ and $\mathcal{CS}_{\delta_{u_i}} = (d  \curvearrowleft w)$.
Otherwise,
let the $u_i$-path be $\mathcal{P}_{u_i}: u_i^1 u_i^2 \ldots u_i^{h_i}$.
Then 
let $\delta_{u_i} = \{P_{ear(w \rightarrow u_i)}\} \cup \bigcup_{j = 1}^{h_i} \delta(u_i^j)$.
By assumption, the invariant holds for $u_i^j, 1 \le j \le h_i$.
Hence, $\mathcal{CS}_{\delta(u_i^j)}, 1 \le j \le h_i,$ has been constructed.
Then
$\mathcal{CS}_{\delta_{u_i}} = P_{ear(w \rightarrow u_i)} \mathcal{CS}_{\delta(u_i^{h_i})}
\mathcal{CS}_{\delta(u_i^{h_i-1})} \ldots \mathcal{CS}_{\delta(u_i^1)}$.

Without loss of generality,
let $\hat{P}(u_1) = \min_{\lessdot} \{ \hat{P}(u_i) \mid 1 \le i \le \tilde{m}\}$,
where $\hat{P}(u_i) = (u_i \curvearrowleft w)$, if $(u_i \curvearrowleft w) \in E \setminus E_T$, and
$\hat{P}(u_i) = P_{(w \rightarrow u_i)}$, if $(w \rightarrow u_i) \in E_T$.
Since $\hat{P}(w) \lessdot \hat{P}(u_i)$, $w$ absorbs $\mathcal{P}_{u_i}, 1 \le i \le \tilde{m},$
resulting in
$\delta(w) = \bigcup_{i=1}^{\tilde{m}} \delta_{u_{i}}
           = \bigcup_{i=1}^{\tilde{m}} (\{ \hat{P}(u_{i}) \} \cup \bigcup_{j=1}^{h_{i}} \delta(u_{i}^{j}))$ and
$\mathcal{CS}_{\delta(w)} = \mathcal{CS}_{\delta_{u_{1}}}  \mathcal{CS}_{\delta_{u_{2}}} \ldots  \mathcal{CS}_{\delta_{u_{\tilde{m}}}}$.

If $w$ has no incoming back-edges with its sink on $\mathcal{P}_w$, then $\mathit{dfs}$ backtracks from $w$ and the invariant holds for $\delta(w)$.
               Otherwise,
let $P_w: (w=) w_0 w_1 \ldots w_k$ and $h$ be the largest index such that there exists
$w \curvearrowleft w_h$.
 Let $\overleftarrow{P}(w_{\ell}) =  \min_{\lessdot} \{ \overleftarrow{P}(w_i) \mid 0 \le i \le h\}$ and
$w_{i_j}, 1 \le j \le \tilde{h}, i_j \notin \{h, \ell \}$ be the remaining vertices
on the path $w_0 w_1 \ldots w_h$, where $\tilde{h} = h-1$ if $\ell \neq h$ and $\tilde{h} = h$ if $\ell = h$.
   By assumption, the invariant holds for $w_i, 0 \le i \le h,$ and hence $\mathcal{CS}_{\delta(w_i)}, 0 \le i \le h,$ are constructed.
         After vertex $w$ absorbed $w_i, 1 \le i \le h$,
 $\mathcal{CS}_{\delta(w)} = \mathcal{CS}_{\delta(w_{\ell})} \ \mathcal{CS}_{\delta(w_h)} \
\mathcal{CS}_{\delta(w_{i_1})} \ldots \mathcal{CS}_{\delta(w_{i_{h-1}})}$  if $\ell \neq h$, and
 $\mathcal{CS}_{\delta(w)} = \mathcal{CS}_{\delta(w_h)} \
\mathcal{CS}_{\delta(w_{i_1})} \ldots \mathcal{CS}_{\delta(w_{i_{h-1}})}$ if $\ell = h,$ and
the invariant holds for $\delta(w)$.
\end{description}

 \vspace{-9pt}
 When the $\mathit{dfs}$  terminates at $r$,
  the $3ecc$s of $G$ and their Mader construction sequences are generated.



\subsubsection{The certifying algorithm}

The following is a pseudo-code of the algorithm which is based on the pseudo-code in~\cite{T07}.  The new instructions are marked with $\bullet$.
   As with~\cite{T07}, for clarity, the algorithm is presented without taking parallel edges into consideration. With a simple modification, the algorithm can handle parallel edges.

 In \textbf{Procedure} \texttt{3-edge-connect-CS}$(w, v)$,
the \textbf{for} loop processes the adjacency list $L[w]$ of vertex $w$.
  The if-part of the first \textbf{if} statement in the \textbf{for} loop deals with
  unvisited vertices leading to non-trivial ears $P$ with $t(P)=w$ or $P = \hat{P}(w)$ while the else-part deals with visited vertices leading to trivial ears $P$ with $t(P)=w$ or $P = \hat{P}(w)$.
\textbf{Procedure} \texttt{Absorb-ear} absorbs the entire $u$-path or $w$-path
  and merges the $\mathcal{CS}$s of the supervertices involved to update $\mathcal{CS}_{\delta(w)}$.
\textbf{Procedure} \texttt{Absorb-path} absorbs the section
  $(w=) w_0 w_1 w_2 \ldots w_h$  of the $w$-path and merges $\mathcal{CS}_{\delta(w_i)}, 0 \le i \le h,$ to update $\mathcal{CS}_{\delta(w)}$.
   \textbf{Procedure} \texttt{Gen}-$\mathcal{CS}$ converts $\mathcal{CS}_{\delta(u)}$ of a $3ecc$ $\sigma(u)$ into a Mader construction sequence of $\acute{G}_{\langle \sigma(u) \rangle}$ by adding  paths or ears to create
 the $K_2^3$-subdivision that leads the construction sequence.

\begin{singlespacing}
\footnotesize

\noindent \textbf{Algorithm} \texttt{Certifying-3-edge-connectivity}

\noindent \textbf{Input:} A connected graph $G=(V,E)$ represented by adjacency lists $L[w], \forall w \in V$

\noindent \textbf{Output:} $\left\{
                             \begin{array}{ll}
                     \mathcal{CS}_{\delta(u)}, u \in V     & \hbox{\ Mader  construction  sequence  for each $u$ representing a  $3ecc$ \ of \ G, } \\
                     \{ \mathcal{C}_{G'} \mid G' \text{ is a 2ecc of } G \}    & \hbox{\ a cactus representation of the cut-pairs for each 2$ecc$ $G'$ of $G$, \ and }\\
                     Bridges         & \hbox{ \ the bridges in $G$ }
                             \end{array}
                           \right.$

\noindent \textbf{begin}

 \textbf{for every} $u \in V$ \textbf{do} $dfs(u) := 0; \ parent(u) := \perp; \ lowpt(u) := \infty$;  \ \ \ // initialization; $\perp = $ undefined

\noindent $\bullet$ \hspace{8pt} \hspace{72pt}
    $\hat{P}(u) := \overleftarrow{P}(u) := \perp$; \ $\mathcal{CS}_{\delta(u)} := \perp$;
\ \ \   // \textbf{note:} $\perp \gtrdot P_f, \forall f \in E \backslash E_T$; $\perp  \ \mathcal{CS}_{\delta(u)} = \mathcal{CS}_{\delta(u)} \perp = \mathcal{CS}_{\delta(u)}$

\noindent $\bullet$ \hspace{8pt}  \hspace{72pt}
    $\sigma(u) := \{u\}; \ Inc_u := \emptyset;  \ \mathcal{P}_u := u$;

 $cnt := 1$; \ \ \  // $\mathit{dfs}$ number counter //  \ \ \
 $Bridges := \emptyset$ \ \ \  // to store the bridges in $G$

 \texttt{3-edge-connect-CS}$(r, \perp)$;

\noindent $\bullet$  \hspace{8pt}  $\mathcal{CS}_{\delta(r)} :=  \hat{P}(r) \mathcal{CS}_{\delta(r)}$; \ \ \ // Finalize $\mathcal{CS}_{\delta(r)}$; Theorem~\ref{CS-for-3ecc}$(i)$

\noindent \textbf{end.}\\

%

\noindent \textbf{Procedure} \texttt{3-edge-connect-CS}$(w, v)$

\noindent \textbf{begin}

 $dfs(w) := cnt; \ cnt := cnt + 1; \ parent(w) := v$;
 $lowpt(w) := dfs(w)$; \ \ \ \ \   // initialization


\noindent $\bullet$ \hspace{8pt}
\textbf{if} $(w \neq r)$ \textbf{then} $ear(v \rightarrow w) := \perp$;

\vspace{3pt}
 \textbf{for each} $u \in L[w]$ \textbf{do}
     \ \ \  \hspace{100pt} // pick the next vertex $u$ in the adjacency list of $w$


 \hspace{20pt} \textbf{if} $(dfs(u) = 0)$ \textbf{then}
  \ \ \ \hspace{80pt} // $u$ is unvisited


  \hspace{40pt} \texttt{3-edge-connect-CS}$(u,w)$;

  \hspace{40pt} \textbf{if}  $(s(\overleftarrow{P}(u)) = u \vee \overleftarrow{P}(u) = \perp)$ \textbf{then}    \ \ \ \hspace{80pt} // equivalent to $(deg_{\hat{G}_u}(u) \le 2)$

     \hspace{50pt}    \texttt{Gen-}$\mathcal{CS}( w, u, \mathcal{P}_u)$;
  \ \  \hspace{131pt} // eject super-vertex $u$ from $\mathcal{P}_u$;
                            finalize $\mathcal{CS}_{\delta(u)}$

     \hspace{40pt} \textbf{if} $( \hat{P}(w) \lessdot \hat{P}(u) )$ \textbf{then}
\ \ \  \hspace{127pt} // equivalent to $(lowpt(w) \le lowpt(u))$ in ~\cite{T07}

\hspace{60pt} \texttt{Absorb-ear($w, \hat{P}(u), w + \mathcal{P}_u$)}
   \ \ \ \  \hspace{60pt}  // absorb the entire $u$-path; + stands for concatenation

  \hspace{40pt} \textbf{else}
   \ \ \ \ \ // $( \hat{P}(u) \lessdot \hat{P}(w) )$;
   \ \ \hspace{117pt} // equivalent to $(lowpt(w) > lowpt(u))$ in ~\cite{T07}

\hspace{68pt} \texttt{Absorb-ear($w, \hat{P}(w), \mathcal{P}_w$)};
    \  \hspace{68pt} // absorb the $w$-path

\hspace{68pt} $\mathcal{P}_w := w + \mathcal{P}_u$; \ $\hat{P}(w) := \hat{P}(u)$;
    \ \  \hspace{60pt}   // equivalent to $(lowpt(w) := lowpt(u))$ in ~\cite{T07}

\noindent $\bullet$ \hspace{8pt}
 \hspace{68pt}  \textbf{if} $(w \neq r)$ \textbf{then} $ear(v \rightarrow w) := ear(w \rightarrow u)$;

 \hspace{20pt} \textbf{else if} $(dfs(u) < dfs(w) \wedge u \neq parent(w))$ \textbf{then}
   \ \  \hspace{48pt}  // $u \curvearrowleft w$ is an outgoing back-edge of $w$

    \hspace{60pt} \textbf{if} $( (u \curvearrowleft w) \lessdot \hat{P}(w) )$ \textbf{then}
  \ \   \hspace{93pt}  // equivalent to $(dfs(u) < lowpt(w))$ in ~\cite{T07}

    \hspace{80pt}   \texttt{Absorb-ear($w, \hat{P}(w), \mathcal{P}_w$)};

    \hspace{80pt}      $\mathcal{P}_w := w$; \ \ $\hat{P}(w) := u \curvearrowleft w$;
   \ \   \hspace{65pt}  // equivalent to $(lowpt(w) := dfs(u))$ in ~\cite{T07}

\noindent $\bullet$ \hspace{8pt}      \hspace{80pt}    \textbf{if} $(w \neq r)$ \textbf{then}
                           $ear(v \rightarrow w) := (u \curvearrowleft w)$;

\hspace{60pt} \textbf{else} \texttt{Absorb-ear($w, u \curvearrowleft w, w$)};
     \  \hspace{63pt} // there is no $u$-path


\noindent $\bullet$ \hspace{8pt}   \hspace{35pt} \textbf{else}
   $Inc_w := Inc_w \cup \{(w \curvearrowleft u)\}$;
\  \hspace{100pt}   // save incoming back-edge in $Inc_w$

\vspace{3pt}
\noindent \hspace{15pt}       \textbf{if} $( (\mathcal{P}_w \neq w) \wedge (Inc_w \neq \emptyset) )$ \textbf{then}
                       \texttt{Absorb-path}$(w, \mathcal{P}_w, Inc_w)$;
\ \ \ \hspace{12pt} // dealing with incoming back-edges

\vspace{3pt}
\noindent \textbf{end.} /* of Procedure \texttt{3-edge-connect-CS} */

\vspace{9pt}

\noindent \textbf{Procedure} \texttt{Gen-$\mathcal{CS}( w, u, \mathcal{P}_u)$}  \ \ \  

\noindent \textbf{begin} \ \ \ // $\mathcal{P}_u: (u=)u_0u_1\ldots u_h$; \
 eject super-vertex $u$ from $\mathcal{P}_u$;
                            finalize $\mathcal{CS}_{\sigma(u)}$

\textbf{output}($\sigma(u)$);   \hspace{125pt} // ouput the $3ecc$ $\sigma(u)$



\noindent $\bullet$  \hspace{8pt}    \textbf{if} $(s(\hat{P}(u)) = u  \vee  (\hat{P}(u) = \perp)) $  \textbf{then} \ \ \   \ \ \
\hspace{25pt} // $deg_{\hat{G}_u}(u) = 1$, i.e. $(w,u)$ is a bridge;
 $ \hat{P}(u) = \perp \Rightarrow \sigma(u) = \{u\}$

\noindent $\bullet$  \hspace{8pt}   \hspace{15pt}
     $\mathcal{CS}_{\delta(u)} := \hat{P}(u) \mathcal{CS}_{\delta(u)}$;
  \hspace{70pt} // Finalize $\mathcal{CS}_{\delta(u)}$ based on Lemma~\ref{CS-for-3ecc}$(i)$

\noindent $\bullet$  \hspace{8pt}   \hspace{15pt}
    $\hat{P}(u) := ear(w \rightarrow u) := \perp; \ \mathcal{P}_u := nil;$
  \hspace{10pt} // $\mathcal{P}_u = nil$ indicates $\mathcal{P}_u$ does not exist

\noindent $\bullet$  \hspace{8pt}   \hspace{15pt}
    $Bridges := Bridges \cup \{(w,u)\}$;   \hspace{32pt} // $(w,u)$ is a bridge


\noindent $\bullet$  \hspace{8pt}    \textbf{else} \ \ \ // $(deg_{\hat{G}_u}(u) = 2)$

\noindent $\bullet$  \hspace{23pt} \textbf{if} $( \mathcal{P}_u = u)$ \textbf{then}
\ \ \    \hspace{116pt}   // cut-pair is  $\{(w \rightarrow u),(d \curvearrowleft \ddot{u})\}$, where $(d \curvearrowleft \ddot{u}) = ear(w \rightarrow u)$

\noindent $\bullet$  \hspace{8pt}  \hspace{30pt}
     $\ddot{u} := s(ear(w \rightarrow u))$; \ $d := t(ear(w \rightarrow u))$;
 \ \ \    \hspace{26pt}
 // determine $\ddot{u}$ and $d$

\noindent $\bullet$  \hspace{8pt}  \hspace{30pt}
     $\hat{P}(u) := (d \curvearrowleft w)$; \ $\mathcal{P}_u := nil$;
 \ \ \    \hspace{77pt}
 // replace $(d \rightsquigarrow_{P_{ear(w \rightarrow u)}} w)$ with virtual edge $(d \curvearrowleft w)$

\noindent $\bullet$  \hspace{24pt}
    \textbf{else} $\ddot{u} := parent(u_1)$;  \
     \ \ \    \hspace{125pt} //  cut pair is $\{(w \rightarrow u),(\ddot{u} \rightarrow u_1)\}$, where
  $\ddot{u} = parent(u_1)$

\noindent $\bullet$  \hspace{8pt}  \hspace{30pt}  $parent(u_1) := w$; \ $ear(w \rightarrow u_1) := ear(w \rightarrow u)$;
\ \ \  \hspace{3pt}  // replace $w \rightsquigarrow_T u_1$ with virtual edge $(w \rightarrow u_1)$

\noindent $\bullet$  \hspace{8pt}  \hspace{30pt}  $\mathcal{P}_u := \mathcal{P}_u - u$;
   \ \ \   \hspace{135pt} // remove $u$ from $\mathcal{P}_u$

\noindent $\bullet$  \hspace{8pt}  \hspace{15pt}
   \textbf{if} $(u \neq \ddot{u})$ \textbf{then}
 \ \ \   \hspace{120pt} // if $u = \ddot{u}$, $\mathcal{CS}_{\delta(u)}$ is already constructed by
                          Theorem~\ref{CS-for-3ecc}$(ii)(a)$

\noindent $\bullet$  \hspace{8pt}  \hspace{30pt}
   \textbf{if} $(t(\overleftarrow{P}(u)) \neq u)$ \textbf{then}
        $\mathcal{CS}_{\delta(u)} := ( u \rightsquigarrow_T \ddot{u} ) \ (u, \ddot{u}) \ \mathcal{CS}_{\delta(u)}$
 \ \ \     //  Finalize $\mathcal{CS}_{\delta(u)}$ ; Lemma~\ref{CS-for-3ecc}$(ii)(b)$ first case

\noindent $\bullet$  \hspace{8pt}  \hspace{30pt}
   \textbf{else}  $\mathcal{CS}'_{\delta(u)} :=  \mathcal{CS} \  \mathcal{CS}_{\delta(u_0)}$ such that $\mathcal{CS}_{\delta(u)} = \mathcal{CS}_{\delta(u_0)} \ \mathcal{CS}$;

\noindent $\bullet$  \hspace{8pt}  \hspace{46pt}
$\mathcal{CS}_{\delta(u)} := ( u \rightsquigarrow_T \ddot{u} ) \ (u, \ddot{u}) \ \mathcal{CS}'_{\delta(u)}$
 \ \ \  \hspace{60pt} \ \ \     //  Finalize $\mathcal{CS}_{\delta(u)}$;  Lemma~\ref{CS-for-3ecc}$(ii)(b)$ second case

\noindent \textbf{end};


\vspace{9pt}

\noindent \textbf{Procedure} \texttt{Absorb-ear$(w, \hat{P},  \mathcal{P}$)}  \ \ \     // absorb the entire $\mathcal{P}$ which is $\mathcal{P}_w$ or $w + \mathcal{P}_u$
with $\hat{P}$ being $\hat{P}(w)$ or $\hat{P}(u)$, respectively

\noindent \textbf{begin} \ \ \
/* $\mathcal{P}: (w =) x_0 x_1 x_2 \ldots x_{k-1} x_k$ such that
 $x_i.next = x_{i+1}, 0 \le i < k$ and
 $x_k.next = \perp$.

\noindent $\bullet$  \hspace{10pt}   $\mathcal{CS} := \perp$;
                                         \ \ \ \ \ \hspace{99pt} // create construction sequence for $\hat{P} \cup \bigcup_{i=1}^k\delta(x_i)$

\noindent $\bullet$ \hspace{10pt}  $x := head(\mathcal{P})$; \ \ \ \ \ \hspace{75pt} // $head(\mathcal{P}) = x_0$

\noindent $\bullet$ \hspace{8pt}
\textbf{while} $(x.next \neq \perp)$ \textbf{do}
  \hspace{60pt} // $x.next$ exists and is to be absorbed by $w$


\noindent $\bullet$ \hspace{8pt}
\hspace{20pt}   $x := x.next$;  \ \ \ \hspace{100pt}  // get next vertex on $\mathcal{P}$

\noindent $\bullet$ \hspace{8pt}
\hspace{20pt}   $\sigma(w) := \sigma(w) \cup \sigma(x)$;  \ \ \ \hspace{60pt}  // $w$ absorbs $x_i$

\noindent $\bullet$ \hspace{8pt}
\hspace{20pt} $\mathcal{CS} := \mathcal{CS}_{\delta(x)} \ \mathcal{CS}$; \ \ \ \ \
\hspace{72pt}  // append $\mathcal{CS}$ to $\mathcal{CS}_{\delta(x)}$

\noindent $\bullet$  \hspace{8pt} $\mathcal{CS} := \hat{P} \ \mathcal{CS}$;
                                         \ \ \ \ \ \hspace{115pt} // $\hat{P}$ is the anchor

\noindent $\bullet$  \hspace{8pt}
\textbf{if} $(\hat{P} \lessdot  \overleftarrow{P}(w))$ \textbf{then}
 \ \ \ \ \ \hspace{90pt} // $\hat{P}$ is the new $\overleftarrow{P}(w)$, so $\mathcal{CS}$ leads the construction sequence

\noindent $\bullet$  \hspace{8pt} \hspace{20pt}   $\mathcal{CS}_{\delta(w)} := \mathcal{CS} \ \mathcal{CS}_{\delta(w)}$;
  \ $\overleftarrow{P}(w) := \hat{P}$

\noindent $\bullet$  \hspace{8pt}
\textbf{else}
     $\mathcal{CS}_{\delta(w)} := \mathcal{CS}_{\delta(w)} \ \mathcal{CS}$;
     \ \ \ \ \ \hspace{65pt} // $\hat{P}$ is not $\overleftarrow{P}(w)$, append $\mathcal{CS}$ to $\mathcal{CS}_{\delta(w)}$

\noindent \textbf{end}.

\vspace{6pt}

\noindent \textbf{Procedure} \texttt{Absorb-path($w, \mathcal{P}_w, Inc$)}  \ \ \ \ \ // absorb a section of the $w$-path $\mathcal{P}_w$

\noindent \textbf{begin}  \ \ \ //   $\mathcal{P}_w: (w=) w_0 w_1 \ldots w_k, k \ge 1$;
$w_i.next = w_{i+1}, 0 \le i < k$ and
 $w_k.next = \perp$.


\noindent $\bullet$ \hspace{8pt} \hspace{10pt} $h := 0$; \ $\hat{w} := w_0$;

\noindent $\bullet$ \hspace{8pt} \hspace{10pt} \textbf{for each} $( (w \curvearrowleft x) \in Inc )$ \textbf{do} \ \ \    \hspace{30pt}     // determine the lowest ancestor $w_h$ of $x$ on $\mathcal{P}_w$

\noindent $\bullet$ \hspace{8pt} \hspace{25pt}
    \ \textbf{while} $((\hat{w}.next \neq \perp) \wedge (\hat{w}.next \preceq x))$ \textbf{do} \{$  h := h+1$; \ $\hat{w} := \hat{w}.next$\};

\noindent $\bullet$ \hspace{8pt} \hspace{10pt}
$\overleftarrow{P}(w_{\ell}) := \min_{\lessdot} \{ \overleftarrow{P}(w_j) \mid 0 \le j \le h \}$;

\noindent $\bullet$ \hspace{8pt} \hspace{10pt}
\textbf{if} $(\overleftarrow{P}(w_{\ell}) \neq \overleftarrow{P}(w_h))$ \textbf{then}
        $\mathcal{CS}_{\delta(w)} := \mathcal{CS}_{\delta(w_{\ell})} \mathcal{CS}_{\delta(w_h)}$
   \textbf{else} $\mathcal{CS}_{\delta(w)} := \mathcal{CS}_{\delta(w_h)}$;
  \ \ \   // Create $\mathcal{CS}_{\delta(w)}$ based on Lemma~\ref{P1}

\noindent $\bullet$ \hspace{8pt} \hspace{10pt}
    \textbf{for} $j := 0$ \textbf{step} 1 \textbf{to} $h-1$ \textbf{do}

\noindent $\bullet$ \hspace{8pt} \hspace{25pt}
    \textbf{if} $((\overleftarrow{P}(w_{\ell}) \neq \overleftarrow{P}(w_j))$ \textbf{then}
    $\mathcal{CS}_{\delta(w)} := \mathcal{CS}_{\delta(w)} \ \mathcal{CS}_{\delta(w_j)}$;


\vspace{3pt}
\hspace{10pt}     \textbf{for} $j := 1$ \textbf{step} 1 \textbf{to} $h$  \textbf{do}
     $\sigma(w) := \sigma(w) \cup \sigma(w_j)$;      
\ $\mathcal{P}_w := \mathcal{P}_w - w_j$;
\ \ \  \hspace{20pt}   // $w$ absorbs $w_j, 1 \le j \le h$

\noindent \textbf{end}.

\normalsize
\end{singlespacing}


\vspace{9pt}

%
%

\noindent \textbf{ Remark:}
 When the condition $deg_{\hat{G}_{u}}(u)=2$ is detected,
 as a result of ejecting $\sigma(u)$, the two corresponding cut-edges are removed and
 a new (virtual) edge is created in $\hat{G}_{u} \setminus \{u\}$ to connect
the two end-vertices of the cut-pairs that are not in $\sigma(u)$.
  Fortunately, we do not need to update the adjacency lists involved to accommodate the changes.
 We just update $\hat{P}(u)$ as it contains the cut-pairs and their end-vertices.
  This is done in \textbf{Procedure} \texttt{Gen}-$\mathcal{CS}$ as follows:

\vspace{-9pt}
\begin{itemize}
  \item $\{(w \rightarrow u), (d \curvearrowleft u)\}$ is the cut-pair in
  $\hat{G}_u$:
         Then $\{(w \rightarrow u), (d \curvearrowleft \ddot{u})\}$ is the cut-pair in $G$ and
  the new edge to be added is $(d \curvearrowleft w)$.
   Since $\hat{P}(u)$ consists of $(w \rightarrow u), u \rightsquigarrow_T \ddot{u}$ and
   $d \curvearrowleft \ddot{u}$ which are replaced by $(d \curvearrowleft w)$,
 we thus update $\hat{P}(u)$ with $(d \curvearrowleft w)$.


   Let $(d \curvearrowleft w_x)$ be the embodiment of $(d \curvearrowleft w)$ when the incoming back-edges of $d$ are examined at vertex $d$.
   Since $L[d]$ was not updated to replace node $\ddot{u}$ with node $w$ when vertex $w$ was
the current vertex of the $\mathit{dfs}$,
   it is $(d \curvearrowleft \ddot{u})$ instead of $(d \curvearrowleft w)$ that will be encountered
when $L[d]$ is processed.
   However, as $(d \curvearrowleft w)$ is an embodiment of $(d \curvearrowleft \ddot{u})$,
 $(d \curvearrowleft w_x)$ is also an embodiment of $(d \curvearrowleft \ddot{u})$.
   Hence, $w_x$ can be correctly identified using $(d \curvearrowleft \ddot{u})$.

\vspace{-6pt}
\item $\{ (w \rightarrow u), (u \rightarrow u_1)\}$ is the cut-pair in
  $\hat{G}_u$:
   Then $\{(w \rightarrow u), (\ddot{u} \rightarrow u_1) \}$ is the cut-pair in $G$ and
 the section $(w \rightarrow u) u \rightsquigarrow_T \ddot{u} (\ddot{u }\rightarrow u_1)$ on $\hat{P}(u) (= \hat{P}(u_1))$ is replaced by $(w \rightarrow u_1)$.
   We thus let $parent(u_1) = w$ and $ear(w \rightarrow u_1) = ear(w \rightarrow u)$.
\end{itemize}



We shall prove that the invariant holds for $\delta(w)$, for every vertex $w$.

 If $deg_G(w) \le 2$, $\delta(w)=\emptyset$ and no $\mathcal{CS}_{\delta(w)}$ is needed.
   Let $deg_{G}(w) > 2$.
   First, consider the leaves of $T$.

\vspace{-3pt}
\begin{lem}\label{P-1}
  Let $w$ be a leaf in $T$ such that $deg_{G}(w) > 2$.
       Let
   $P_1, P_2, \ldots, P_{deg_G(w)-3}$ be the ears in
  $\delta(w) \setminus \{\overleftarrow{P}(w)\}$
in arbitrary order.
  Then $\mathcal{CS}_{\delta(w)} = \overleftarrow{P}(w) P_1 P_2 \ldots P_{deg_G(w)-3}$.
$($Figure~4$($a$))$.
\end{lem}

\vspace{-6pt}
\noindent \textbf{Proof:}
  Since $w$ is a leaf, $\sigma(w) = \{w\}$. Therefore, $\delta(w) =
\{ P \mid P \text{ is an ear } \wedge t(P)=w\} =
  \{ f \mid f=(u \curvearrowleft w) \in E \setminus E_T \} \setminus \{ \hat{f} \}$, where $P_{\hat{f}} = \hat{P}(w)$.
   Clearly, after $\overleftarrow{P}(w)$ is added to the construction sequence,  $w$ becomes a branch vertex.
 The remaining ears (back-edges) $P_1, P_2, \ldots, P_{deg_G(w)-3}$ in  $\delta(w) \setminus \{\overleftarrow{P}(w)\}$ can be added to the
construction sequence as Mader paths by operation $(i)$ or $(ii)$.
   The lemma follows.   \ \ \ \ \  $\blacksquare$

 \begin{figure}[h!]
 \centering
\includegraphics[width=5.5in]{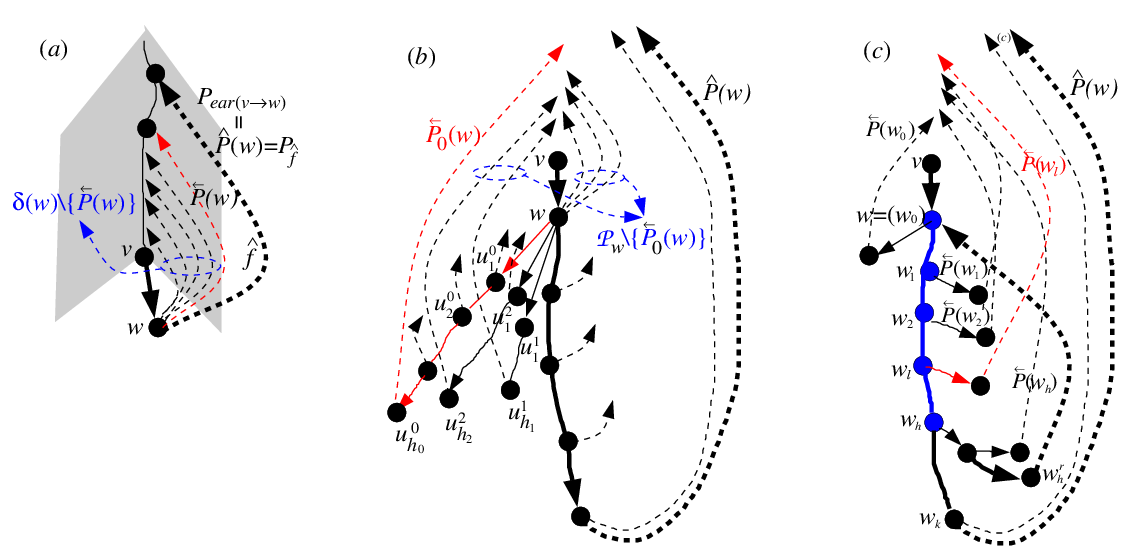}
 \caption[ ]{$\overleftarrow{P}(w)$, where $(a):$ $w$ is a leaf;
             $(b)$, $(c):$ $w$ is an internal vertex.}
 \end{figure}\label{Fig4.eps}


\vspace{6pt}
Next, consider the internal vertices.
   Let $w$ be an internal vertex of $T$ with $deg_G(w) > 2$.
In processing $L[w]$, vertices that are children of $w$  or form an outgoing back-edge of $w$ with $w$ are processed first.
   The vertices that form an incoming back-edge of $w$ with $w$ are considered later.


\vspace{-6pt}
\begin{lem} \label{P0}
Let $w$ be an internal vertex of $T$ such that $deg_{G}(w) > 2$.

Let $u \in L[w]$.

\noindent $(i)$ If $u$ is a child of $w$ and
$u_1, u_2, \ldots, u_{h}$ is the $u$-path after the $\mathit{dfs}$
 backtracked from $u$ to $w$ and $u$ is ejected if $deg_{\hat{G}_{u}}(u)=2$.
Suppose the invariant holds for each $u_j, 1 \le j \le h$.

Let $\delta_u = \{ P_{ear(w \rightarrow u)} \} \cup  \bigcup_{j=1}^h \delta(u_j)$.
Then  $\mathcal{CS}_{\delta_u} = P_{ear(w \rightarrow u)} \
 \mathcal{CS}_{\delta(u_{h})} \mathcal{CS}_{\delta(u_{h-1})}   \ldots \mathcal{CS}_{\delta(u_1)}$.                                   

\noindent $(ii)$ If $(u \curvearrowleft w) \in E \setminus E_T$,
let $\delta_u = \{ (u \curvearrowleft w) \}$.
Then  $\mathcal{CS}_{\delta_u} = (u \curvearrowleft w)$.

\end{lem}

\vspace{-6pt}
\noindent \textbf{Proof:}

\noindent $(i)$    After $P_{ear(w \rightarrow u)}$ is added to the construction sequence, $w$ becomes a branch vertex.

For each $\overleftarrow{P}(u_j), 1 \le j \le h,$
since  $t(\overleftarrow{P}(u_j)) \in \sigma(u_j)$
and by Lemma~\ref{Tsin09}$(ii)$,
  $s(P_{ear(w \rightarrow u)}) \preceq s(\overleftarrow{P}(u_j)) \preceq w \prec t(\overleftarrow{P}(u_j))$,
 $\overleftarrow{P}(u_j)$ can be added to the construction sequence
by operation $(ii)$ or $(iii)$.
 As the invariant holds for $u_j$,
 the remaining ears in $\delta(u_j)$ can be added to the construction sequence  after $\overleftarrow{P}(u_j)$.
Hence, $\mathcal{CS}_{\delta_u} = P_{ear(w \rightarrow u)} \
 \mathcal{CS}_{\delta(u_{h})} \mathcal{CS}_{\delta(u_{h-1})}   \ldots \mathcal{CS}_{\delta(u_1)}$.

\noindent
$(ii)$ Obvious.
  \ \ \ \ \  $\blacksquare$

Now, consider the incoming back-edges of $w$.


\begin{lem}\label{P1}
    Let $w$ be an internal vertex of $T$.
  Let $\hat{G}$ be the graph to which $G$ has been transformed
after the child edges and outgoing back-edges of $w$ are processed
and
$w (=w_0), w_1, w_2, \ldots, w_k (k \ge 1)$ be the $w$-path after the child $u$ of $w$ on the path is ejected if
$deg_{\hat{G}_u} (u) = 2$.

  Suppose the invariant holds for $w_i, 0 \le i \le k$.


  Let $w \curvearrowleft w_h$ be the incoming back-edge of $w$ in $\hat{G}$ with the largest index $h$.

  Let
 $\overleftarrow{P}(w_{\ell}) =  \min_{\lessdot}\{ \overleftarrow{P}(w_i) \mid 0 \le i \le h\}$, and

\hspace{15pt}  $\overleftarrow{P}(w_{i_1}) \overleftarrow{P}(w_{i_2}) \ldots \overleftarrow{P}(w_{i_{\tilde{h}}})$ be the ears in $\{\overleftarrow{P}(w_i) \mid 0 \le i \le h \} \setminus \{\overleftarrow{P}(w_{\ell}), \overleftarrow{P}(w_h)\}$ in arbitrary order, where $\tilde{h} = h$ or $h-1$ depending on whether $w_{\ell} = w_h$.

   Then, after vertex $w$ absorbed $w_i, 1 \le i \le h$,
   $\overleftarrow{P}(w) = \overleftarrow{P}(w_{\ell})$ and

  $\mathcal{CS}_{\delta(w)} = \left\{
                                \begin{array}{ll}
                                  \mathcal{CS}_{\delta(w_{\ell})} \mathcal{CS}_{\delta(w_h)} \mathcal{CS}_{\delta(w_{i_1})} \mathcal{CS}_{\delta(w_{i_2})} \ldots \mathcal{CS}_{\delta(w_{i_{\tilde{h}}})}, & \hbox{\text{if } $\ell \neq h$;} \\
                                   \mathcal{CS}_{\delta(w_h)}  \mathcal{CS}_{\delta(w_{i_1})} \mathcal{CS}_{\delta(w_{i_2})} \ldots \mathcal{CS}_{\delta(w_{i_{\tilde{h}}})}, & \hbox{ \text{if } $\ell = h$.}
                                \end{array}
                              \right.
$ $($Figure~4$($c$))$.

\end{lem}

\noindent \textbf{Proof:}
   By assumption, the invariant holds for $w_i, 0 \le i \le k$.
   Hence, the invariant holds for $w_{\ell}$.
  It follows that after $\overleftarrow{P}(w_{\ell})$ is added to the construction sequence,
 the remaining ears in $\delta(w_{\ell})$ can be added to the sequence as Mader paths.
   This produces the construction sequence $\mathcal{CS}_{\delta(w_{\ell})}$.

If $\overleftarrow{P}(w_h) \neq \overleftarrow{P}(w_{\ell})$,
   then $s(\overleftarrow{P}(w_{\ell})) \preceq s(\overleftarrow{P}(w_h)) \preceq t(\overleftarrow{P}(w_{\ell})) \prec t(\overleftarrow{P}(w_h))$
which implies that
   ear $\overleftarrow{P}(w_h)$ can be added as a Mader path by operation $(ii)$ or $(iii)$
after $\overleftarrow{P}(w_{\ell})$.
   Since the invariant holds for $w_h$,
   the remaining ears in $\delta(w_h)$, which includes $P_{w \curvearrowleft w_h^{\tau}}$, where
  $w \curvearrowleft w_h$ is an embodiment of $w \curvearrowleft w_h^{\tau}$,
can be added to the sequence as Mader paths. This produces the construction sequence $\mathcal{CS}_{\delta(w_{\ell})} \mathcal{CS}_{\delta(w_{h})}$.

   After that, $w$ becomes a branch vertex.
      The ears $\overleftarrow{P}(w_{i_j}), 1 \le j \le \tilde{h},$  can be added to the sequence as Mader path by operation $(ii)$ or $(iii)$ (if $i_j \neq 0$),
or by operation $(i)$ or $(ii)$ (if $i_j = 0$).
    Since the invariant holds for  $w_{i_j}, 1 \le j \le \tilde{h}$,   after ear $\overleftarrow{P}(w_{i_j})$ is added, the remaining ears  in $\delta(w_{i_j})$
can be added to the construction sequence as Mader paths.
We thus have:

\vspace{6pt}
\noindent  $\left\{
                                \begin{array}{ll}
                                  \mathcal{CS}_{\delta(w_{\ell})} \mathcal{CS}_{\delta(w_h)} \mathcal{CS}_{\delta(w_{i_1})} \mathcal{CS}_{\delta(w_{i_2})} \ldots \mathcal{CS}_{\delta(w_{i_{\tilde{h}}})}, & \hbox{\text{if } $\ell \neq h$;} \\
                                   \mathcal{CS}_{\delta(w_h)}  \mathcal{CS}_{\delta(w_{i_1})} \mathcal{CS}_{\delta(w_{i_2})} \ldots \mathcal{CS}_{\delta(w_{i_{\tilde{h}}})}, & \hbox{ \text{if } $\ell = h$}
                                \end{array}
                              \right.
$
is a construction sequence for $\bigcup_{i=0}^h \delta(w_i)$.

\vspace{6pt}
 After vertex $w$ absorbed $w_i, 1 \le i \le h$,
$\hat{\delta}(w) = \bigcup_{i=0}^h \hat{\delta}(w_i)$ which implies that
 $\delta(w) = \bigcup_{i=0}^h \hat{\delta}(w_i) \setminus \{ \hat{P}(w)\}
            = \bigcup_{i=0}^h ( \hat{\delta}(w_i) \setminus \{ \hat{P}(w)\} )$.
Since $w_i, 1 \le i \le h,$ lies on the $w$-path, $\hat{P}(w_i) = \hat{P}(w_0) = \hat{P}(w)$.
It follows that
 $\delta(w) = \bigcup_{i=0}^h (\hat{\delta}(w_i) \setminus \{ \hat{P}(w_i)\})
= \bigcup_{i=0}^h  \delta(w_i)$.
     The above construction sequence is thus $\mathcal{CS}_{\delta(w)}$.

 Since $\overleftarrow{P}(w_{\ell}) =  \min_{\lessdot}\{ \overleftarrow{P}(w_i) \mid 0 \le i \le h\}$
and
   $\overleftarrow{P}(w_i) = \min_{\lessdot}\delta(w_i)$,
by the transitivity of $\lessdot$,
   $\overleftarrow{P}(w_{\ell}) = \min_{\lessdot}\bigcup_{i=0}^h \delta(w_i) = \min_{\lessdot}\delta(w)$
   which implies that $\overleftarrow{P}(w) = \overleftarrow{P}(w_{\ell})$.
The lemma thus follows.
 \ \ \ \ \  $\blacksquare$


\begin{lem}\label{CS-for-3ecc}
   Let $w \in V$.
  When the $\mathit{dfs}$ backtracks from $w$ to its parent $v$ or terminates execution if $w=r$,
suppose the invariant holds for $w$.
\vspace{-12pt}
\begin{description}
  \item[$(i)$] if $deg_{\hat{G}_w}(w) = 1$ or $w = r$, then $\hat{P}(w) \mathcal{CS}_{\delta(w)}$ is a construction sequence of $\acute{G}_{\langle \sigma(w) \rangle}$; (Figure 3$(e),(f)$)

\vspace{-9pt}
  \item[$(ii)$] if $deg_{\hat{G}_w}(w) = 2$, let $P_w: (w=) w_0 w_1 \ldots w_k$ be the $w$-path and $\ddot{w} = parent(w_1)$ or $\ddot{w} = t(ear(v \rightarrow w))$.

\vspace{-9pt}
   \begin{description}
     \item[$(a)$]  If $w = \ddot{w}$, then $\mathcal{CS}_{\delta(w)}$ is a construction sequence of $\acute{G}_{\langle \sigma(w) \rangle}$.  (Figure 3$(c),(d)$)

     \item[$(b)$]  If $w \neq \ddot{w}$, then  (Figure 3$(a),(b)$)
           \begin{itemize}
             \item if $t(\overleftarrow{P}(w)) \neq w$,
       $(w \rightsquigarrow_T \ddot{w}) (w, \ddot{w}) \mathcal{CS}_{\delta(w)}$ is a construction sequence of $\acute{G}_{\langle \sigma(w) \rangle}$,
             where $(w, \ddot{w})$ is a new (virtual) edge.

             \item   If $t(\overleftarrow{P}(w)) = w$,
                    then $\mathcal{CS}_{\delta(w)} = \mathcal{CS}_{\delta(w_{\ell})} \mathcal{CS}_{\delta(w_h)} \mathcal{CS}_{\delta(w_{i_1})} \mathcal{CS}_{\delta(w_{i_2})} \ldots \mathcal{CS}_{\delta(w_{i_{h-1}})}$, and
     $(w \rightsquigarrow_T \ddot{w}) (w, \ddot{w}) \mathcal{CS}'_{\delta(w)}$ is a construction sequence of $\acute{G}_{\langle \sigma(w) \rangle}$,
where $\mathcal{CS}'_{\delta(w)} = \mathcal{CS}_{\delta(w_h)}$ $\mathcal{CS}_{\delta(w_{i_1})} \mathcal{CS}_{\delta(w_{i_2})} \ldots \mathcal{CS}_{\delta(w_{i_{h-1}})}\mathcal{CS}_{\delta(w_{\ell})}$.

           \end{itemize}

   \end{description}
\end{description}
\end{lem}

\vspace{-12pt}
\noindent \textbf{Proof:}

\vspace{-12pt}
\begin{description}
  \item[$(i)$] If $deg_{\hat{G}_w}(w) = 1$, the parent-edge of $w$ is a bridge which implies that $lowpt(w) = dfs(w)$.
   By Lemma~\ref{ears-and-lowpt}$(a)$, $s(\hat{P}(w)) = w$ which implies that
$\hat{P}(w)$ is a cycle with $t(\hat{P}(w)) = w$.
   Hence, $\hat{P}(w) \in \hat{\delta}(w)$.
  By Lemma~\ref{ears-and-lowpt}$(b)$, $t(\overleftarrow{P}(w))$ lies on $\hat{P}(w)$.
  Clearly, $s(\overleftarrow{P}(w)) = w$ and $\overleftarrow{P}(w)$ is not a cycle.
   Hence, $t(\overleftarrow{P}(w)) \neq w$  and
 $\hat{P}(w)$ and $\overleftarrow{P}(w)$ form a $K_2^3$-subdivision (Figure~3$(c)$).

  By assumption, $\mathcal{CS}_{\delta(w)}$ is a construction sequence for $\delta(w)$.
 As $\overleftarrow{P}(w)$ is the first ear in $\mathcal{CS}_{\delta(w)}$,
   $\hat{P}(w) \ \mathcal{CS}_{\delta(w)}$ is a construction sequence for $\hat{\delta}(w)$ starting with the aforementioned $K_2^3$-subdivision.
 Since after $w$ is ejected, $\sigma(w)$ is a $3ecc$,
  $\hat{P}(w) \ \mathcal{CS}_{\delta(w)}$ is thus a construction sequence for $\acute{G}_{\langle \sigma(w) \rangle}$.
The proof for the case where $w = r$ is similar (Figure~3$(f)$).


\vspace{-6pt}
  \item[$(ii)$] if $deg_{\hat{G}_w}(w) = 2$,
     then $P_{ear(v \rightarrow w)} = \hat{P}(w)$  which implies that  $t(\hat{P}(w)) \prec w$.
   Hence, $\hat{P}(w) \notin \hat{\delta}(w)$ which implies that $\hat{\delta}(w) = \delta(w)$.

\vspace{-6pt}
   \begin{description}
      \item[$(a)$]   $w = \ddot{w}$:
  Then $w$ has no incoming back-edges originated from the $w$-path~(Figure~3(c),(d)) which implies that $w$ is the only vertex on $\hat{P}(w)$ that is in $\sigma(w)$.
 Since $t(\overleftarrow{P}(w))$ lies on $\hat{P}(w)$ and $\overleftarrow{P}(w) \in \delta(w)$,
   therefore $t(\overleftarrow{P}(w)) = w$ which implies that
$\overleftarrow{P}(w) = P_{ear(w \rightarrow u)}$ for some child $u$ of $w$.
  By Lemma~\ref{P0},
$\mathcal{CS}_{\delta_u} = P_{ear(w \rightarrow u)} \
 \mathcal{CS}_{\delta(u_{h})} \mathcal{CS}_{\delta(u_{h-1})}   \ldots \mathcal{CS}_{\delta(u_1)}$
 $ = \overleftarrow{P}(w) \ \mathcal{CS}_{\delta(u_{h})} \mathcal{CS}_{\delta(u_{h-1})}
 \ldots \mathcal{CS}_{\delta(u_1)}$,
where
   $u_1, u_2, \ldots, u_{h}$ is the $u$-path after the $\mathit{dfs}$
 backtracked from $u$ to $w$ and $u$ is ejected if $deg_{\hat{G}_{u}}(u)=2$.
By assumption, the invariant holds for $w$.
 Since $\overleftarrow{P}(w)$ is the anchor of $\delta(w)$,
therefore
$\mathcal{CS}_{\delta(w)} = \mathcal{CS}_{\delta_u} \mathcal{CS}$, where
$\mathcal{CS}$ is some chain of construction sequences.
  Since the anchor of $\mathcal{CS}_{\delta(u_h)}$ is $\overleftarrow{P}(u_h)$,
  $\mathcal{CS}_{\delta(w)}$ thus has $\overleftarrow{P}(w) \overleftarrow{P}(u_h)$
as the two leading ears.
  Since the parent-edge of $w$ is a cut-edge, $s(\overleftarrow{P}(w)) = w$ which implies that
  $\overleftarrow{P}(w)$ is a cycle.
 By  Lemma~\ref{ears-and-lowpt}$(b)$, $t(\overleftarrow{P}(u_{h}))$ lies on $\overleftarrow{P}(w)$.
    Since $t(\overleftarrow{P}(u_{h})) \in \sigma(u_{h})$,
    $t(\overleftarrow{P}(u_{h})) \neq w$.
 By Lemma~\ref{Tsin09}$(ii)$, $s(\overleftarrow{P}(u_{h})) = w$.
      Hence, $\overleftarrow{P}(w)$ and $\overleftarrow{P}(u_{h})$ form a $K_2^3$-subdivision
and  $\mathcal{CS}_{\delta(w)}$ starts with the $K_2^3$-subdivision.
    After $w$ is ejected, $\sigma(w)$ is a $3ecc$,
 $\mathcal{CS}_{\delta(w)}$ is thus a construction sequence of $\acute{G}_{\langle \sigma(w) \rangle}$.

     \item[$(b)$]   $w \neq \ddot{w}$: Then $w$ must have absorbed a section of $\mathcal{P}_w$.
Let the section be $(w=) w_0 w_1 \ldots w_h$. By assumption, the invariant holds for
 $w_i, 0 \le i \le h$.
 Hence,
        by Lemma~\ref{P1},

  $\mathcal{CS}_{\delta(w)} = \left\{
                                \begin{array}{ll}
                                  \mathcal{CS}_{\delta(w_{\ell})} \mathcal{CS}_{\delta(w_h)} \mathcal{CS}_{\delta(w_{i_1})} \mathcal{CS}_{\delta(w_{i_2})} \ldots \mathcal{CS}_{\delta(w_{i_{\tilde{h}}})}, & \hbox{\text{if } $\ell \neq h$;} \\
                                   \mathcal{CS}_{\delta(w_h)}  \mathcal{CS}_{\delta(w_{i_1})} \mathcal{CS}_{\delta(w_{i_2})} \ldots \mathcal{CS}_{\delta(w_{i_{\tilde{h}}})}, & \hbox{ \text{if } $\ell = h$}
                                \end{array}
                              \right.
$, where
  $\{ w_{i_1}, w_{i_2}, \ldots, w_{i_{\tilde{h}}}\} = \{ w_0, w_1, \ldots, w_h \} \setminus \{ w_{\ell}, w_h\}$,
 is a construction sequence for $\delta(w)$ and hence
    for $\hat{\delta}(w)$ as $\hat{\delta}(w) = \delta(w)$.
   If $w_{\ell} \neq w$, then
      $w \prec w_{\ell}$.
By Lemma~\ref{ears-and-lowpt}$(b)$, $t(\overleftarrow{P}(w_{\ell}))$ lies on $\hat{P}(w_{\ell})$ which is $\hat{P}(w)$.
   As $t(\overleftarrow{P}(w_{\ell})) \in \sigma(w_{\ell})$,
     $w \prec t(\overleftarrow{P}(w_{\ell})) \preceq \ddot{w}$ which implies that
     $t(\overleftarrow{P}(w_{\ell}))$ lies on $w \rightsquigarrow_T \ddot{w}$.
     Moreover, as $(v \rightarrow w)$ is a cut-edge,
$s(\overleftarrow{P}(w_{\ell})) = w$.
     Hence,
     the path $w \rightsquigarrow_T \ddot{w}$, $\overleftarrow{P}(w_{\ell})$,
          and the virtual edge $(w, \ddot{w})$ form a $K_2^3$-subdivision.
   Note that as $\hat{P}(w) \notin \hat{\delta}(w)$ and the path $w \rightsquigarrow_T \ddot{w}$ lies on $\hat{P}(w)$,
    the  $K_2^3$-subdivision and the ears in $\hat{\delta}(w)$, excluding $\overleftarrow{P}(w)$, are disjoint.
   Since $\overleftarrow{P}(w_{\ell}) (= \overleftarrow{P}(w))$ is
the anchor of $\mathcal{CS}_{\delta(w)}$, after $w$ is ejected, $\sigma(w)$ becomes a $3ecc$ and
   $(w \rightsquigarrow_T \ddot{w}) (w, \ddot{w}) \mathcal{CS}_{\delta(w)}$ is a construction sequence of $\acute{G}_{\langle \sigma(w) \rangle}$.

    If $w_{\ell} = w $,
    then $s(\overleftarrow{P}(w_{\ell})) = t(\overleftarrow{P}(w_{\ell})) = w$.
     $\overleftarrow{P}(w_{\ell})$ is thus a cycle.
    By Lemma~\ref{P1},
    $\mathcal{CS}_{\delta(w)} = \mathcal{CS}_{\delta(w_{\ell})} \mathcal{CS}_{\delta(w_h)} \mathcal{CS}_{\delta(w_{i_1})} \mathcal{CS}_{\delta(w_{i_2})} \ldots \mathcal{CS}_{\delta(w_{i_{\tilde{h}}})}$ as $\ell \neq h$.
Unfortunately,  $(w \rightsquigarrow_T \ddot{w}) (w, \ddot{w}) \mathcal{CS}_{\delta(w)}$ is not a construction sequence of $\acute{G}_{\langle \sigma(w) \rangle}$
as
$\overleftarrow{P}(w_{\ell})$ (the anchor of $\mathcal{CS}_{\delta(w)}$),
$(w \rightsquigarrow_T \ddot{w})$ and $(w, \ddot{w})$ do not form a $K_2^3$-subdivision.
    Consider the sequence $\mathcal{CS}'_{\delta(w)} = \mathcal{CS}_{\delta(w_h)} \mathcal{CS}_{\delta(w_{i_1})} \mathcal{CS}_{\delta(w_{i_2})} \ldots \mathcal{CS}_{\delta(w_{i_{\tilde{h}}})}\mathcal{CS}_{\delta(w_{\ell})}$.
     Clearly, $\mathcal{CS}'_{\delta(w)}$ is also a construction sequence of $\delta(w)$ as after adding $\mathcal{CS}_{\delta(w_h)} \mathcal{CS}_{\delta(w_{i_1})}$
     $\mathcal{CS}_{\delta(w_{i_2})} \ldots \mathcal{CS}_{\delta(w_{i_{\tilde{h}}})}$,
    $w$ becomes a branch vertex;
     the anchor $\overleftarrow{P}(w_{\ell})$ of $\mathcal{CS}_{\delta(w_{\ell})}$ can thus be added to
     the construction sequence by operation $(i)$ following by the remaining ears in  $\delta(w_{\ell})$.
       Since $h \neq \ell$,  by an argument similar to the above case, it is easily verified that
$\overleftarrow{P}(w_h)$ (the anchor of $\mathcal{CS}'_{\delta(w)}$),
$w \rightsquigarrow_T \ddot{w}$, and $(w, \ddot{w})$ form a $K_2^3$-subdivision.
          Hence, $(w \rightsquigarrow_T \ddot{w}) (w, \ddot{w}) \mathcal{CS}'_{\delta(w)}$ is a construction sequence of $\acute{G}_{\langle \sigma(w) \rangle}$.
     \ \ \ \ \  $\blacksquare$
   \end{description}
\end{description}

The construction sequence $\mathcal{CS}_{\delta(w)}$ can be conveniently represented by a linked list.
 The ordering of the ears in the list obeys that of the ears in $\mathcal{CS}_{\delta(w)}$.
  Hence, $\overleftarrow{P}(w)$ is the \emph{first node} in the list.
   An ear $P_f$ in the list can be conveniently represented by $f$ as $P_f$ can be easily determined through $f$ and the arrays $parent$ and $ear$.
 The node for ear $P_f$ is created when $f$ is encountered.
To efficiently handle the situation described in Case $(ii)(b)$ of Lemma~\ref{CS-for-3ecc}, where  $\mathcal{CS}_{\delta(w)}$ is
to be replaced by $\mathcal{CS}'_{\delta(w)}$,
  a pointer is kept in the first node of $\mathcal{CS}_{\delta(w)}$ pointing at
 the first node of $\mathcal{CS}_{\delta(w_h)}$.
  The pointer is created when $\mathcal{CS}_{\delta(w_h)}$ is appended to $\mathcal{CS}_{\delta(w_{\ell})}$.


\begin{thm}\label{correctness}
\textbf{Algorithm} Certifying-3-edge-connectivity constructs a Mader construction sequence for every 3-edge-connected component of $G$.
\end{thm}
\noindent \textbf{Proof:} (By induction on the  height of $T$ to prove that the invariant holds for all vertices)

   At a leaf $w$,
 the else-part of the \textbf{if} statement in the \textbf{for} loop of
\textbf{Procedure} \texttt{3-edge-connect-CS}$(w,v)$  is executed for each $u \in L[w]$.
It is easily verified that when $L[w]$ is completely processed,
$\hat{P}(w), \overleftarrow{P}(w)$ are correctly computed and $CS_{\delta(w)}$ satisfies Lemma~\ref{P-1}.
Hence,  the invariant holds for $w$.



Let $w$ be an internal vertex such that $deg_{G}(w) > 2$.

Let $u_1, u_2, \ldots, u_m$ be the vertices in $L[w]$ listed in the order they are visited by the $\mathit{dfs}$ such that
either $(w \rightarrow u_i) \in E_T$ or
$(u_i \curvearrowleft w) \in E \setminus E_T, 1 \le j \le m$.

Let $P_i(w) = P_{ear(w \rightarrow u_i)}$ or $P_i(w) = (u_i \curvearrowleft w)$  whichever exists, and


$(i)$ if $P_i(w) = P_{ear(w \rightarrow u_i)}$,
let the $u_i$-path be $\mathcal{P}_{u_i}: u_i^1 u_i^2 \ldots u_i^{h_i}$ after $u_i$ is ejected if $deg_{\hat{G}_{u_i}}(u_i) \le 2$ and
 $\delta_{u_i} = \{ P_{ear(w \rightarrow u_i)} \} \cup \bigcup_{j=1}^{h_{i}} \delta(u_{i}^{j}))$;
$(ii)$ if $P_i(w) = (u_i \curvearrowleft w)$,
let $\delta_{u_i} = \{(u_i \curvearrowleft w)\}$.

Suppose the invariant holds for $u^i_j, 1 \le j \le h_i, 1 \le i \le m$.


We shall apply induction on $q$ to prove the following assertion:

After processing the vertices $u_1, u_2, \ldots, u_q (q \le m)$,

$\hat{P}(w) =  P_{\hat{\ell}}(w) = \min_{\lessdot} \{ P_i(w) \mid 1 \le i \le q \}$,

$\overleftarrow{P}(w) = P_{\ell}(w) = \min_{\lessdot} (\{ P_i(w) \mid 1 \le i \le q \wedge i \neq \hat{\ell} \}$,

$\delta(w) = \bigcup \{\delta_{u_{i}} \mid 1 \le i \le q \wedge i \neq \hat{\ell} \}$,
and

$\mathcal{CS}_{\delta(w)} = \mathcal{CS}_{\delta_{u_{\ell}}}  \mathcal{CS}_{\delta_{u_{i_1}}} \ldots  \mathcal{CS}_{\delta_{u_{i_{q-2}}}}$,
where
 $\{ u_{i_1}, u_{i_2}, \ldots, u_{i_{q-2}} \} = \{u_1, u_2, \ldots, u_q\} \setminus \{ u_{\ell}, u_{\hat{\ell}} \}$ such that
$\mathcal{CS}_{\delta_{u_{\ell}}} = \overleftarrow{P}(w) \mathcal{CS}_{\delta(u^{h_{\ell}}_{\ell})} \mathcal{CS}_{\delta(u_{\ell}^{h_{\ell-1}})} \ldots \mathcal{CS}_{\delta(u_{\ell}^1)}$ and

$\mathcal{CS}_{\delta_{u_{i_j}}} = P_{i_j}(w) \mathcal{CS}_{\delta(\tilde{u}^{\tilde{h}_{j}}_{j})} \mathcal{CS}_{\delta(\tilde{u}_j^{\tilde{h}_j-1})} \ldots \mathcal{CS}_{\delta(\tilde{u}_j^1)}, 1 \le j \le q-2$, where $\tilde{h}_j = h_{i_j}$ if $\tilde{u}_j = u_{i_j}$.


\noindent When $q = 1$,

$(i)$ if $(w \rightarrow u_1) \in E_T$,
the then-part of the \textbf{if} statement in the \textbf{for} loop
in \textbf{Procedure} \texttt{3-edge-connect-CS} is executed.
   When the $\mathit{dfs}$ backtracks from $u_1$ to $w$,
If $deg_{\hat{G}_{u_1}}(u_1) \le 2$,   \textbf{Procedure} \texttt{Gen-$\mathcal{CS}$} is invoked
and $u_1$ is ejected.
By the induction hypothesis, the invariant holds for $u_1$.
Hence, $\mathcal{CS}_{\delta(u_1)}$ can be used to generate a
 Mader construction sequence for $\acute{G}_{\sigma(u_1)}$ based on Lemma~\ref{CS-for-3ecc}.

 If $deg_{\hat{G}_{u_1}}(u_1) = 1$,  $(w,u_1)$ is a bridge and
$\hat{P}(u_1) = \perp$.
The assertion vacuously holds true.

If $deg_{\hat{G}_{u_1}}(u_1) = 2$
and  $\mathcal{P}_{u_1}: u_1$,
then the cut pair is $\{(w \rightarrow u_1),(d \curvearrowleft u_1)\}$.
Let $\{(w \rightarrow u_1),(d \curvearrowleft \ddot{u_1})\}$
be the corresponding cut-pair in $G$
(note that $(d \curvearrowleft u_1)$ is an embodiment of $(d \curvearrowleft \ddot{u_1})$).
Clearly, $ear(w \rightarrow u_1) = (d \curvearrowleft \ddot{u_1})$.
Hence, $P_1(w) = P_{ear(w \rightarrow u_1)} = P_{(d \curvearrowleft \ddot{u_1})}$.
Since after $u_1$ is ejected,
the path $q \rightsquigarrow_{P_{(d \curvearrowleft \ddot{u_1})}} w$
 is replaced by the virtue edge $(d \curvearrowleft w)$,
$(d \curvearrowleft w)$ thus becomes $P_1(w)$.
  Hence, we have $\hat{P}(u_1) = P_1(w) = (d \curvearrowleft w)$ and
$\mathcal{P}_{u_1} = nil$.
Since
$\hat{P}(w) = \perp$,
the else-part of the third \textbf{if} statement is executed
resulting in         
$\hat{P}(w) = (d \curvearrowleft w),
\overleftarrow{P}(w) = \perp, \delta(w) = \emptyset$, and $\mathcal{CS}_{\delta(w)} = \perp$.
The assertion thus holds.


If $deg_{\hat{G}_{u_1}}(u_1) = 2$
and  $\mathcal{P}_{u_1}: u_1 u_1^1 \ldots u_1^{h_1}$,
then the cut pair is $\{ (w \rightarrow u_1),(u_1 \rightarrow u_1^1) \}$.
After $u_1$ is ejected,
 $\mathcal{P}_{u_1}: u_1^1 u_1^2 \ldots u_1^{h_1} (h_1 \ge 1)$.
Since
$\hat{P}(u_1) = P_1(w)$ and $\hat{P}(w) = \perp, \hat{P}(u_1)\lessdot \hat{P}(w)$ and
the else-part of the third \textbf{if} statement is executed
resulting in                
 $\hat{P}(w) = P_1(w),
\overleftarrow{P}(w) = \perp, \delta(w) = \emptyset$, and $\mathcal{CS}_{\delta(w)} = \perp$.
The assertion clearly holds.

$(ii)$ If $(u_1 \curvearrowleft w) \in E \setminus E_T$, the argument is same as the above case when
$deg_{\hat{G}_{u_1}}(u_1) = 2$ and  $\mathcal{P}_{u_1}: u_1$, and after
$(d \curvearrowleft w)$ becomes $P_1(w)$.
Hence, we have                 
$\hat{P}(w) = (u_1 \curvearrowleft w),
\overleftarrow{P}(w) = \perp, \delta(w) = \emptyset$, $\mathcal{CS}_{\delta(w)} = \perp$, and
the assertion holds.


\noindent Suppose the assertion holds for $q - 1$.

$(i)$  If $(w \rightarrow u_q) \in E_T$, \textbf{Procedure} \texttt{3-edge-connect-CS} is invoked.
   When the $\mathit{dfs}$ backtracks from $u_q$ to $w$,
if $deg_{\hat{G}_{u_q}}(u_q) \le 2$, $u_q$ is ejected and
  $\mathcal{CS}_{\delta(u_q)}$ is used to generate a
 Mader construction sequence for $\acute{G}_{\sigma(u_q)}$ based on Lemma~\ref{CS-for-3ecc}.
  If $deg_{\hat{G}_{u_q}}(u_q) =1$ (i.e. $(w \rightarrow u_q)$ is a bridge), then $P_q(w) = \perp$ and
the assertion obviously holds for $q$.
 Otherwise, let
$\mathcal{P}_{u_q}: u_q^1 u_q^2 \ldots u_q^{h_q} (h_q \ge 1)$.
Then, $\delta_{u_{q}} =  \{ P_q(w) \} \cup \bigcup_{j=1}^{h_{q}} \delta(u_{q}^{j})$ (note that $\bigcup_{j=1}^{h_{q}} \delta(u_{q}^{j}) = \emptyset$ for the degenerated case $P_q(w) = (d \curvearrowleft w)$).
   By assumption, the invariant holds for $u_q^j, 1 \le j \le h_q,$ implying that
$\mathcal{CS}_{\delta(u_q^j)}, 1 \le j \le h_q$ are computed.

\vspace{-6pt}
\begin{itemize}
  \item  If $\hat{P}(w) \lessdot \hat{P}(u_q) (=P_q(w))$,
  then
$\hat{P}(w) = \min_{\lessdot}\{P_i(w) \mid 1 \le i \le q - 1\}$
 implies
    $\hat{P}(w) = \min_{\lessdot}\{P_i(w) \mid 1 \le i \le q\}$.
  Moreover, \textbf{Procedure} \texttt{Absorb-ear} is invoked.
On exiting the \textbf{while} loop,
 $\sigma(w) = \sigma(w) \cup \bigcup_{j=1}^{h_q} \sigma(u_q^j)$ and
as a result of absorbing $\mathcal{P}_{u_q}$,
$\delta(w) = \delta(w) \cup \delta_{u_q} =
\bigcup \{ \delta_{u_i} \mid 1 \le i \le q-1 \wedge i \neq \hat{\ell}\} \cup \delta_{u_q} =
\bigcup \{ \delta_{u_i} \mid 1 \le i \le q \wedge i \neq \hat{\ell}\}$,
where $\delta_{u_q} = \{ P_{ear(w \rightarrow u_q) }\} \cup \bigcup_{j=1}^{h_q} \delta(u_q^j)
 = \{ P_{q}(w) \} \cup \bigcup_{j=1}^{h_q} \delta(u_q^j)$.
By  Lemma~\ref{P0},
$\mathcal{CS}_{\delta_{u_q}} = P_q(w) \mathcal{CS}_{\delta(u_q^{h_q})} \mathcal{CS}_{\delta(u_q^{h_q-1})} \ldots \mathcal{CS}_{\delta(u_q^1)}$.

If $P_q(w) \lessdot \overleftarrow{P}(w)$,
  then
   $\overleftarrow{P}(w)
 = \min_{\lessdot}\{P_i(w) \mid 1 \le i \le q-1 \wedge i \neq \hat{\ell} \}$ implies that $P_q(w)
 = \min_{\lessdot}\{P_i(w) \mid 1 \le i \le q \wedge i \neq \hat{\ell} \}$ and
the then-part of the \textbf{if} statement is executed making
$\overleftarrow{P}(w) = P_q(w)$.
Moreover, By the induction hypothesis,
$\mathcal{CS}_{\delta(w)} = \mathcal{CS}_{\delta_{u_{\ell}}}  \mathcal{CS}_{\delta_{u_{i_1}}} \ldots  \mathcal{CS}_{\delta_{u_{i_{(q-1)-2}}}}$,
where
 $\{ u_{i_1}, u_{i_2}, \ldots, u_{i_{(q-1)-2}} \} = \{u_1, u_2, \ldots, u_{q-1}\} \setminus \{ u_{\ell}, u_{\hat{\ell}} \}$.
Since after the ears in $\mathcal{CS}_{\delta_{u_q}}$ are added to the construction sequence,
 $w$ becomes a branch vertex.
 $P_{\ell}(w)$ can thus be added to the construction sequence by operation $(i)$ or $(ii)$
following by the remaining ears in the current $\mathcal{CS}_{\delta(w)}$.
We thus have $\mathcal{CS}_{\delta(w)}
= \mathcal{CS}_{\delta_{u_q}} \ \mathcal{CS}_{\delta_{u_{\ell}}}  \mathcal{CS}_{\delta_{u_{i_1}}} \ldots  \mathcal{CS}_{\delta_{u_{i_{(q-1)-2}}}}$.
Since the anchor of $\mathcal{CS}_{\delta_{u_q}}$ is
 $P_q(w)  = \overleftarrow{P}(w)$,
   the assertion thus holds.

If $\overleftarrow{P}(w) \lessdot P_q(w)$, the else-part
is executed resulting in
   $\hat{P}(w) = P_{\hat{\ell}}(w) = \min_{\lessdot} \{ P_i(w) \mid 1 \le i \le q \}$,
       $\overleftarrow{P}(w) = P_{\ell}(w)
                 = \min_{\lessdot} \{ P_i(w) \mid 1 \le i \le q \wedge i \neq q\}$,
       $\delta(w) = \bigcup \{ \delta_{u_i} \mid 1 \le i \le q \wedge i \neq q \}$, and
$\mathcal{CS}_{\delta(w)}
= \mathcal{CS}_{\delta_{u_{\ell}}}  \mathcal{CS}_{\delta_{u_{i_1}}} \ldots  \mathcal{CS}_{\delta_{u_{i_{(q-1)-2}}}} \ \mathcal{CS}_{\delta_{u_q}}$
   as
 after the ears in the current $\mathcal{CS}_{\delta(w)}$ are added to the construction sequence,
  $w$ becomes a branch vertex, and
 $P_q(w)$ can be added to the sequence with operation $(i)$ or $(ii)$ following  by
 the remaining ears in $\mathcal{CS}_{\delta_{u_q}}$.
Since the anchor of $\mathcal{CS}_{\delta_{u_{\ell}}}$ is
 $P_{\ell}(w)  = \overleftarrow{P}(w)$,
the assertion thus holds.

  \item  If $\hat{P}(u_q) \lessdot \hat{P}(w)$, the argument is same as the above case except that the current $\mathcal{P}_w$ is absorbed by $w$.
   As a result,
       $\hat{P}(w) = P_q(w) = \min_{\lessdot} \{ P_i(w) \mid 1 \le i \le q \}$,
       $\overleftarrow{P}(w) = P_{\hat{\ell}}(w)
                 = \min_{\lessdot} \{ P_i(w) \mid 1 \le i \le q \wedge i \neq q\}$,
       $\delta(w) = \bigcup \{ \delta_{u_i} \mid 1 \le i \le q \wedge i \neq q \}$, and
$\mathcal{CS}_{\delta(w)}
= \mathcal{CS}_{\delta_{u_{\hat{\ell}}}} \ \mathcal{CS}_{\delta_{u_{\ell}}} \  \mathcal{CS}_{\delta_{u_{i_1}}} \ldots  \mathcal{CS}_{\delta_{u_{i_{(q-1)-2}}}}$
   as
 after the ears in $\mathcal{CS}_{\delta_{u_{\hat{\ell}}}}$ are added to the construction sequence,
 $w$ becomes a branch vertex.
$P_{\ell}(w)$ can thus be added to the sequence with operation $(i)$ or $(ii)$
 following  by the remaining ears in the current
$\mathcal{CS}_{\delta(w)} =
\mathcal{CS}_{\delta_{u_{\ell}}} \  \mathcal{CS}_{\delta_{u_{i_1}}} \ldots  \mathcal{CS}_{\delta_{u_{i_{(q-1)-2}}}}$.
Since $P_{\hat{\ell}}(w) = \overleftarrow{P}(w)$, the assertion thus holds.

\end{itemize}

\vspace{-9pt}
$(ii)$ If $(u_q \curvearrowleft w) \in E \setminus E_T$, the else-part of the \textbf{if} statement is executed.
The remaining argument is same as the above case but much simpler as
 $\mathcal{P}_{u_q} = nil$,  and
$\delta_{u_q} = \{ (u_q \curvearrowleft w) \}$ and
$\mathcal{CS}_{\delta_{u_q}} = (u_q \curvearrowleft w)$
by Lemma~\ref{P0}.

The assertion thus holds for $q$.
 Hence, when $L[w]$ is completely processed,


$\hat{P}(w) =
P_{\hat{\ell}}(w) = \min_{\lessdot} \{ P_i(w) \mid 1 \le i \le m \}
= \min_{\lessdot} \hat{P}_w$,
$\sigma(w) =
\{w\} \cup \bigcup
\{ \bigcup_{j=1}^{h_{i}}\sigma(u_{i}^{j}) \mid 1 \le i \le m \wedge i \neq \hat{\ell} \}$ which consists of the vertices in $V_{T_w} \setminus V_{T_{u_{\hat{\ell}}}}$ which are precisely the vertices absorbed by $w$,
$\delta(w) =
\bigcup \{\delta_{u_{i}} \mid 1 \le i \le m \wedge i \neq \hat{\ell} \}
= \bigcup \{\{ P_i(w) \} \cup \bigcup_{j=1}^{h_{i}} \delta(u_{i}^{j}) \mid 1 \le i \le m \wedge i \neq \hat{\ell} \}
= \{ P_i(w) \mid 1 \le i \le m \wedge i \neq \hat{\ell} \}
  \cup \bigcup \{ \bigcup_{j=1}^{h_{i}} \delta(u_{i}^{j}) \mid 1 \le i \le m \wedge i \neq \hat{\ell} \}$
 which is the set of ears with their sink in $\sigma(w)$,
$\overleftarrow{P}(w) =
P_{\ell}(w) = \min_{\lessdot} (\{ P_i(w) \mid 1 \le i \le m \wedge i \neq \hat{\ell} \}
= \min_{\lessdot} (\{ \min_{\lessdot} \delta_{u_i} \mid 1 \le i \le m \wedge i \neq \hat{\ell} \}$
(since $P_i(w) = \min_{\lessdot}\delta_{u_i})
 = \min_{\lessdot} ( \bigcup \{ \delta_{u_i} \mid 1 \le i \le m \wedge i \neq \hat{\ell} \})
 = \min_{\lessdot} \delta(w)$,
  and

$\mathcal{CS}_{\delta(w)} = \mathcal{CS}_{\delta_{u_{\ell}}}  \mathcal{CS}_{\delta_{u_{i_1}}} \ldots  \mathcal{CS}_{\delta_{u_{i_{m-2}}}}$,
where
 $\{ u_{i_1}, u_{i_2}, \ldots, u_{i_{m-2}} \} = \{u_1, u_2, \ldots, u_m\} \setminus \{ u_{\ell}, u_{\hat{\ell}} \}$ such that
$\mathcal{CS}_{\delta_{u_{\ell}}} = \overleftarrow{P}(w) \mathcal{CS}_{\delta(u^{h_{\ell}}_{\ell})} \mathcal{CS}_{\delta(u_{\ell}^{h_{\ell-1}})} \ldots \mathcal{CS}_{\delta(u_{\ell}^1)}$ and

$\mathcal{CS}_{\delta_{u_{i_j}}} = P_{i_j}(w) \mathcal{CS}_{\delta(\tilde{u}^{\tilde{h}_{j}}_{j})} \mathcal{CS}_{\delta(\tilde{u}_j^{\tilde{h}_j-1})} \ldots \mathcal{CS}_{\delta(\tilde{u}_j^1)}, 1 \le j \le m-2$, where $\tilde{u}_j = u_{i_j}$ and $\tilde{h}_j = h_{i_j}$.


If $\mathcal{P}_w = w$ or $Inc_w = \emptyset$,  the \textbf{if} statement following the \textbf{for} loop is not executed and the $\mathit{dfs}$ backtracks to $v$ with
$\mathcal{CS}_{\delta(w)}$. Hence, the invariant holds for vertex $w$.

Otherwise, \textbf{Procedure} \texttt{Absorb-path} is invoked.
Let $\mathcal{P}_w$ be $(w=)w_0 w_1 \ldots w_k (k \ge 1)$.
 The first \textbf{for} loop determines the largest index $h$ such that
 there is a back-edge $(w \curvearrowleft x) \in Inc$ of which $(w \curvearrowleft w_h)$ is an embodiment.
 Then $\overleftarrow{P}(w_{\ell}) = \min_{\lessdot}\{\overleftarrow{P}(w_j) \mid 0 \le j \le h \}$ is computed.
 By assumption, the invariant holds for $w_i, 0 \le i \le h$.
Hence, by Lemma~\ref{P1},
  the \textbf{if} statement and the \textbf{for} loop following it
  generates $CS_{\delta(w)}$.
The invariant thus holds for vertex $w$.
Since vertex $w$ has absorbed the path $(w=) w_0 w_1 \ldots w_h$,
  the last \textbf{for} loop updates $\sigma(w) = \bigcup_{j=0}^h \sigma(w_j)$ and
shortens $P_w$ to $(w=)w_0 w_{h+1} \ldots w_k$.

  Finally, if $w=r$,
when execution of \textbf{Procedure} \texttt{3-edge-connect-CS$( r, \perp)$} terminates,
the last instruction in \textbf{Algorithm}  \texttt{Certifying-3-edge-connectivity}  makes
 $\hat{P}(r) \mathcal{CS}_{\delta(r)}$ the Mader construction sequence of $\acute{G}_{\sigma(r)}$ based on Lemma~\ref{CS-for-3ecc}$(i)$.
The theorem thus follows.
 \ \ \ \ \ \  $\blacksquare$

 A complete example is given in the Appendix (Figure~7).


  To assure that the algorithm runs in linear-time, we maintain the following data-structures:

$(i)$ An array $parent[w], w \in V \setminus \{r\}$, such that $parent[w]$ is the parent vertex of $w$ in $T$:
 this array allows an ear $P_f$ of length $k$ to be traced in $O(k)$ time starting from its back-edge $f$.
$(ii)$ An array $ear[w], w \in V \setminus \{r\}$, such that $ear[w] = ear(parent[w] \rightarrow w)$:  this array allows the ear to which a tree edge belongs to be determined in $O(1)$ time.
  Both arrays can be created in $O(|V|)$ time during the depth-first search.
   The adjacency lists, $L[w], w \in V$,  can clearly be constructed in $O(|V|+|E|)$ time.
    $Inc_w$ and the $\mathcal{CS}$'s are represented by linked lists.

\begin{thm}\label{time}
Algorithm Certifying-3-edge-connectivity takes $O(|V|+|E|)$ time.

\end{thm}

\noindent \textbf{Proof:}
   \textbf{Algorithm} \texttt{Certifying-3-edge-connectivity} is an extension of
\textbf{Algorithm} \texttt{3-edge-} \texttt{connectivity} of~\cite{T07}.
  The extension includes new instructions for generating $\mathcal{CS}_{\delta(w)}, \hat{P}(w), \overleftarrow{P}(w), Inc_w,$ $w \in V$.
   Since \textbf{Algorithm} \texttt{3-edge-}\texttt{connectivity} takes $O(|V|+|E|)$ time~\cite{T07},
it suffices to show that the extension takes $O(|V|+|E|)$ time.

In \textbf{Procedure} \texttt{3-edge-connect-CS} (whose counterpart is \textbf{Procedure} \texttt{3-edge-connect} in~\cite{T07}),
 the new instructions for initialization clearly takes $O(1)$ time.
The instructions for generating $Inc_w, w \in V,$ takes $O(\sum_{w \in V}|Inc_w|)= O(|E|-|E_T|)$ time. 
\textbf{Procedure} \texttt{Absorb-ear} increases the time spent on absorbing the entire $u$-path or
$w$-path by a constant factor.
In \textbf{Procedure} \texttt{Absorb-path}, the \textbf{for each} statement takes $O(h + |Inc_w|)$ time
(note that verifying ancestor relation can be done in $O(1)$ time~\cite{T07}).
Determining $\overleftarrow{P}(w_{\ell})$ takes $O(h)$ time.
 The \textbf{for} loops take $O(h)$ time. The remaining new instructions takes $O(1)$ time.
   Since each tree-edge is absorbed at most once, $\sum_{w \in V} ( O(h)) = O|E_T|$.
Hence, in total, the procedure takes $\sum_{w \in V} ( O(h + |Inc_w|) + O(h) + O(1) ) = O(|E_T|) + O(|E|-|E_T|) + O(|E_T|) + O(|V|) = O(|E|)$ time.
\textbf{Procedure} \texttt{Gen}-$\mathcal{CS}$ clearly takes $O(1)$ time.
 Since  $\dot{G}_{\langle \sigma(w) \rangle}, w \in V,$ are disjoint and
 each virtual edge in $\dot{G}_{\langle \sigma(w) \rangle}$ corresponds to a cut-edge in $G$,
 generating their Mader construction sequences takes $O(|V|+|E|)$ time.
 Hence, the new instructions take $O(|V|+|E|)$ time.
  \textbf{Algorithm} Certifying-3-edge-connectivity thus runs in $O(|V|+|E|)$ time.
\ \ \ \ \ $\blacksquare$

\subsection{Generating a cactus representation (negative certificate)}


Let $G=(V,E)$ be a 2-edge-connected undirected graph containing cut-pairs.
A \emph{cactus representation} of the cut-pairs of $G$
consists of an undirected graph $\mathcal{C}_G = (V_{\mathcal{C}},E_{\mathcal{C}})$,
 where the elements in $V_{\mathcal{C}}$ are called \emph{node}s, and a function $\Phi: V \rightarrow V_{\mathcal{C}}$
 such that the cut-pairs of $G$ are precisely the preimages of the cut-pairs of $\mathcal{C}_G$.
  Specifically,
$\forall X \subseteq V_{\mathcal{C}}, (X,\overline{X})$ is a cut-pair in $\mathcal{C}_G$ if and only if
  $(\Phi^{-1}(X), \overline{\Phi^{-1}(X)}$) is a cut-pair in $G$
(Note that $(X,\overline{X})$ is the pair of edges with one end-vertex in $X$ and the other in $\overline{X}$, where $\overline{X}  = V_{\mathcal{C}} \setminus X$).
Nagamochi et al.~\cite{NI08} pointed out that the nodes in $V_{\mathcal{C}}$ are precisely the 3eccs of $G$ which
   can be determined by contracting the latter.
Since \textbf{Algorithm} \texttt{3-edge-connectivity} uses contraction to determine the 3eccs,
it can be easily modified to produce
a cactus representation of the cut-pairs for each 2-edge-connected component of $G$.

   Since if $\{e,e'\}$ and $\{e',e''\}$ are cut-pairs, $\{e,e''\}$ is also a cut-pair~\cite{TWO92,T09},
  the set of all cut-edges can be partitioned into a collection of disjoint subsets
$\mathcal{E}_i, 1 \le i \le \kappa,$ such that every two edges from the same subset form a cut-pair and no two edges from different subsets form a cut-pair.
   Moreover, as the two edges of each cut-pair lie on the same ear~\cite{TWO92,T09},
  the cut-edges in each $\mathcal{E}_i$ lie on the same ear $P$ and hence
 can be lined up along $P$ in an order $e_1 e_2 \ldots e_{|\mathcal{E}_i|}$, where $e_1 = (x_1 \hookrightarrow y_1)$ or $e_1 = (x_1 \rightarrow y_1)$, and $e_j = (x_j \rightarrow y_j), 2 \le j \le |\mathcal{E}_i|$,
 such that $x_{j+1} \preceq y_j, 1 \le j < |\mathcal{E}_i|$.
   Note that we use $(x \hookrightarrow y)$ instead of $(y \curvearrowleft x)$ to represent back-edges in this section and
   that the orientation of the tree edges on $P$ follows that of $P$ which is from the child to the parent.
  Each $\mathcal{E}_i$ is called a \emph{cut-edge chain} and the cut-edge $e_1$ is called the generator of $\mathcal{E}_i$~\cite{TWO92,T09}.
  From the above discussion, each $\mathcal{E}_i$ gives rise to a unique cycle in the cactus $\mathcal{C}_G$. Hence, it remains to show how to convert $\mathcal{E}_i$ to a cactus cycle.

  Let $P$ be the ear containing $\mathcal{E}_i$.
Then, either

$(i)$ $P: s(P)= x \overset{f}{\hookrightarrow} y \rightsquigarrow_T x_1 \overset{e_1}{\rightarrow} y_1
     \rightsquigarrow_T x_2 \overset{e_2}{\rightarrow} y_2
     \rightsquigarrow_T  \cdots  \rightsquigarrow_T
     x_{|\mathcal{E}_i|} \overset{e_{|\mathcal{E}_i|}}{\rightarrow} y_{|\mathcal{E}_i|} \rightsquigarrow_T t(P)$, or

$(ii)$ $P: s(P)= x_1 \overset{e_1}{\hookrightarrow} y_1 \rightsquigarrow_T x_2 \overset{e_2}{\rightarrow} y_2
     \rightsquigarrow_T  \cdots  \rightsquigarrow_T
     x_{|\mathcal{E}_i|} \overset{e_{|\mathcal{E}_i|}}{\rightarrow} y_{|\mathcal{E}_i|} \rightsquigarrow_T t(P)$.

Note that
the generator $e_1$ is a tree-edge in Case $(i)$ and a back-edge in Case $(ii)$.
   We shall modify \textbf{Algorithm} \texttt{3-edge-connectivity} to generate a cycle,
$\sigma(x_1) \sigma(x_2)  \cdots \sigma(x_{|\mathcal{E}_i|}) \sigma(x_1)$,
 for $\mathcal{C}_G$ based on $P$ and $\mathcal{E}_i$  as follows.

First, consider Case $(i)$  (Figure~5$(i)$).
When the $\mathit{dfs}$ backtracks from $x_2$ to $y_2$, either $x_2 = y_1$ or $x_2$ has absorbed $y_1$ (directly or indirectly).
Hence, $\mathcal{P}_{x_2}$ is $x_2 x_1 v_1 \ldots v_h$.
Since $deg_{\hat{G}}(x_2) = 2$, $x_2$ is ejected and $\sigma(x_2)$ becomes a $3ecc$
which is a node in the cactus cycle corresponding to $\mathcal{E}_i$.
   This implies that the path $\sigma(x_1)\sigma(x_2)\sigma(y_2)$ exists in $\mathcal{C}_G$.
This path is represented by its internal node $\sigma(x_2)$ which is attached to $x_1$
as $x_1.tchain$.
When the $\mathit{dfs}$ backtracks from $x_3$ to $y_3$,
  either $x_3 = y_2$ or $x_3$ has absorbed $y_2$.
$\mathcal{P}_{x_3}$ is thus $x_3 x_1 v_1 \ldots v_h$ and $\sigma(x_3)=\sigma(y_2)$.
   Hence, $x_1.tchain$ represents
 the path $\sigma(x_1)\sigma(x_2)\sigma(x_3)$.
   Since $deg_{\hat{G}}(x_3) = 2$, $x_3$ is ejected and $\sigma(x_3)$ becomes an internal  node of the path $\sigma(x_1)\sigma(x_2)\sigma(x_3)\sigma(y_3)$ in $\mathcal{C}_G$.
Hence, $x_1.tchain$ becomes $\sigma(x_2) \sigma(x_3)$.
   Similarly,
when the $\mathit{dfs}$ backtracks from $x_{|\mathcal{E}_i|}$ to $y_{|\mathcal{E}_i|}$,
  $\mathcal{P}_{x_{|\mathcal{E}_i|}}$ is $x_{|\mathcal{E}_i|} x_1 v_1 \ldots v_h$.
Since $deg_{\hat{G}}(x_{|\mathcal{E}_i|}) = 2$, $x_{|\mathcal{E}_i|}$ is ejected and $\sigma(x_{|\mathcal{E}_i|})$ becomes an internal node of the path $\sigma(x_1)\sigma(x_2)\sigma(x_3) \ldots \sigma(x_{|\mathcal{E}_i|}) \sigma(y_{|\mathcal{E}_i|})$ in $\mathcal{C}_G$.
   Hence, $x_1.tchain$ becomes $\sigma(x_2) \sigma(x_3) \ldots \sigma(x_{|\mathcal{E}_i|})$.
  Then, one of the following two cases follows.

 \begin{figure}
 \centering
 \includegraphics[width=6.5in]{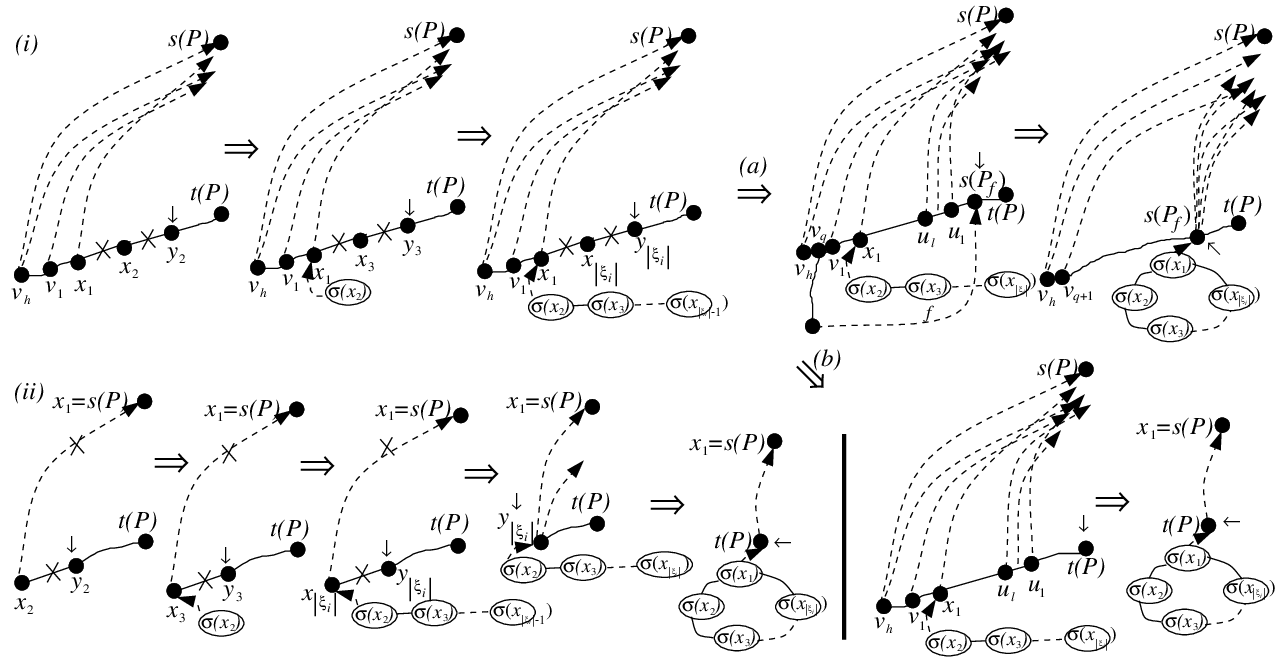}
 \caption[ ]{.}
 \end{figure}\label{Fig6.eps}

\vspace{-6pt}
\begin{description}
  \item[$(a)$]   
  There exists an ear $P'$ such that $s(P') \preceq y_{|\mathcal{E}_i|}$ and
$x_1 \preceq t(P')$: (Figure 5$(i)(a)$).
   Let $P_f$ be the one with $s(P_f)$ closest to $y_{|\mathcal{E}_i|}$.
 After the $\mathit{dfs}$ backtracks to $s(P_f)$, when the back-edge $f$ is encountered,
let $\mathcal{P}_{s(P_f)}$ be $s(P_f) u_1 u_2 \ldots u_{\ell} x_1 (=v_0) v_1 v_2 \ldots v_h, (\ell, h \ge 0),$
(note that $y_{|\mathcal{E}_i|} \in \sigma(u_{\ell})$)
and $t(P_f) \in \sigma(v_q)$.
   Then, $s(P_f)$ absorbs $u_1 u_2 \ldots u_{\ell} x_1 v_1 \ldots v_{q}$ which includes $x_1$ and $u_{\ell}$ (hence, $y_{|\mathcal{E}_i|}$).
  As a result, $\sigma(s(P_f)) = \sigma(x_1)= \sigma(y_{|\mathcal{E}_i|})$.
  Hence, the path $\sigma(x_2) \sigma(x_3) \ldots \sigma(x_{|\mathcal{E}_i|})$ kept in $x_1.tchain$ and $\sigma(x_1)$ form a cactus cycle
$\sigma(x_1) \sigma(x_2) \sigma(x_3) \ldots \sigma(x_{|\mathcal{E}_i|}) \sigma(x_1)$ in $\mathcal{C}_G$. The cycle is attached to $s(P_f)$.
   If $s(P_f)$ is absorbed by another vertex at a later stage, the cycle is attached to that vertex  until a vertex $z$ to which the cycle is attached is ejected.
Then, as $\sigma(z) = \sigma(x_1)$,
$\sigma(z) \sigma(x_2) \sigma(x_3) \ldots \sigma(x_{|\mathcal{E}_i|}) \sigma(z)$ is a cactus cycle in $\mathcal{C}_G$.

\vspace{-6pt}   
  \item[$(b)$] There does not exist such an ear $P'$ : (Figure~5$(i)(b)$)
     After the $\mathit{dfs}$ backtracks to $t(P)$,
   let $t(P) \rightsquigarrow_T y$ be transformed into
  $t(P) u_1 u_2 \ldots u_{\ell} x_1 (=x_0) v_1 v_2 \ldots v_h, \ell, h \ge 0,$
  (note that $y_{|\mathcal{E}_i|} \in \sigma(u_{\ell})$)
when $t(P)$ is to absorb it.
     Then, after $t(P)$ absorbed $x_1$ and $u_{\ell}$ (hence, $y_{|\mathcal{E}_i|}$),
  $\sigma(x_1) = \sigma(y_{|\mathcal{E}_i|})$ which implies that
   $x_1.tchain$ and $\sigma(x_1)$ form a cactus cycle
$\sigma(x_1) \sigma(x_2) \sigma(x_3) \ldots \sigma(x_{|\mathcal{E}_i|}) \sigma(x_1)$ in $\mathcal{C}_G$.
  The rest of the argument is same as that for Case $(a)$.
\end{description}

\vspace{-9pt}
 Next, consider Case $(ii)$ (Figure~5$(ii)$).
This case is similar to Case $(i)(b)$ except after ejecting $x_2$,
the node $\sigma(x_2)$  is attached to $y_2$ instead of $x_1$ as $y_2.bchain$.
    Then, $y_2.bchain$ is transferred along $y_2 \rightsquigarrow_T t(P)$ through the absorb operation until
   it is transferred to $x_3$, at which it becomes $x_3.bchain$.
  After the $\mathit{dfs}$ backtracks to $y_3$, $x_3$ is ejected and $x_3.bchain$ is extended to
$\sigma(x_2) \sigma(x_3)$ which then becomes $y_3.bchain$.
 This process is repeated for $x_i, 4 \le i \le |\mathcal{E}_i|$.
When the $\mathit{dfs}$ backtracks to $y_{|\mathcal{E}_i|}$,
 after  $x_{|\mathcal{E}_i|}$ is ejected, $x_{|\mathcal{E}_i|}.bchain$ is extended to
$\sigma(x_2) \sigma(x_3) \ldots \sigma(x_{|\mathcal{E}_i|})$ which then becomes $y_{|\mathcal{E}_i|}.bchain$.
 After the $\mathit{dfs}$ backtracks to $t(P)$,
 when $t(P)$ absorbs $u_{\ell}$ such that
    $y_{|\mathcal{E}_i|} \in \sigma(u_{\ell})$ and
    $u_{\ell}.bchain = y_{|\mathcal{E}_i|}.bchain$,
$\sigma(t(P)) = \sigma(y_{|\mathcal{E}_i|})$
which implies that $y_{|\mathcal{E}_i|}.bchain$ represents the path
$\sigma(x_1) \sigma(x_2) \sigma(x_3) \ldots \sigma(x_{|\mathcal{E}_i|}) \sigma(t(P))$.
 Since $(x_1 =) s(P)$ and $t(P)$ are 3-edge-connected, $\sigma(x_1) =  \sigma(t(P))$.
 Hence, $y_{|\mathcal{E}_i|}.bchain$ and $\sigma(x_1)$ form a cactus cycle
 $\sigma(x_1) \sigma(x_2) \sigma(x_3) \ldots \sigma(x_{|\mathcal{E}_i|}) \sigma(x_1)$ in $\mathcal{C}_G$.
 The remaining argument is same as that for Case $(i)(a)$.


The above modifications can be easily incorporated into \textbf{Algorithm} \texttt{Certifying-3-edge-connectivity}.
The following is a pseudo-code of the modified algorithm. For clarity,
instructions that are not related to the construction of $\mathcal{C}_G$ have been omitted.
The new instructions for generating the cactus are marked by $\bullet$.

At each vertex $x$, the following variables are maintained:

\vspace{-12pt}
\begin{itemize}
  \item $x.tchain$: a chain of nodes ($\sigma(x_i)$'s) attached to $x$ corresponding to a cut-edge-chain whose generator is the parent edge of $x$, and
      $\sigma(x) + x.tchain$ is a path in $\mathcal{C}_G$;
\vspace{-9pt}
  \item $x.bchain$: a chain of nodes ($\sigma(x_i)$'s) attached to $x$ corresponding to a cut-edge-chain with back-edge generator such that $x.bchain + \sigma(x)$ is a path in $\mathcal{C}_G$;
\vspace{-9pt}
  \item $C.cycle(x)$: contains the cactus cycles attached to $x$.

\end{itemize}

   A node $\sigma(x)$ in a cactus cycles is represented by $x$.
   When a vertex $u$ is ejected,
 if $deg_{\hat{G}_u}(u) = 1$, then
  a cactus representation of the cut-pairs in $G_{\langle \sigma(u) \rangle}$ is constructed.
If $deg_{\hat{G}_u}(u) = 2$, $\sigma(u)$ becomes a node in $\mathcal{C}_G$.
 \textbf{Procedure} \texttt{Gen-$\mathcal{CS}$} is called to append $\sigma(u)$ to $u_1.tchain$ if both cut-edges are tree-edge, where $u_1$ is the vertex following $u$ in the $u$-path before $u$ is ejected, or to $u.bchain$, otherwise. In the latter case, the updated $u.bchain$ is
temporary kept at $bchain_t$ because if $\mathcal{P}_w = w$, $w.bchain$ might have contained a cactus chain.
If $\mathcal{P}_u \lessdot \mathcal{P}_w$, then $bchain_t$ becomes $w.bchain$  after the current $w.bchain$ (if exists) is turned into a cactus cycle and attached to $w$;
if  $\mathcal{P}_w \lessdot \mathcal{P}_u$, $bchain_t$ is turned into a cactus cycle and attached to $w$~(Figure~6$(i)$).
  When vertex $w$ absorbs $\mathcal{P}_w$ or $\mathcal{P}_u$, \textbf{Procedure}  \texttt{Absorb-ear} is called to
transfer all cactus cycles attached to each absorbed vertex to $w$, and convert
  the $tchain$ (if exists) of each absorbed vertex as well as the $bchain$ (if exists) of the last absorbed vertex into cactus cycles and attach them to $w$~(Figure~6$(ii)$).
\textbf{Procedure} \texttt{Absorb-path} is similar to \textbf{Procedure}  \texttt{Absorb-ear}
 but is called to absorb a section of $\mathcal{P}_w$ only.
 An example is given in the Appendix (Figure~8).

\begin{figure}[h!]
\centering
\includegraphics[width=6.5in]{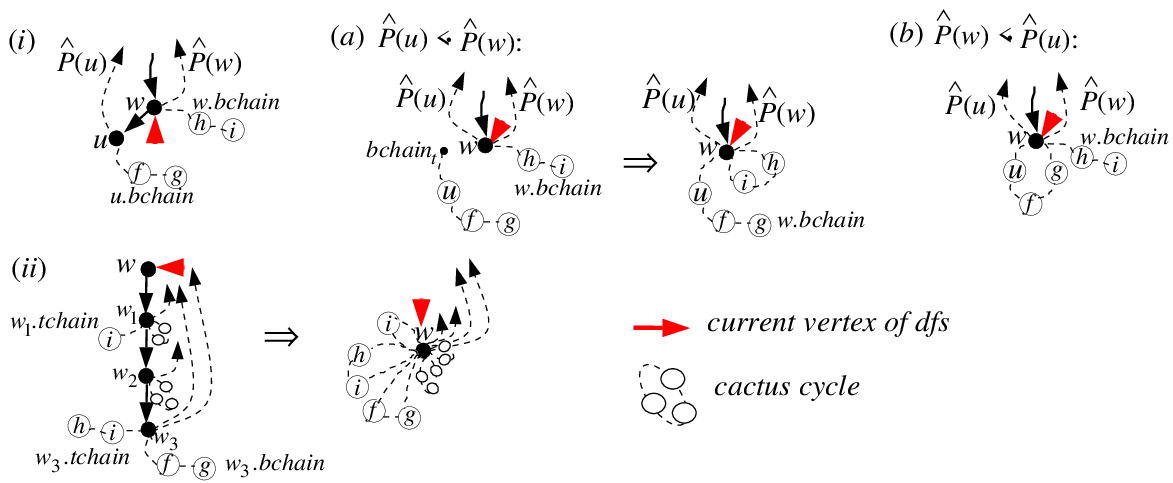}

\caption[ ]{
}
 \end{figure}\label{Fig7.eps}


\begin{singlespacing}
\footnotesize

\noindent \textbf{Algorithm} \texttt{Certifying-3-edge-connectivity}

\noindent \textbf{Input:} A connected graph $G=(V,E)$ represented by adjacency lists $L[w], \forall w \in V$

\noindent \textbf{Output:} $\left\{
                             \begin{array}{ll}
                     \mathcal{CS}_{\delta(u)}, u \in V     & \hbox{\ Mader  construction  sequence  for each $u$ representing a  $3ecc$ \ of \ G, } \\
                     \{ \mathcal{C}_{G'} \mid G' \text{ is a 2ecc of } G \}    & \hbox{\ a cactus representation of the cut-pairs for each 2$ecc$ $G'$ of $G$, \ and }\\
                     Bridges         & \hbox{ \ the bridges in $G$ }
                             \end{array}
                           \right.$

\noindent \textbf{begin}

 \textbf{for every} $u \in V$ \textbf{do} $dfs(u) := 0; \ parent(u) := \perp; \ lowpt(u) := \infty$;  \ \ \ // initialization; $\perp = $ undefined

\noindent $\bullet$ \hspace{8pt} \hspace{72pt}
    $\hat{P}(u) := \overleftarrow{P}(u) := \perp$;
\ \ \   // \textbf{note:} $\perp \gtrdot P_f, \forall f \in E \backslash E_T$;

\noindent $\bullet$ \hspace{8pt}  \hspace{72pt}
    $\sigma(u) := \{u\}; \ Inc_u := \emptyset;  \ \mathcal{P}_u := u$;

 $cnt := 1$; \ \ \  // $\mathit{dfs}$ number counter   \ \ \

 \texttt{3-edge-connect-CS}$(r, \perp)$;

\noindent $\bullet$
    \  \textbf{for each} (cycle $xQx \in C.cycle(r) \wedge x \neq r)$ \textbf{do}

\noindent $\bullet$ \hspace{8pt} \hspace{15pt}      Convert $xQx$ to $rQr$;
  \hspace{15pt}  \ \ \ // make $r$ the starting and ending node of the cycle; note: $\sigma(r) = \sigma(x)$

\noindent \textbf{end.}\\

%

\noindent \textbf{Procedure} \texttt{3-edge-connect-CS}$(w, v)$

\noindent \textbf{begin}

 $dfs(w) := cnt; \ cnt := cnt + 1; \ parent(w) := v$;
 \ $lowpt(w) := dfs(w)$;


\noindent $\bullet$ \hspace{8pt}
$w.tchain := w.bchain := bchain_t := \perp$; \ $C.cycle(w) := \emptyset$;
\ \ \ \ \ \hspace{25pt}  // initialization


\vspace{3pt}
 \textbf{for each} $u \in L[w]$ \textbf{do} \ \ \
\hspace{174pt} // pick the next vertex $u$ in the adjacency list of $w$


 \hspace{15pt} \textbf{if} $(dfs(u) = 0)$ \textbf{then} \ \ \ \hspace{160pt}   // $u$ is unvisited

  \hspace{30pt} \texttt{3-edge-connect-CS}$(u,w)$;

  \hspace{30pt} \textbf{if}  $(s(\overleftarrow{P}(u)) = u \vee \overleftarrow{P}(u) = \perp)$ \textbf{then} \ \ \ \ \
  \ \ \ \hspace{77pt} // equivalent to $(deg_{\hat{G}_u}(u) \le 2)$

\noindent $\bullet$ \hspace{8pt}  \hspace{45pt}    \texttt{Gen-}$\mathcal{CS}( w, u, \mathcal{P}_u, bchain_t)$;
  \ \  \hspace{100pt}
 // create a node $\sigma(u)$ in the cactus

     \hspace{30pt} \textbf{if} $( \hat{P}(w) \lessdot \hat{P}(u) )$ \textbf{then}
\ \ \  \hspace{135pt} // equivalent to $(lowpt(w) \le lowpt(u))$ in ~\cite{T07}

 \hspace{45pt} \texttt{Absorb-ear($w, u, \hat{P}(u), w + \mathcal{P}_u, bchain_t$)}
  \  \ \ \ \hspace{30pt} // $w$ absorbs the entire $u$-path

  \hspace{30pt} \textbf{else}
   \ \ \ \ \ // $( \hat{P}(u) \lessdot \hat{P}(w) )$;
   \ \ \hspace{126pt} // equivalent to $(lowpt(w) > lowpt(u))$ in ~\cite{T07}

 \hspace{45pt}  \texttt{Absorb-ear($w, u, \hat{P}(w), \mathcal{P}_w, \perp$)};
 \hspace{68pt}  \ \ \  // $w$ absorbs the entire $w$-path

\noindent $\bullet$ \hspace{8pt} \hspace{45pt}
   $w.bchain := bchain_t$; \  $bchain_t := \perp$;
    \ \ \hspace{65pt}  // transfer $bchain_t$ to $w.bchain$

\hspace{45pt} $\mathcal{P}_w := w + \mathcal{P}_u$;\  $\hat{P}(w) := \hat{P}(u)$;
   \ \  \hspace{132pt}  

\hspace{45pt} \textbf{if} $(w \neq r)$ \textbf{then} $ear(v \rightarrow w) := ear(w \rightarrow u)$;

 \hspace{15pt} \textbf{else if} $(dfs(u) < dfs(w) \wedge u \neq parent(w))$ \textbf{then}
   \ \  \hspace{50pt}  // $u \curvearrowleft w$ is an outgoing back-edge of $w$

    \hspace{60pt} \textbf{if} $( (u \curvearrowleft w) \lessdot \hat{P}(w) )$ \textbf{then}
  \ \   \hspace{90pt}  // equivalent to $(dfs(u) < lowpt(w))$ in ~\cite{T07}

\noindent $\bullet$ \hspace{8pt}     \hspace{80pt}   \texttt{Absorb-ear($w, u, \hat{P}(w), \mathcal{P}_w, \perp$)};
\ \ \ \hspace{30pt} // $w$ absorbs the entire  $w$-path

    \hspace{80pt}  $P_w := w$; \  $\hat{P}(w) := u \curvearrowleft w$;
   \ \   \hspace{65pt}  // equivalent to $(lowpt(w) := dfs(u))$ in ~\cite{T07}

\hspace{80pt} \textbf{if} $(w \neq r)$ \textbf{then} $ear(v \rightarrow w) :=
(u \curvearrowleft w)$;

\hspace{60pt} \textbf{else} \texttt{Absorb-ear($w, u, u \curvearrowleft w,   nil, \perp$)};
     \  \hspace{5pt} // there is no $u$-path; \ no chain is to be updated

  \hspace{30pt} \textbf{else} $Inc_w := Inc_w \cup \{(w \curvearrowleft u)\}$;
\   \hspace{96pt}   // save incoming back-edge

\vspace{3pt}
\noindent \hspace{15pt}       \textbf{if} $( (\mathcal{P}_w \neq nil) \wedge (Inc_w \neq \emptyset) )$ \textbf{then}
                       \texttt{Absorb-path}$(w, \mathcal{P}_w, Inc_w)$;
\ \ \ \  // dealing with incoming back-edges

\vspace{3pt}



\vspace{3pt}
\noindent \textbf{end.} /* of Procedure \texttt{3-edge-connect-CS} */

\vspace{9pt}
\noindent \textbf{Procedure} \texttt{Gen}-$\mathcal{CS}(w, u, \mathcal{P}_u, bchain_t)$

\noindent \textbf{begin} \ \ \ // Create a cactus node $\sigma(u)$  and attach it to the corresponding $tchain$ or $bchain$.

\noindent $\bullet$ \hspace{8pt}        \textbf{for each} (cycle $xQx \in C.cycle(u) \wedge x \neq u)$ \textbf{do}

\noindent $\bullet$ \hspace{8pt} \hspace{15pt}      Convert $xQx$ to $uQu$;
  \hspace{65pt}  \ \ \ // make $u$ the starting and ending node of the cycle; note: $\sigma(u) = \sigma(x)$


    \textbf{if} $( \hat{P}(u) = \perp \vee s(\hat{P}(u)) = u )$ \textbf{then}
 \hspace{35pt} \ \ \   // $(w,u)$ is a bridge (i.e. $deg_{\hat{G}_u}(u) = 1$)

\noindent $\bullet$ \hspace{8pt} \hspace{15pt}   // Construction of a cactus representation for the
2-edge-connected component containing $\sigma(u)$ is complete;

\hspace{15pt}   $\hat{P}(u) := ear(w \rightarrow u) := \perp; \ \mathcal{P}_u := nil$;
\hspace{5pt} // update $\hat{P}(u)$ and $\mathcal{P}_u$ accordingly


  \textbf{else}  \ \ \ \ \ // $deg_{\hat{G}_u}(u) = 2$

 \hspace{15pt}     \textbf{if} ($\mathcal{P}_u = u$) \textbf{then}
  \hspace{85pt}  \ \ \   // $\mathcal{P}_u: u$, i.e. the generator of the cut-edge chain is a back-edge.

\noindent $\bullet$ \hspace{8pt}  \hspace{30pt}        $bchain_t := u.bchain + u$;
  \hspace{32pt}   \ \ \  // extend $u.bchain$ to include $u$ and keep it in $bchain_t$ temporarily.

\hspace{8pt}  \hspace{20pt}
  $d := t(ear(w \rightarrow u))$;
  $\hat{P}(u) := (d \curvearrowleft w)$;
  $\mathcal{P}_u := nil$;   \hspace{85pt}

 \hspace{15pt}       \textbf{else}
 \ \ \ \hspace{130pt} //  $\mathcal{P}_u: u u_1 \ldots u_{\ell}, \ell \ge 1$, i.e. the generator of the cut-edge chain is a tree-edge

\noindent $\bullet$ \hspace{8pt}  \hspace{32pt}           $u_1.tchain := u_1.tchain  + u$;
 \hspace{15pt}   \ \ \  // extend $u_1.tchain$ to include $u$, where $\{(u_1 \rightarrow u), (u \rightarrow w)\}$ is the cut-pair

\hspace{8pt}  \hspace{20pt} $parent(u_1) := w$; \ $ear(w \rightarrow u_I) := ear(w \rightarrow u)$;

\hspace{8pt}  \hspace{20pt}
    $\mathcal{P}_u := \mathcal{P}_u - u$;   \hspace{70pt}

\vspace{3pt}
\noindent \textbf{end.}




\vspace{9pt}

\noindent \textbf{Procedure} \texttt{Absorb-ear$(w, u, \hat{P},  \mathcal{P}, bchain_t$)}  \ \ \     // absorb the entire $\mathcal{P}$ which is either $\mathcal{P}_w$ or $w + \mathcal{P}_u$

\noindent \textbf{begin} \ \ \
/* $\mathcal{P}: (w=)x_0 x_1 x_2 \ldots x_{k-1} x_k$ is either a $\mathcal{P}_w$ or $w + \mathcal{P}_u$;
 $x_i.next = x_{i+1}, 0 \le i < k$ and
 $x_k.next = \perp$.


\noindent \textbf{if} $(\mathcal{P} \neq nil)$ \textbf{then}

\noindent $\bullet$ \hspace{8pt}
\textbf{if} $( bchain_t \neq \perp )$ \textbf{then}
  \hspace{150pt} // $\mathcal{P} = u$; $u.bchain$ exists and is kept in $bchain_t$;.

\noindent $\bullet$ \hspace{8pt}
\hspace{20pt}   \emph{C.cycle}$(w)$ := \emph{C.cycle}$(w) \cup
\{ w + bchain_t + w \}$;    \ \ \ \ \  \hspace{15pt}   // convert $bchain_t$ to a cactus cycle and attach it to $w$

\noindent $\bullet$ \hspace{8pt}
\hspace{20pt}   $bchain_t := \perp$;

\noindent $\bullet$ \hspace{8pt}
\textbf{else}   \hspace{60pt} // absorb the entire $\mathcal{P}$

\hspace{10pt}  $x := head(\mathcal{P})$; \ \ \ \ \ \hspace{150pt} // $head(\mathcal{P}) = x_0 (=w)$

\hspace{10pt}  \textbf{while} $(x.next \neq \perp)$ \textbf{do}
  \hspace{132pt} // $x.next$ exists and is to be absorbed by $w$

\hspace{20pt}   $x := x.next$;  \ \ \ \hspace{152pt}  // get next vertex on $\mathcal{P}$



\noindent $\bullet$ \hspace{8pt}
\hspace{20pt}   \emph{C.cycle}$(w)$ := \emph{C.cycle}$(w) \cup $ \emph{C.cycle}$(x)$;
 \ \ \ \ \  \hspace{53pt}  // Transfer all cactus cycles attached to $x$ to $w$

\noindent $\bullet$ \hspace{8pt}
\hspace{20pt}  \textbf{if} ($x$.\emph{tchain} $\neq \perp$) \textbf{then}
 \ \ \ \ \  \hspace{117pt}  // Convert $x.tchain$ to a cactus cycle

\noindent $\bullet$ \hspace{8pt}
\hspace{35pt}   \emph{C.cycle}$(w)$ := \emph{C.cycle}$(w) \cup
\{ w + x.tchain + w \}$;

\noindent $\bullet$ \hspace{18pt}
\textbf{if} $(x.bchain \neq \perp)$ \textbf{then}
  \hspace{135pt} // $x.bchain$ exists,

\noindent $\bullet$ \hspace{8pt}
\hspace{20pt}   \emph{C.cycle}$(w)$ := \emph{C.cycle}$(w) \cup
\{ w + x.bchain + w \}$;    \ \ \ \ \  \hspace{10pt}   // convert $x.bchain$ to a cactus cycle

\hspace{10pt} \textbf{if} $(\hat{P} \lessdot \overleftarrow{P}(w))$ \textbf{then}
           $\overleftarrow{P}(w) := \hat{P}$;




\noindent \textbf{end}.

\vspace{6pt}

\noindent \textbf{Procedure} \texttt{Absorb-path($w, \mathcal{P}_w, Inc$)}  \ \ \ \ \ // absorb a section of the $w$-path $\mathcal{P}_w$

\noindent \textbf{begin}  \ \ \ //   $\mathcal{P}_w: (w=) w_0 w_1 \ldots w_k, k \ge 1$;
$w_i.next = w_{i+1}, 0 \le i < k$ and
 $w_k.next = \perp$.


 $h := 0$; \ $\hat{w} := w$;

 \textbf{for each} $( (w \curvearrowleft x) \in Inc )$ \textbf{do} \ \ \    \hspace{30pt}     // determine the lowest ancestor $w_h$ of $x$ on $\mathcal{P}_w$

  \hspace{15pt}  \textbf{while} $(\hat{w}.next \neq \perp) \wedge (\hat{w}.next \preceq x))$ \textbf{do} $h := h+1$;
\ $\hat{w} := \hat{w}.next$;



    \textbf{for} $j := 1$ \textbf{step} 1 \textbf{to} $h$ \textbf{do}


 \noindent $\bullet$ \hspace{8pt}
\hspace{15pt}   \emph{C.cycle}$(w)$ := \emph{C.cycle}$(w) \cup $ \emph{C.cycle}$(w_j)$;
 \ \ \ \ \  \hspace{53pt}  // Transfer all cactus cycles attached to $w_j$ to $w$

 \noindent $\bullet$ \hspace{8pt}
\hspace{15pt}  \textbf{if} ($w_j$.\emph{tchain}) $\neq \perp$ \textbf{then}
 \ \ \ \ \  \hspace{117pt}  // Convert $w_j.tchain$ to a cactus cycle

\noindent $\bullet$ \hspace{8pt}
\hspace{30pt}   \emph{C.cycle}$(w)$ := \emph{C.cycle}$(w) \cup
\{ w + w_j.tchain + w \}$;





\noindent \textbf{end}.

\normalsize
\end{singlespacing}

%
%

\begin{lem}\label{CORRECTNESS-cactus}
 Let $w \in V \setminus \{r\}$. When the $\mathit{dfs}$ backtracks from vertex $w$ to its parent vertex, let the $w$-path be $\mathcal{P}_w: w(=w_0) w_1 w_2 \ldots w_k$.

For each cut-edge chain $\mathcal{E}: e_1e_2 \ldots e_{|\mathcal{E}|}$,
where $e_1 =( x_1 \hookrightarrow y_1)$ or $(x_1 \rightarrow y_1)$, $e_i = (x_i \rightarrow y_i); y_i \prec x_i \preceq y_{i-1}, 2 \le i \le |\mathcal{E}|$,  such that
$w \preceq y_2$, either

\vspace{-9pt}
\begin{description}
  \item[$(a)$]
  there exists a cactus cycle $\sigma(x_1) \ldots \sigma(x_{|\mathcal{E}|}) \sigma(x_1)$
 attached to a vertex $x$  such that $\sigma(x) = \sigma(x_1)$, where $x$ is some $w_j, 0 \le j \le k$, or $x_1 = x$ and $x$  has been ejected by the eject-absorb operation; or

\vspace{-6pt}
  \item[$(b)$]
there exists a path
 $\sigma(x_2) \ldots \sigma(x_{\ell})$, where $w \prec x_{\ell} \wedge
(x_{\ell + 1} \preceq  w \vee \ell = |\mathcal{E}|)$,
    attached to some $w_j, 1 \le j \le k,$ as $w_j.tchain$
 such that $w_j = x_1$  if the generator $e_1 $ is a tree-edge, or
   attached to $w_k$ as $w_k.bchian$ such that $\sigma(w_k) = \sigma(y_{\ell})$
 if the generator $e_1$ is a back-edge.
\end{description}

\end{lem}

\vspace{-9pt}
\noindent \textbf{Proof:}
   (By induction on the height of $w$ in $T$)

 If $w$ is a leaf, the assertion vacuously holds true.

 Let $w$ be an internal vertex.
When the $\mathit{dfs}$ backtracks from a child $u$ to $w$, if $deg_{\hat{G}_u}(u) \le 2$,
   \textbf{Procedure} \texttt{Gen}-$\mathcal{CS}$ is invoked.
By the induction hypothesis, the assertion holds for $u$.
By  $(a)$, each cut-edge chain in $G_{\langle \sigma(u) \rangle}$ has been transformed to
a cactus cycle $x Q x$ ($Q$ is a path) which is attached to $u$ or to a vertex $x$ that has been rejected.
   Then for those cycles $x Q x$ attached to $u$ such that $x \neq u$,
since $\sigma(x) = \sigma(u)$,
the \textbf{for} loop replaces the starting and ending vertex $x$ with $u$,
resulting in $uQu$.
   Then, if $deg_{\hat{G}_u}(u) = 1$, a cactus representation of the cut-pairs in
$G_{\langle \sigma(u) \rangle}$ is created.

   If $deg_{\hat{G}_u}(u) = 2$, then either $\{ (w \rightarrow u),(u \rightarrow u_1) \}$ or $\{ (w \rightarrow u),(d \curvearrowleft u) \}$ is a cut-pair.

 In the former case, $(u \rightarrow w)$ is a cut-edge of a cut-edge chain
$\mathcal{E}: e_1 e_2 \ldots e_{\ell} \ldots e_{|\mathcal{E}|}$ such that $e_i = (x_i \rightarrow y_i), 1 \le i \le |\mathcal{E}|,$ and
$e_{\ell} = (x_{\ell} \rightarrow y_{\ell}) = (u \rightarrow w)$.
Then $u_1 = x_1$ and $u_1.tchain = \sigma(x_2) \sigma(x_3) \ldots \sigma(x_{\ell - 1})$,
After executing the \textbf{else} part of the second \textbf{if} statement,
 $u_1.tchain =  \sigma(x_2) \sigma(x_3) \ldots \sigma(x_{\ell})$.
 \textbf{Procedure} \texttt{Absorb-ear} is then invoked.
$\bullet$
If $\hat{P}(w) \lessdot \hat{P}(u)$, $\hat{P}(u)$ terminates at $w$.
Hence,
 $w$ absorbs  $\mathcal{P}_u: u_1 \ldots u_h (h \ge 1)$
 resulting in
$\sigma(w) = \sigma(u_i), 1 \le i \le h$, and
 the cactus cycle attached to $u_i$ are transferred to $w$.
 Moreover,
 as $\sigma(u_i) = \sigma(u_{i-1}), 1 < i \le h,$
 If $u_i.tchain$ is not null,  $u_i.tchain$  and $\sigma(w)$ form a cactus cycle.
  The cycle $w + u_i.tchain + w$ is thus created and attached to $w$.
If $w_h.bchain$ is not null, it is also turned into a cactus cycle $w + w_h.bchain + w$ and
attached to $w$.
$\bullet$
If $\hat{P}(u) \lessdot \hat{P}(w)$,  $\hat{P}(w)$ terminates at $w$, and
vertex $w$ absorbs the $w$-path $\mathcal{P}_w: w_1 \ldots w_k$.
The process is same as that for absorbing $\mathcal{P}_u$ above.

 In the latter case, $(u \rightarrow w)$ is a cut-edge of a cut-edge chain
$\mathcal{E}: e_1 e_2 \ldots e_{\ell} \ldots e_{|\mathcal{E}|}$ such that
$e_1 = (x_1 \hookrightarrow y_1),  e_i = (x_i \rightarrow y_i), 2 \le i \le |\mathcal{E}|,$  and
$e_{\ell} = (x_{\ell} \rightarrow y_{\ell}) = (u \rightarrow w)$.
Since $(b)$ holds for $u$,
$u.bchain = \sigma(x_2) \sigma(x_3) \ldots \sigma(x_{\ell - 1})$.
After executing the \textbf{then} part of the second \textbf{if} statement,
$u.bchain + \sigma(x_{\ell}) = \sigma(x_2) \sigma(x_3) \ldots \sigma(x_{\ell})$ is created
which is stored at $bchain_t$.
 \textbf{Procedure} \texttt{Absorb-ear} is then invoked.
$\bullet$
If $\hat{P}(w) \lessdot \hat{P}(u)$, since $bchain_t$ is not null,
the \textbf{then} part of the first \textbf{if} statement is executed
 and
the cycle $w + bchain_t + w$ is created and attached to $w$.
$\bullet$
If $\hat{P}(u) \lessdot \hat{P}(w)$,  $\hat{P}(w)$ terminates at $w$.
Hence,
vertex $w$ absorbs the $w$-path $\mathcal{P}_w: w_1 \ldots w_k$.
The process is same as that for absorbing $\mathcal{P}_u$ above.
After that $bchain_t$ is transferred to $w.bchain$.

If $deg_{\hat{G}_u}(u) > 2$, then by Lemma~\ref{Tsin09}$(ii)$, there exists $(x \curvearrowleft u)$ with
$x \preceq w$  which implies that $(w \rightarrow u)$ does not form a cut-pair with an edge on $\mathcal{P}_u$.
Hence, $u.tchain = \perp$ and $u.bchain$ remains unchanged.
The remaining argument is same as the above case.

When an $(u \curvearrowleft w)$ is encountered, the argument is same as the above case when $u$ is a child of $w$ but much simpler as $\mathcal{P}_u$ is $nil$ and $bchain_t$ is not involved.

Hence, when $L[w]$ is completely processed, both $(a)$ and $(b)$ hold for $w$.


Finally, if $( \mathcal{P}_w \neq nil ) \wedge ( Inc_w \neq \emptyset )$, then
\textbf{Procedure} \texttt{Absorb-path} is invoked
and vertex $w$ absorbs  $w_i, 1 \le i \le h,$ on $\mathcal{P}_w$.
The process is same as that of absorbing the entire $\mathcal{P}_w$ except that $w_k.bchain$ is not involved if $h = k$.
  Hence, when the $\mathit{dfs}$ backtracks from $w$, the assertion holds for vertex $w$.
\ \ \ \ \  $\blacksquare$

\begin{thm}\label{correctness-cactus-thm}
 \textbf{Algorithm} \texttt{Certifying 3-edge-connectivity} generates a cactus representation of the cut-pairs for
 each  2-edge-connected components of $G$.

\end{thm}

\vspace{-9pt}
\noindent \textbf{Proof:}
 Let the parent edge of $w$ be a bridge. Then $deg_{\hat{G}_w}(w) = 1$.
By Lemma~\ref{CORRECTNESS-cactus}, when the $\mathit{dfs}$ backtracks from $w$ to its parent,
Condition $(a)$ or $(b)$ holds for every cut-pair chain with $w \preceq y_2$.
 Since $deg_{\hat{G}_w}(w) = 1$,
   $\mathcal{P}_w = w$ which implies that $k = 0$.
Hence, only $(a)$ holds for every cut-pair chain in $G_{\langle \sigma(w) \rangle}$.
After the \textbf{for} loop in
 \textbf{Procedure} \texttt{Gen-$\mathcal{CS}$} relabels
 the starting and ending vertex of each cactus cycle attached to $w$ with  $\sigma(w)$,
a cactus representation of the cut-edges in $G_{\langle \sigma(w) \rangle}$ is constructed.

 For the root $r$, when execution of \textbf{Procedure} \texttt{3-edge-connect-CS$(r,\perp)$} terminates,
 $\mathcal{P}_r = r$ which implies that only $(a)$ holds for every cut-pair chain in $G_{\langle \sigma(r) \rangle}$.
Hence, after the last \textbf{for} loop relabels the starting and ending vertex of each cactus cycle attached to $r$ with  $\sigma(r)$,
a cactus representation of the cut-edges in $G_{\langle \sigma(r) \rangle}$ is constructed.
 \ \ \ \ \ $\blacksquare$

\vspace{-6pt}
\begin{thm}
\textbf{Procedure} \texttt{3-edge-connect-CS} runs in $O(|V|+|E|)$ time.
\end{thm}

\vspace{-9pt}
\noindent \textbf{Proof:}
  The time complexity follows from the fact that the new instructions increase the run time of
Procedures \texttt{3-edge-connect-CS}, \texttt{Absorb-ear} and \texttt{Absorb-path}
by a constant factor, and
the total time spent on renaming the starting and ending nodes of the cactus cycles in \textbf{Procedure}  \texttt{Gen-$\mathcal{CS}$} is bounded by the total size of the cactus cycles which is clearly $O(|E|)$.
 \ \ \ \ \  $\blacksquare$

 The two algorithms presented can be easily combined into one
so that the resulting \textbf{Algorithm} \texttt{Certifying-} \texttt{3-edge-connectivity}
 generates the $3ecc$s, the Mader construction sequences for the $3ecc$s, the bridges,  and
a cactus representation of the cut-pairs for each of the 2-edge connected components of $G$ simultaneously in $O(|V|+|E|)$ time by making one pass over the input graph.

\section{Conclusion}

We presented a linear-time certifying algorithm for recognizing 3-edge-connected graphs.
The algorithm does not require the input graph $G$ to be 2-edge-connected and makes only one pass
over $G$ to seamlessly generate the following outputs:
  the $3ecc$s of $G$ each of which is accompanied by a Mader construction sequence serving as a positive certificate, a cactus representation of the cut-pairs of each 2-edge-connected component of $G$, and the bridges.
  Clearly, if $G$ is 3-edge-connected, only one Mader construction sequence is generated and no cactus representations nor bridges are generated.
To verify the certificates, the methods of Mehlhorn et al.~\cite{MNS17} can be used.

In~\cite{NT14}, it is reported that to check the condition `$deg_{G_u}(u) = 2$',
 it is more efficient (in terms of execution time and implementation) to compute the
degrees of the vertices directly
 than to maintain, at each vertex, a list of embodiments of the ears absorbed by the vertex~\cite{T07}.
  In our algorithm, as we must compute $\overleftarrow{P}(u)$ for each vertex $u$, and
 `$deg_{G_u}(u) \le 2$' if and only if `($\overleftarrow{P}(u) = \perp) \vee (s(\overleftarrow{P}(u)) = u)$',
computing the degrees of the vertices is unnecessary.





\begin{singlespacing}
\begin{small}

\end{small}

\newpage

\vspace{12pt}
 \begin{figure}[h!]
\includegraphics[width=6in]{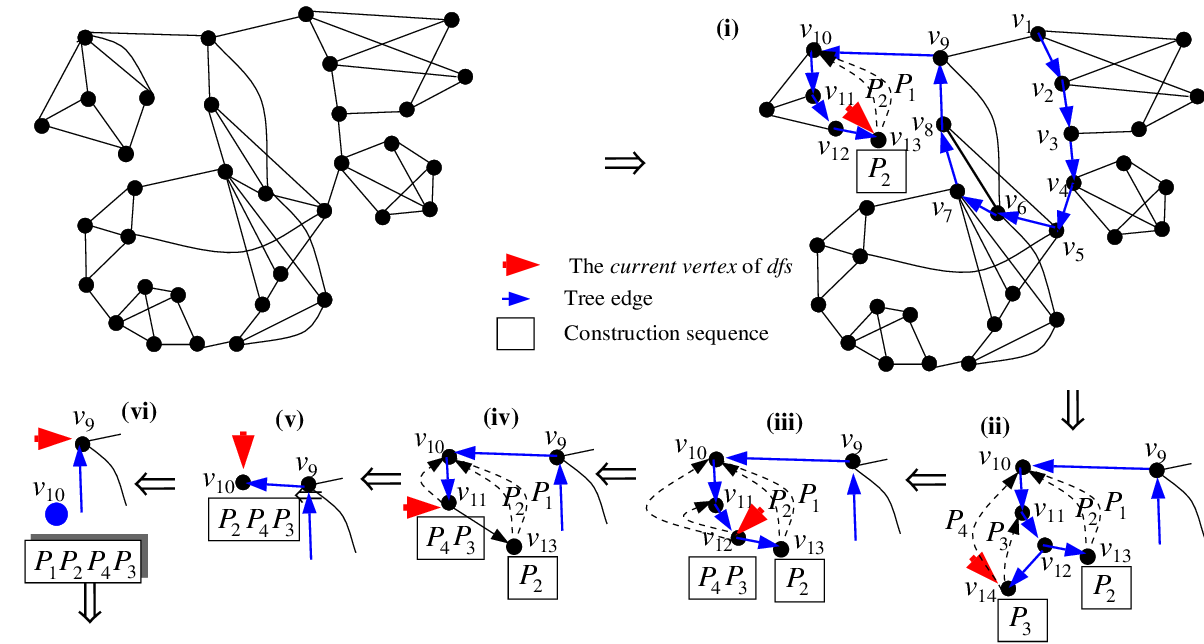}
\caption[ ]{}

\vspace{9pt}
\small
\textbf{(i)}
A $\mathit{dfs}$ starts from $v_1$ and traverses to $v_{13}$ at which:
 $\hat{P}(v_{13}) = P_1$; 
 $\mathcal{CS}_{\delta(v_{13})} = \underline{P_2}$
(\textbf{note:} for clarity, we use $P_i$ to denote $P_{f_i}$ and label the back-edge $f_i$ with $P_i$; we underline $\overleftarrow{P}(w)$ in $\mathcal{CS}_{\delta(w)}$ which is always the first ear).

\textbf{(ii)}
 $\mathit{dfs}$ backtracks to $v_{12}$ and advances to $v_{14}$ at which:
 $\hat{P}(v_{14}) = P_4$;
$\mathcal{CS}_{\delta(v_{14})} = \underline{P_3}$.

\textbf{(iii)}
  $\mathit{dfs}$ backups to $v_{12}$:
 as $P_1 \lessdot P_4$, $v_{12}$ absorbs $v_{14}$ giving $\mathcal{CS}_{\delta(v_{12})} = \underline{P_4} P_3$; $\hat{P}(v_{12}) = P_1$;

\textbf{(iv)}
$\mathit{dfs}$ backups to $v_{11}$:
   as $w_{\ell} = w_h = v_{12}$, $v_{11}$ absorbs $v_{12}$ giving  $\mathcal{CS}_{\delta(v_{11})} = \underline{P_4} P_3$,  $\hat{P}(v_{11}) = P_1$.

\textbf{(v)}
$\mathit{dfs}$  backups to $v_{10}$:
  as $w_{\ell} = w_h = v_{13}$,    $v_{10}$ absorbs $v_{11}, v_{13}$, giving
  $\mathcal{CS}_{\delta(v_{10})} = \underline{P_2} P_4 P_3$; $\hat{P}(v_{10}) = P_1$.

\textbf{(vi)}
$\mathit{dfs}$ backups to $v_{9}$: As $deg_{\hat{G}_{v_{10}}}(v_{10}) = 1$,
$(v_9,v_{10})$ is a bridge and $v_{10}$ is ejected.
 As $\hat{P}(v_{10}) = P_1$, $\hat{P}(v_{10})\mathcal{CS}_{\delta(v_{10})} = P_1 P_2 P_4 P_3$ is a construction sequence of
   $\acute{G}_{\delta(v_{10})}$.

\end{figure}\label{Fig-5-1.eps}

%
 \begin{figure}
\includegraphics[width=6in]{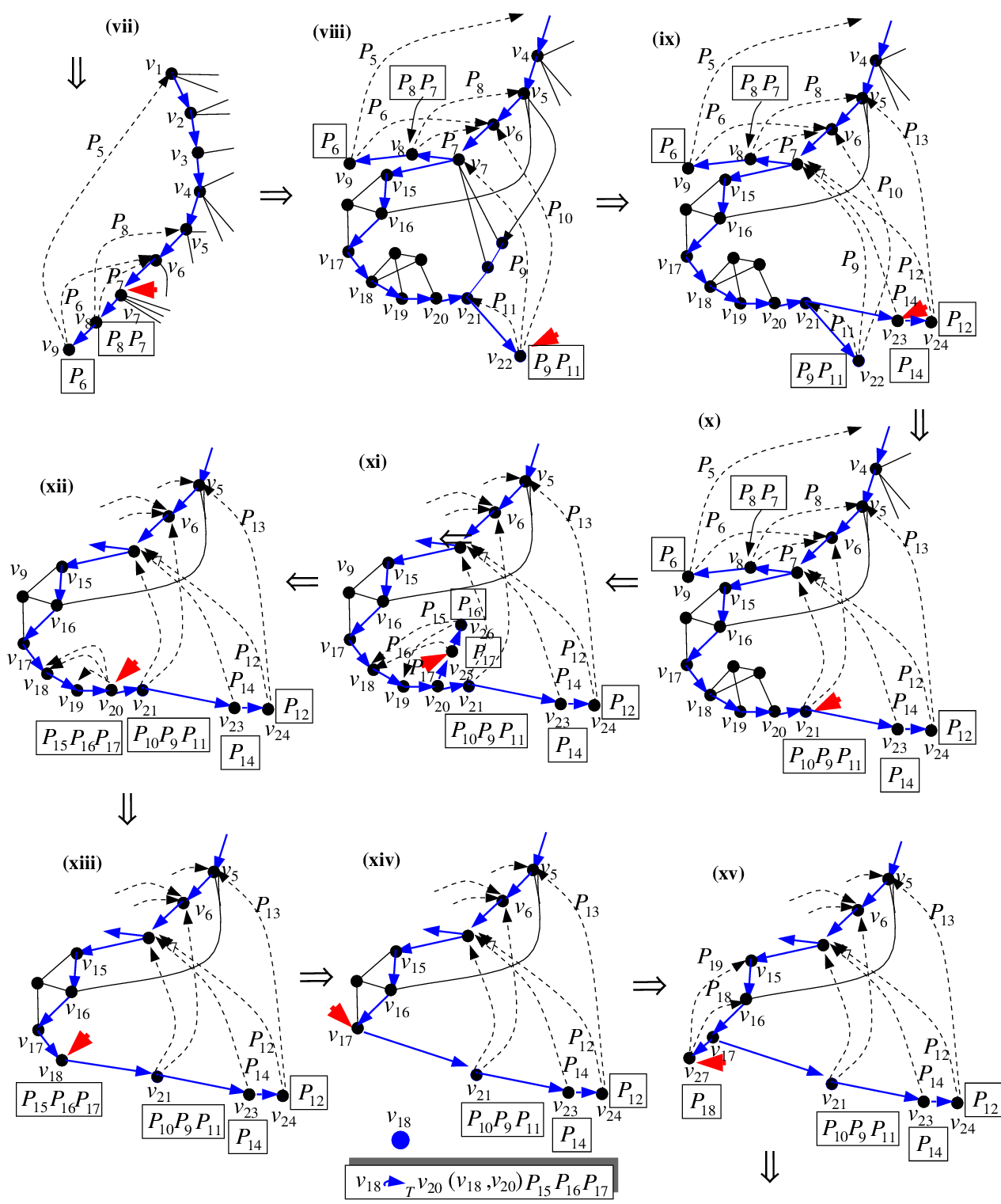}

\vspace{9pt}
\small

\textbf{(vii)}
 At $v_9$, $\mathcal{CS}_{\delta(v_9)} = \underline{P_6}$, $\hat{P}(v_{9}) = P_{5}$.
  The  $\mathit{dfs}$ backups to $v_8$: $\mathcal{CS}_{\delta(v_8)} = \underline{P_8} P_7$,
$\hat{P}(v_{8}) = P_{5}$.
\textbf{(viii)}
      $\mathit{dfs}$ backups to $v_7$ and then advances to $v_{22}$:
     $\hat{P}(v_{22}) = P_{10}; \mathcal{CS}_{\delta(v_{22})} = \underline{P_9} P_{11}$.
\textbf{(ix)}
   $\mathit{dfs}$ backups to $v_{21}$ and then advances to $v_{24}$:
 $\hat{P}(v_{24}) = P_{13}; \mathcal{CS}_{\delta(v_{24})} = \underline{P_{12}}$.
\textbf{(x)}
    $\mathit{dfs}$ backups to $v_{23}$: $\hat{P}(v_{23}) = P_{13}; \mathcal{CS}_{\delta(v_{23})} = \underline{P_{14}}$;
    then backups to $v_{21}$: as $P_{13} \lessdot P_{10}$, $v_{21}$ absorbs $v_{22}$ giving $\mathcal{CS}_{\delta(v_{21})} = \underline{P_{10}} P_9 P_{11};
\hat{P}(v_{21}) = P_{13}$.
\textbf{(xi)}
    $\mathit{dfs}$ backups to $v_{20}$ and advances to $v_{26}$:
  $\hat{P}(v_{26}) = P_{15}; \mathcal{CS}_{\delta(v_{26})} = \underline{P_{16}}$.
    $\mathit{dfs}$ backups to $v_{25}$:
$\hat{P}(v_{25}) = P_{15}; \mathcal{CS}_{\delta(v_{25})} = \underline{P_{17}}$.
\textbf{(xii)}
   $\mathit{dfs}$ backups to $v_{20}$: as $P_{13} \lessdot P_{15}$, $v_{20}$ absorbs $v_{25}, v_{26}$, giving $\mathcal{CS}_{\delta(v_{20})} = \underline{P_{15}} P_{16} P_{17};
\hat{P}(v_{20}) = P_{13}$.
\textbf{(xiii)}
 $\mathit{dfs}$ backups to $v_{19}$:
   as $w_{\ell} = w_h = v_{20}$, $v_{19}$ absorbs $v_{20}$ giving  $\mathcal{CS}_{\delta(v_{19})} = \underline{P_{15}} P_{16} P_{17};
\hat{P}(v_{19}) = P_{13}$.
   Then $\mathit{dfs}$ backups to $v_{18}$:
  as $w_{\ell} = w_h = v_{19}$, $v_{18}$ absorbs $v_{19}$ giving $\mathcal{CS}_{\delta(v_{18})} = \underline{P_{15}} P_{16} P_{17};
\hat{P}(v_{18}) = P_{13}$.
\textbf{(xiv)}
    The $\mathit{dfs}$  backups to $v_{17}$:
   as $deg_{\hat{G}_{v_{18}}}(v_{18}) = 2$,
$v_{18}$ is ejected; as $u = v_{18}$, $\ddot{u} =  parent(v_{21}) = v_{20}$,
 $( v_{18} \rightsquigarrow_T v_{20} ) (v_{18}, v_{20}) P_{15} P_{16} P_{17}$ is a construction sequence of
   $\acute{G}_{\delta(v_{18})}$;
a new (virtual) edge $(v_{17} \rightarrow v_{21})$ is added to the remaining graph,
and $\hat{P}(v_{17}) = P_{13}$.
\textbf{(xv)}
     $\mathit{dfs}$ advances to $v_{27}$; $\hat{P}(v_{27}) = P_{19}$,
$\mathcal{CS}_{\delta(v_{27})} = \underline{P_{18}}$. ~ ~ ~ ~ ~ ~ ~ ~ ~ ~

\end{figure}\label{Fig-5-2.eps}

 \begin{figure}
\includegraphics[width=6in]{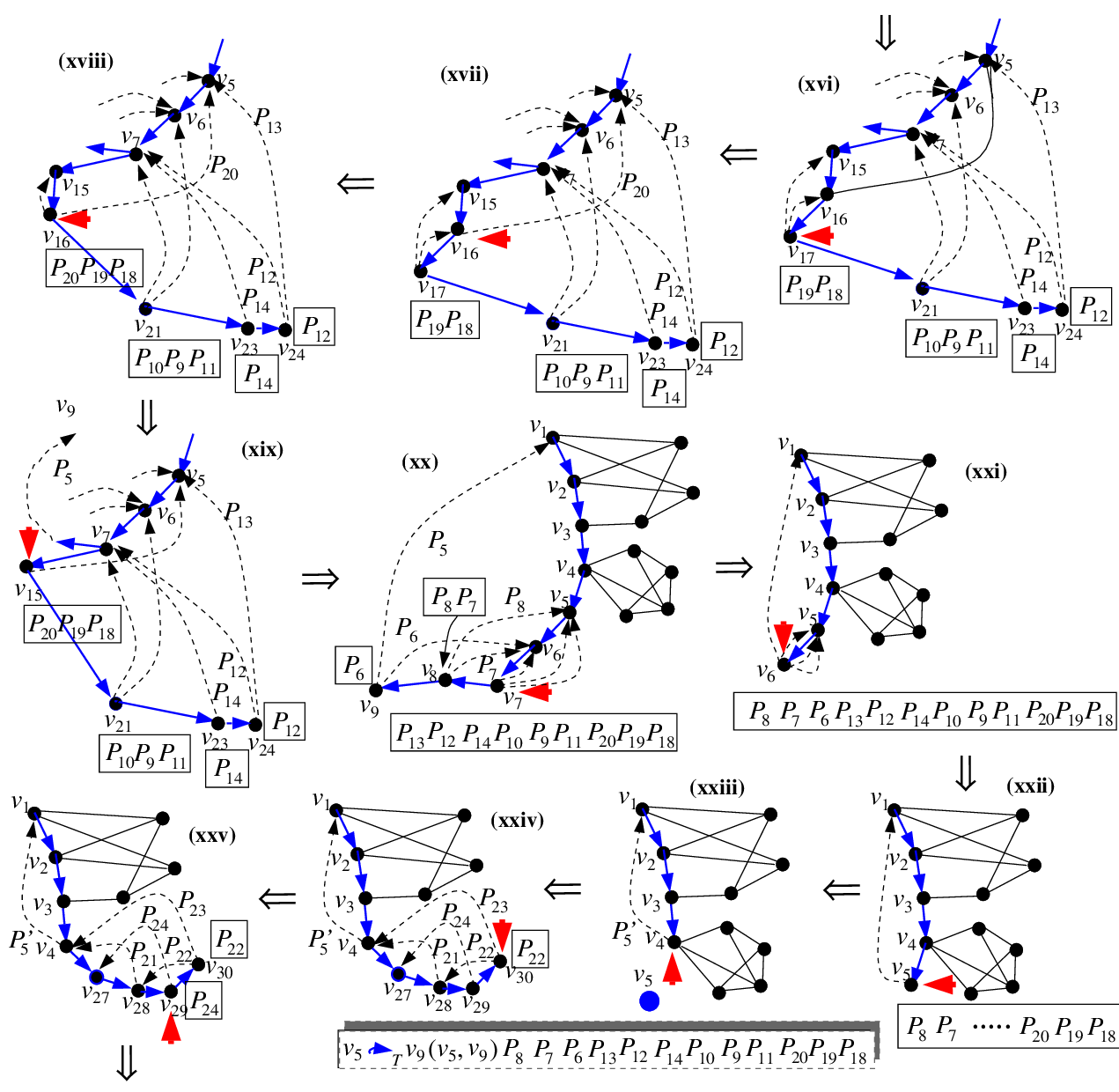}

\small

 \textbf{(xvi)}
$\mathit{dfs}$ backtups to $v_{17}$:
   as $P_{13} \lessdot P_{19}$, $v_{17}$ absorbs $v_{27}$ giving $\mathcal{CS}_{\delta(v_{17})} = \underline{P_{19}} P_{18}$ and
$\hat{P}(v_{17}) = P_{13}$.

\textbf{(xvii)}
     $\mathit{dfs}$  backups to $v_{16}$: 
first, $(v_5 \curvearrowleft v_{16})$ is processed giving $\mathcal{CS}_{\delta(v_{16})} = \underline{P_{20}}$.

\textbf{(xviii)}
Then, the incoming back-edge $(v_{16} \curvearrowleft v_{17})$ is processed.
  As $w_h = v_{17}, w_{\ell} = v_{16}$, $v_{16}$ absorbs $v_{17}$ giving  $\mathcal{CS}_{\delta(v_{16})} = \underline{P_{20}} P_{19} P_{18};
\hat{P}(v_{16}) = P_{13}$.

\textbf{(xix)}
     $\mathit{dfs}$ backups to $v_{15}$: as $w_h = w_{\ell} = v_{16}$,
    $v_{15}$ absorbs $v_{16}$ giving
$\mathcal{CS}_{\delta(v_{15})} = \underline{P_{20}} P_{19} P_{18};
\hat{P}(v_{15}) = P_{13}$.

\textbf{(xx)}
     $\mathit{dfs}$  backups to $v_{7}$: as $P_{5} \lessdot P_{13}$, $v_{7}$ absorbs the $v_{15}$-path:~$v_{15} v_{21} v_{23} v_{24}$, giving
$\mathcal{CS}_{\delta(v_{7})} =
\hat{P}(v_{15}) \mathcal{CS}_{\delta(v_{24})} \mathcal{CS}_{\delta(v_{23})} \mathcal{CS}_{\delta(v_{21})} \mathcal{CS}_{\delta(v_{15})} =
\underline{P_{13}} P_{12} P_{14} P_{10} P_9 P_{11} P_{20} P_{19} P_{18}$ and
$\hat{P}(v_{7}) = P_{5}$.

\textbf{(xxi)}
    $\mathit{dfs}$  backups to $v_{6}$:
as $P_8 = \min_{\lessdot}\{ P_6, P_8, P_{13} \}$,
  $w_{\ell} = v_8, w_h = v_9$, and
  $v_6$ absorbs the path: $v_7 v_8 v_9$ giving
$\mathcal{CS}_{\delta(v_{6})} =
\mathcal{CS}_{\delta(v_{8})} \mathcal{CS}_{\delta(v_{9})} \mathcal{CS}_{\delta(v_{7})} =
\underline{P_8} P_7 P_6 P_{13} P_{12} P_{14} P_{10} P_9 P_{11} P_{20} P_{19} P_{18}$ and
$\hat{P}(v_{6}) = P_{5}$.

\textbf{(xxii)}
    $\mathit{dfs}$ backups to $v_{5}$:
as $w_h = w_{\ell} = v_6$,
 $v_{5}$ absorbs $v_{6}$ giving
$\mathcal{CS}_{\delta(v_5)} = \underline{P_8} P_7 P_6 P_{13} P_{12} P_{14} P_{10} P_9 P_{11} P_{20} P_{19} P_{18}$ and
$\hat{P}(v_{5}) = P_{5}$.

\textbf{(xxiii)}
  $\mathit{dfs}$  backups to $v_4$:
as $deg_{\hat{G}_{v_5}}(v_5) = 2$, $v_5$ is ejected.
  Since $w = v_5$, and  $\hat{P}(v_{5}) = P_{5} = P_{ear(v_4 \rightarrow v_5)} = P_{v_1 \curvearrowleft v_9}$
implies that $\ddot{w} = t(v_1 \curvearrowleft v_9) = v_9$,
 $(v_5 \rightsquigarrow_T v_9) (v_5, v_9) P_8 P_7 P_6 P_{13} P_{12} P_{14} P_{10} P_9 P_{11} P_{20} P_{19} P_{18}$ is a construction sequence of
   $\acute{G}_{\delta(v_{5})}$.
A new (virtual) edge $P_5' = (v_1 \curvearrowleft  v_4)$ is added to the remaining graph and
  $\hat{P}(v_4) = P_5'$.

\textbf{(xxiv)}
  $\mathit{dfs}$ advances to $v_{30}$ during which the back-edges (trivial ears)
 $P_{21}, P_{22}, P_{23}$ are encountered in the listed order.
 Since $P_{23} \lessdot P_{22}$,
 $\hat{P}(v_{30}) = P_{23}$, $\mathcal{CS}_{\delta(v_{30})} = \underline{P_{22}}$.

\textbf{(xxv)}
    $\mathit{dfs}$ backups to $v_{29}$ at which the back-edge (trivial ears)
$P_{24}$ is encountered.
  Since $P_{23} \lessdot P_{24}$,  $\hat{P}(v_{29}) = P_{23}$, $\mathcal{CS}_{\delta(v_{29})} = \underline{P_{24}}$.


\end{figure}\label{Fig-5-3.eps}

\newpage
 \begin{figure}
\includegraphics[width=6in]{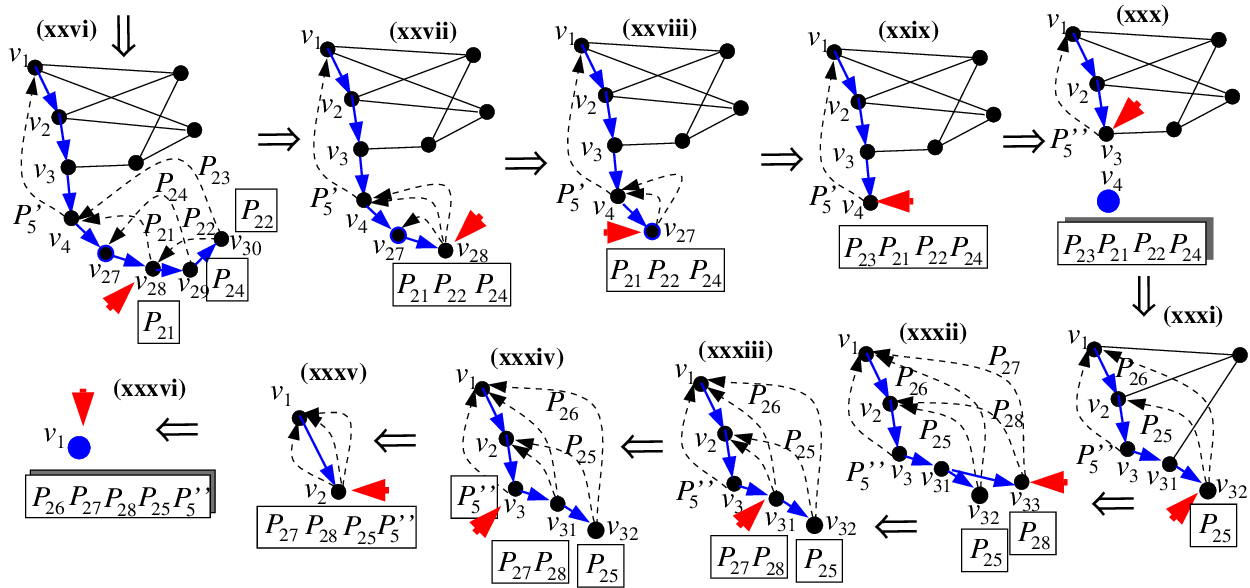}


\small

\textbf{(xxvi)}
    $\mathit{dfs}$ backtracks to $v_{28}$:
Since $P_{23} \lessdot P_{21}$,
$\hat{P}(v_{28}) = P_{23}$, $\mathcal{CS}_{\delta(v_{28})} = \underline{P_{21}}$.

\textbf{(xxvii)}
Then, as $P_{21} = \min_{\lessdot}\{ P_{21}, P_{22}, P_{24} \}$,
$w_{\ell} = v_{28}, w_h = v_{30}$, and
$v_{28}$ absorbs the $v_{28}$-path: $v_{28} v_{29} v_{30}$  giving
$\mathcal{CS}_{\delta(v_{28})} = \underline{P_{21}} P_{22} P_{24}$
 and  $\hat{P}(v_{28}) = P_{23}$.

\textbf{(xxviii)}
    $\mathit{dfs}$ backups to $v_{27}$: As $w_{\ell} = w_h = v_{28}$,
$v_{27}$ absorbs the $v_{27}$-path: $v_{27} v_{28}$ giving
 $\mathcal{CS}_{\delta(v_{27})} = \underline{P_{21}} P_{22} P_{24}$ and
$\hat{P}(v_{27}) = P_{23}$.

\textbf{(xxix)}
   The $\mathit{dfs}$ backups to $v_{4}$: as $(\hat{P}(v_4) =) P_5' \lessdot P_{23} (= \hat{P}(v_{27}))$,
 $v_4$ absorbs $v_{27}$ giving
$\mathcal{CS}_{\delta(v_{4})} = \hat{P}(v_{27})\mathcal{CS}_{\delta(v_{27})} =
  \underline{P_{23}} P_{21} P_{22} P_{24}$.

\textbf{(xxx)}
     $\mathit{dfs}$ backups to $v_{3}$:
     as $deg_{\hat{G}_{v_{4}}}(v_{4}) = 2$,
$v_{4}$ is ejected. As $u = \ddot{u} = v_{4}$,
 $\mathcal{CS}_{\delta(v_{4})} = P_{23} P_{21} P_{22} P_{24}$ is a construction sequence of
   $\acute{G}_{\delta(v_{4})}$.
A new edge $P_5'' = (v_{1} \curvearrowleft  v_{3})$ is added to the remaining graph, and
  $\hat{P}(v_3) = P_5''$.

\textbf{(xxxi)}
   $\mathit{dfs}$ advances to $v_{32}$ at which $P_{25}, P_{26}$ are encountered: $\hat{P}(v_{32}) = P_{26}$, $\mathcal{CS}_{\delta(v_{32})} = \underline{P_{25}}$.

\textbf{(xxxii)}
   $\mathit{dfs}$ backups to $v_{31}$     at which $\hat{P}(v_{31}) = P_{26}$
 and then advances to $v_{33}$
 at which $P_{27}, P_{28}$ are encountered: $\hat{P}(v_{33}) = P_{27}$, $\mathcal{CS}_{\delta(v_{33})} = \underline{P_{28}}$.

\textbf{(xxxiii)}
     $\mathit{dfs}$ backups to $v_{31}$:
 as $(\hat{P}(v_{31}) = ) P_{26} \lessdot P_{27}  (= \hat{P}(v_{33}))$, $v_{31}$ absorbs $v_{33}$ giving
$\mathcal{CS}_{\delta(v_{31})} = \hat{P}(v_{33}) \mathcal{CS}_{\delta(v_{33})} =
\underline{P_{27}} P_{28}$, and $\hat{P}(v_{31}) = P_{26}$.

\textbf{(xxxiv)}
   $\mathit{dfs}$ backups to $v_{3}$:
as $(\hat{P}(v_{31}) =) P_{26} \lessdot P_5'' (= \hat{P}(v_3))$,
$\mathcal{CS}_{\delta(v_{3})}  = \underline{P_5''}$ and
 $\hat{P}(v_3) = P_{26}$.

\textbf{(xxxv)}
  $\mathit{dfs}$ backups to  $v_{2}$:
   as $P_{27} = \min_{\lessdot} \{P_{25}, P_{27}, P_5''\}$,
   $v_{\ell} = v_{31}, v_h = v_{32}$, and
$v_2$ absorbs the path $v_2 v_3 v_{31} v_{32}$ giving
$\mathcal{CS}_{\delta(v_{2})}  = \underline{P_{27}} P_{28} P_{25} P_5''$, and
$\hat{P}(v_2) =  P_{26}$.

\textbf{(xxxvi)}
   $\mathit{dfs}$ backups to $v_{1}$: $\hat{P}(v_1) = P_{26}$.
As $w_h = v_{\ell} = v_{2}$,
$v_1$ absorbs the path $v_1 v_2$ giving
$\mathcal{CS}_{\delta(v_{1})} = \underline{P_{27}} P_{28} P_{25} P_5''$.
Since $v_1 = r$,
 $\hat{P}(v_1) \mathcal{CS}_{\delta(v_{1})} = P_{26} P_{27} P_{28} P_{25} P_5''$
is a Mader construction sequence of
   $\acute{G}_{\delta(v_{1})}$.  \ $\square$

\normalsize

 \end{figure}\label{Fig-5-4.eps}

\newpage

 \begin{figure}[h!]
\includegraphics[width=6.5in]{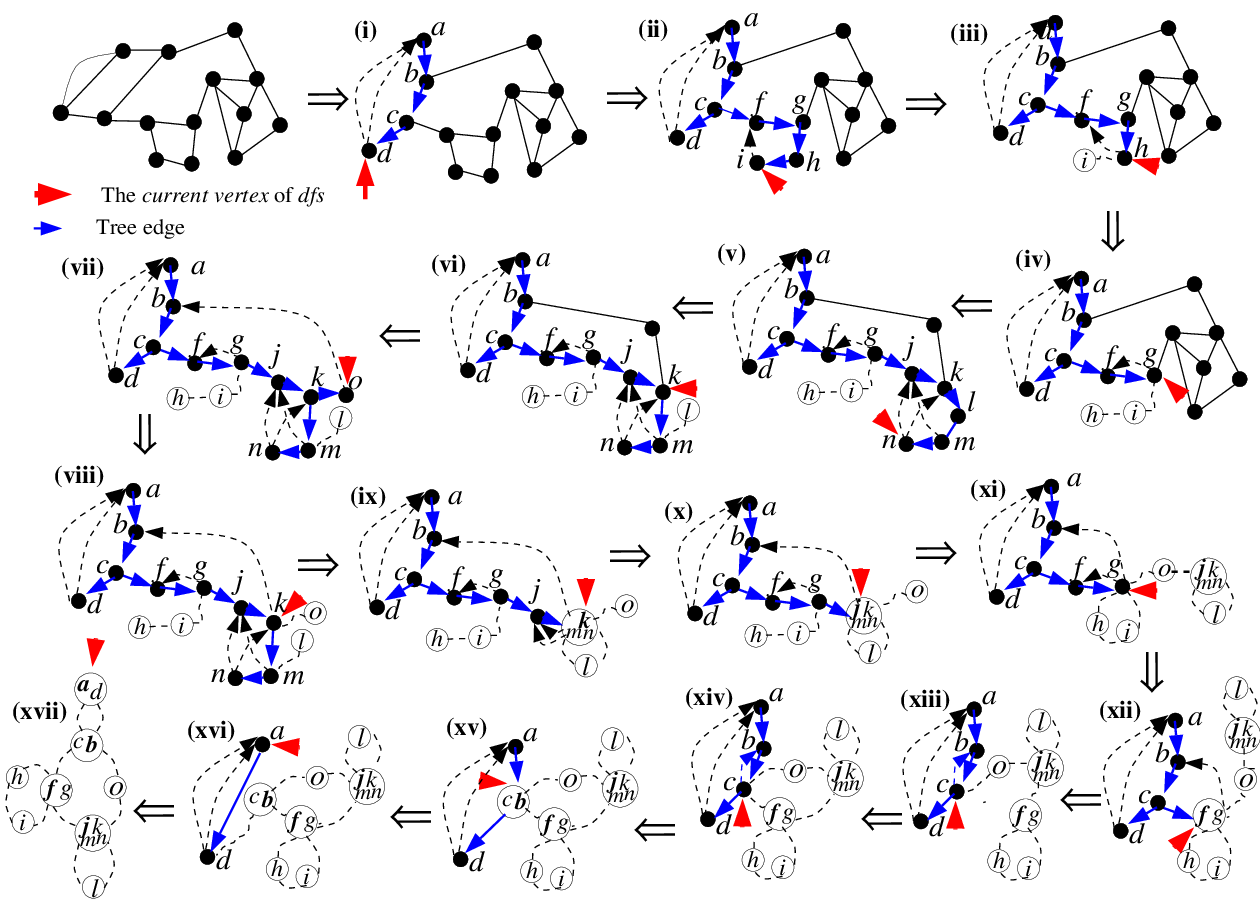}

 \caption[ ]{}

\vspace{9pt}

\textbf{(i), (ii)}
$\mathit{dfs}$ starts from $a$, advances to $d$,
then
backtracks to $c$ and advances to $i$,

\textbf{(iii)}
$\mathit{dfs}$ backtracks from $i$ to  $h$:  eject $i$, $h.bchain$:$\{i\}$
   (note: \{~\} denotes $\sigma()$).

\textbf{(iv)}
$\mathit{dfs}$ backtracks to $g$: eject $h$, $g.bchain$:$\{i\}$-$\{h\}$.

\textbf{(v)}
   $\mathit{dfs}$ advances to $n$.

\textbf{(vi)}
$\mathit{dfs}$ backtracks to $k$: eject $\ell$, $m.tchain$: $\{\ell\}$.

\textbf{(vii), (viii)}
   $\mathit{dfs}$ advances to $o$, then backtracks to $k$: eject $o$, $k.bchain$: $\{o\}$.

\textbf{(ix)}
 Vertex $k$ absorbs $m$-path, $\sigma(k) = \{k,m,n\}$, cycle $\{k\}$-$\{\ell\}$-$\{k\}$ attached to $k$.

\textbf{(x)}
   $\mathit{dfs}$  backtracks to $j$, $j$ absorbs $k$, $\sigma(j)=\{j,k,m,n\}$,
  attach cycle $\{k\}$-$\{\ell\}$-$\{k\}$ to $j$,
 $j.bchain$: $\{o\}$.

\textbf{(xi)}
   $\mathit{dfs}$  backtracks to $g$: eject $j$,
rename cycle $\{k\}$-$\{\ell\}$-$\{k\}$ attached to $j$ as $\{j\}$-$\{\ell\}$-$\{j\}$,
$bchain_t$: $\{o\}$-$\{j\}$.
Since $\hat{P}(j) \lessdot \hat{P}(g)$,
convert $g.bchain$ into cycle $\{g\}$-$\{i\}$-$\{h\}$-$\{g\}$, transfer $bchain_t$ to $g.bchain$: $\{o\}$-$\{j\}$.

\textbf{(xii)}
   $\mathit{dfs}$  backtracks to $f$: $f$ absorbs $g$,
$\sigma(f) = \{f,g\}$, $f.bchain:= \{o\}$-$\{j\}$,
  attach cycle $\{g\}$-$\{i\}$-$\{h\}$-$\{g\}$ to $f$.

\textbf{(xiii)}
   $\mathit{dfs}$  backtracks to $c$: eject $f$, rename cycle $\{g\}$-$\{i\}$-$\{h\}$-$\{g\}$ attached to $f$ as $\{f\}$-$\{i\}$-$\{h\}$-$\{f\}$, $bchain_t: \{o\}$-$\{j\}$-$\{f\}$.

\textbf{(xiv)}
 Since $\hat{P}(c) \lessdot \hat{P}(f)$,
convert $bchain_t$ into cycle  $\{c\}$-$\{o\}$-$\{j\}$-$\{f\}$-$\{c\}$ and attach to $c$.

\textbf{(xv)}
   $\mathit{dfs}$  backtracks to $b$: $b$ absorbs $c$, $\sigma(b) = \{b,c\}$, attach
cycle  $\{c\}$-$\{o\}$-$\{j\}$-$\{f\}$-$\{c\}$ to $b$.

\textbf{(xvi)}
   $\mathit{dfs}$  backtracks to $a$: ejects $b$,
rename cycle  $\{c\}$-$\{o\}$-$\{j\}$-$\{f\}$-$\{c\}$ attached to $b$  as
$\{b\}$-$\{o\}$-$\{j\}$-$\{f\}$-$\{b\}$,  $d.tchain$: $\{b\}$.

\textbf{(xvii)}
Then $a$ absorbs $d$, $\sigma(a)=\{a,d\}$, convert $d.tchain$ into cycle $\{a\}$-$\{b\}$-$\{a\}$.
   The cactus representation of the cut-pair consists of cycles:
$\{a\}$-$\{b\}$-$\{a\}$,
$\{b\}$-$\{o\}$-$\{j\}$-$\{f\}$-$\{b\}$,
$\{f\}$-$\{i\}$-$\{h\}$-$\{f\}$, and
$\{j\}$-$\{\ell\}$-$\{j\}$. $\square$
 ~ ~ ~ ~ ~ ~ ~ ~ ~ ~ ~ ~ ~ ~ ~ ~ ~ ~ ~ ~ ~  ~ ~


 \end{figure}\label{Fig8.eps}

\end{singlespacing}
\end{document}